\documentclass[11pt,letterpaper]{article}
\usepackage[T1]{fontenc}
\usepackage{lmodern,amsmath,amssymb,amsthm,mathtools,booktabs,array,microtype}
\usepackage[margin=1.02in]{geometry}
\usepackage{xcolor,hyperref}
\definecolor{linkink}{RGB}{15,63,89}
\hypersetup{colorlinks=true,linkcolor=linkink,citecolor=linkink,urlcolor=linkink,
 pdftitle={From Partial Access to Coherence-Weighted Quantum Graphs},
 pdfauthor={Frank Barrows},bookmarksnumbered=true}
\newtheorem{theorem}{Theorem}[section]
\newtheorem{proposition}[theorem]{Proposition}
\newtheorem{lemma}[theorem]{Lemma}
\newtheorem{corollary}[theorem]{Corollary}
\theoremstyle{definition}
\newtheorem{example}[theorem]{Example}
\newcommand{\A}{\mathcal A}
\newcommand{\B}{\mathcal B}
\newcommand{\N}{\mathcal N}
\newcommand{\M}{\mathcal M}
\newcommand{\HH}{\mathcal H}
\newcommand{\LL}{\mathcal L}
\newcommand{\RR}{\mathcal R}
\newcommand{\Tr}{\operatorname{Tr}}
\newcommand{\tr}{\operatorname{tr}}
\newcommand{\id}{\operatorname{id}}
\newcommand{\rank}{\operatorname{rank}}
\newcommand{\ran}{\operatorname{ran}}
\newcommand{\Span}{\operatorname{span}}
\newcommand{\supp}{\operatorname{supp}}
\newcommand{\spec}{\operatorname{spec}}
\newcommand{\Ad}{\operatorname{Ad}}
\newcommand{\MD}{\operatorname{MD}}
\newcommand{\ev}{\operatorname{ev}}
\newcommand{\norm}[1]{\left\lVert #1\right\rVert}
\newcommand{\ket}[1]{\lvert #1\rangle}
\newcommand{\bra}[1]{\langle #1\rvert}
\newcommand{\ff}{\mathfrak f}
\newcommand{\Ws}{W^{\mathrm s}}
\newcommand{\Wr}{W^{\mathrm r}}
\newcommand{\Ds}{\Delta^{\mathrm s}}
\newcommand{\Dr}{\Delta^{\mathrm r}}
\newcommand{\one}{\mathbf 1}
\newcommand{\CZ}{\mathbb C}
\newcommand{\ZZ}{\mathbb Z}\newcommand{\EE}{\mathbb E}
\newcommand{\Talg}{\mathcal T_\alpha}

\newcommand{\cov}{\operatorname{Cov}}
\newcommand{\Rea}{\operatorname{Re}}
\newcommand{\defeq}{\mathrel{:=}}
\usepackage{xcolor}
\usepackage[most]{tcolorbox}
\usepackage{comment}

\usepackage{xparse}

\definecolor{fbcolorA}{RGB}{170,0,0}
\definecolor{fbbackA}{RGB}{255,235,235}

\definecolor{fbcolorB}{RGB}{0,90,170}
\definecolor{fbbackB}{RGB}{235,245,255}

\definecolor{fbcolorC}{RGB}{0,125,70}
\definecolor{fbbackC}{RGB}{235,250,240}

\definecolor{fbcolorD}{RGB}{130,70,160}
\definecolor{fbbackD}{RGB}{245,238,250}

\newcounter{fbdepth}

\newcommand{\fblinecolor}{%
  \ifcase\value{fbdepth}%
    fbcolorA%
  \or fbcolorA%
  \or fbcolorB%
  \or fbcolorC%
  \else fbcolorD%
  \fi
}

\newcommand{\fbbackcolor}{%
  \ifcase\value{fbdepth}%
    fbbackA%
  \or fbbackA%
  \or fbbackB%
  \or fbbackC%
  \else fbbackD%
  \fi
}

\newcommand{\add}[1]{\textcolor{blue}{#1}}
\newcommand{\adef}{\mathrel{\textcolor{blue}{:=}}}

\definecolor{cutorange}{RGB}{214,96,0}
\newcommand{\plainarg}[1]{#1}
\newcommand{\plainadef}{\mathrel{:=}}
\DeclareRobustCommand{\cut}[1]{{\color{cutorange}\let\add\plainarg\let\adef\plainadef #1}}
\makeatletter\newcommand{\affiliation}[1]{\if\relax\detokenize{#1}\relax\else\expandafter\gdef\expandafter\@author\expandafter{\@author\\[2pt]{\small #1}}\fi}\makeatother
\title{From Partial Access to Coherence-Weighted Quantum Graphs}
\author{Frank Barrows\thanks{Email: \href{mailto:fbarrows@lanl.gov}{fbarrows@lanl.gov}}}
\affiliation{Theoretical Division, Los Alamos National Laboratory, Los Alamos, New Mexico 87545, USA}
\date{}
\begin{document}
\maketitle
\begin{abstract}
The information acquired by an environment constrains the operations and relations that remain accessible
in a quantum system. We describe this connection by introducing a coherence-weighted quantum graph, whose vertices are sectors, orthogonal subspaces, and whose edges are partial isometries between them:  their products specify source, range, and composition, and the products of evolved edges  define positive operator weights, which measure the coherence each edge retains. For sectors whose projections remain fixed by the dynamics, the largest remaining edge weight equals the squared maximal root fidelity of the conditional environment records.  
A family of jumps shared between sectors realizes these fidelities through classical, time-resolved jump records, so the same record fidelities determine both the remaining edge amplitude and the failure of a path relation.  We derive an operator multiplication defect as  a conditional squared prediction error, and the Lindbladian semigroup dynamics from repeated partial access  to independently prepared environments. The algebra of surviving products then constrains every possible edge with given endpoints.   Record fidelities bound the finite-time weights on the unmatched directions for every implementing edge, while a mismatch of retained ranks forces norm-one asymptotic defects for every implementing edge. The weights thereby connect environmental distinguishability to the algebraic relations that can survive decoherence.

\end{abstract}
\begingroup
\small
\tableofcontents
\endgroup
\clearpage

\section{Introduction}\label{sec:intro}
A classical description of the world contains more than a list of states. It contains observables that can be read repeatedly, relations between configurations, and operations on those configurations. To describe the classical world left after decoherence, we therefore ask which of these relations survive the open evolution of a quantum system.

We formulate this question in the observable algebra. A resolution of the identity into physically distinguished projections specifies sectors, the vertices of a graph, and partial isometries between their subspaces specify directed edges. The products of an edge with its adjoint give its source and range projections; products of edges describe paths. This organizes the
operations and their relations as a quantum graph. An automorphism preserves these products, although it may move the sectors themselves. Restricting access to part of the algebra can change this: the reduced evolution need not preserve products.

Environment-induced decoherence  suppresses interference, this suppression can be related to information acquired by the environment and to the selection of stable pointer observables~\cite{Zurek2003}.   Visibility--distinguishability relations quantify this connection~\cite{Englert1996}.
Existing algebraic theory of decoherence identifies  the observables on which the open dynamics remain multiplicative, and their separation from decaying modes, as the decoherence-free algebra
\cite{doi:10.1007/s00023-012-0199-3,doi:10.1142/S0129055X16500033,SassoUmanita2021}, developed for general semigroups by Dhahri, Fagnola and Rebolledo~\cite{DhahriFagnolaRebolledo2010}, building on Evans~\cite{Evans1977}. The same structure appears in quantum information. For a unital channel on densities the noiseless subsystems are read off from the commutant of the algebra generated by the Kraus operators~\cite{KnillLaflammeViola2000,HolbrookKribsLaflamme2003,ChoiKribs2006}, and the multiplicative domain gives the codes that a unitary can correct~\cite{doCJK}.

A second line of work treats relations themselves as noncommutative objects. Weaver's quantum relations~\cite{Weaver} are weak\(^*\)-closed bimodules over a commutant, and the noncommutative graphs of Duan, Severini and Winter~\cite{DuanSeveriniWinter2013} are operator systems. In this work, we use a quantum graph that is compatible with these notions: the span of the edges between two sectors is a quantum relation. What we add is the edge itself rather than its span, because open dynamics acts on edges one at a time and gives each of them operator weights.

The link between what a system keeps and what its environment learns is the subject of complementarity in quantum information. A channel can be approximately reversed exactly when its complementary channel reveals little~\cite{BenyOreshkov2010}, and the information--disturbance trade-off relates how far a channel is from reversible to what its complementary channel reveals~\cite{KretschmannSchlingemannWerner};  Theorem~\ref{thm:information_weights} gives an exact identity of this kind for coherence between sectors using operator algebras. 
Closest to our setting are channels that preserve a decomposition into subspaces, \AA berg's gluings of completely positive maps~\cite{Aberg2004} are the maps on a direct sum of subspaces with prescribed restrictions to the summands; what the restrictions leave free is a matrix in the off-diagonal blocks, which sets the coherence between the subspaces. For two paths with an internal degree of freedom, Oi and \AA berg~\cite{OiAberg2006} show that the largest interference visibility, over all subspace-preserving operations that prepare given internal states in the two paths, is the Uhlmann fidelity of those internal states.
Our aim is to connect these descriptions to notions of observable and accessible subalgebras at the level of individual operations and their products. Our bridge is quantum graphs, with edges representing intersector coherences. The operator weights will measure remaining coherence and the information about the sector held by the environment, while the retained algebra will constrain which quantum relations can survive.

Our starting point is the observation that an environment can acquire a record of the sector occupied by the system.   Prepare a joint state
\(\Psi=c_v\ket v\otimes e_v+c_w\ket w\otimes e_w\), where the system vectors \(\ket v\), \(\ket w\) lie in two different sectors and the normalized environment vectors are  \(e_v\), \(e_w\). The expectation of the system operator \(S=\ket v\bra w\) is
\begin{equation}
 \langle\Psi,(S\otimes I)\Psi\rangle  =c_v^{*}c_w\langle e_v,e_w\rangle .
\end{equation}
 If the environment vectors are orthogonal, any interference between the two paths vanishes, and if they agree its full amplitude remains \cite{Englert1996,Zurek2003}.

For the  lowering operator \(S=\ket g\bra e\) of a two-level system with ground state \(\ket g\) and excited state \(\ket e\), the products \(S^*S=\ket e\bra e\) and \(SS^*=\ket g\bra g\) specify the source and range for this map.

Now suppose the system interacts with an environment while only system observables are accessible. Write \(\Phi_t\) for their Heisenberg evolution, the observable obtained by transporting the original source  is \(\Phi_t(S^*S)\), while the source product  is \(\Phi_t(S)^*\Phi_t(S)\). In a system state \(\rho\) their predictions differ by
\begin{equation}
\Tr\!\left[\rho\bigl(\Phi_t(S^*S)-\Phi_t(S)^*\Phi_t(S)\bigr)\right].
\end{equation}
We ask first how this difference is produced, and ultimately whether any operation with the same two initial endpoints can preserve their relation.

The information role of the operator weights, \(\Phi_t(S)^*\Phi_t(S)\), is the central result.
Theorem~\ref{thm:information_weights} equates the largest remaining coherence between two preparations with the fidelity of their conditional environment records, its proof uses the purification characterization of fidelity \cite{doi:10.1016/0034-4877(76)90060-4} and the cross-operator
identity of weak decoupling duality~\cite[Lemma~20]{HaydenWinter2012}. For a single qubit, the coherence that survives is given by the overlap of the two conditional states of the environment.  Previous work by Oi and \AA berg~\cite{OiAberg2006} fixes the internal states and optimizes the operation, whereas  here the dynamics is given, the edge is optimized, and the fidelity is that of the records held by the environment. What is new is the statement for edges. The largest record fidelity  is  the norm of the reduced dynamics on a corner of the algebra, and its square is the largest weight that any edge between the two sectors retains, hence an upper bound on the weight of every edge; the operator weight of a given edge gives the retained squared amplitude for each preparation. Corollary~\ref{prop:shared_jump_weight} shows how the loss of coherence of an individual edge depends on what the environment can distinguish within the sectors.

The relations preserved in full at all times form a retained algebra $\N$. Under a faithful stationary state, this algebra contains all nondecaying modes, as in the established decoherence decomposition \cite{doi:10.1063/1.5030954,SassoUmanita2021}.
Amato, Facchi and Konderak further clarify the relation between multiplicative and spectral asymptotics through the Choi--Effros product~\cite{AmatoFacchiKonderak2026}. Our multiplication defects, the differences \(\Phi_t(X^*Y)-\Phi_t(X)^*\Phi_t(Y)\), follow the original operator products at finite times, so they retain the source, range and composition of the physical edges throughout the evolution. When $\N$ is commutative, the dynamics converge to a conditional expectation onto a pointer algebra. Matrix blocks in $\N$ instead can retain quantum coherence.

Two endpoints in \(\N\) can have equal rank in the full algebra and different ranks in the central summands of \(\N\). Edges between them then exist, but no edge between them is preserved, and Theorem~\ref{thm:sharp_endpoint_defects} gives the least loss that each must suffer. Theorem~\ref{thm:finite_record_obstruction} connects this dimension mismatch to the information result at finite times; it therefore relates a constraint on all possible implementations to the information actually acquired by the environment. At large times, again for a faithful stationary state, the record fidelities between retained blocks of \(\N\) vanish, and a mismatch of retained ranks forces a norm-one asymptotic source and range defect for every edge between the two endpoints (Corollary~\ref{cor:asymptotic_loss}).

 The weights in the coherence-weighted quantum graphs carry the information: Theorem~\ref{thm:information_weights} and Corollary~\ref{prop:shared_jump_weight} express them through the fidelity of the records held by the environment. The retained algebra then constrains every edge with given endpoints, through the sharp bounds of Theorem~\ref{thm:sharp_endpoint_defects}.

To derive this connection, we begin with partial access to a larger algebra undergoing reversible evolution: only a subalgebra of its observables is accessible. A conditional expectation gives the best accessible prediction of an evolved observable; this prediction is the reduced evolution of an open quantum system, and Theorem~\ref{thm:partial_access} identifies its conditional squared error with the failure to preserve multiplication. A repeated interaction with fresh environments supplies the Markov limit used for the dynamical results \cite{AttalPautrat}. Decoherence is thereby derived in the observable algebra from partial access alone. 

Except where we explicitly pass to finite von Neumann algebras or other normal representations,
the system is finite-dimensional (Theorem~\ref{thm:sharp_endpoint_defects}, Proposition~\ref{prop:retained_accumulation} and Supplement~\ref{supp:von_neumann_extensions}). We use the Heisenberg picture, and write \(X^*\) for the adjoint.

Section~\ref{sec:setup} constructs the reduced dynamics, weights, and multiplication defects. Section~\ref{sec:representations} identifies their information content and computes the classical records of the common-reservoir model.
Section~\ref{sec:asymptotic_algebra} derives the dissipation balance and the retained algebra. Section~\ref{sec:sharp_loss} gives the sharp comparison and the finite-time record bound, and ends by reading the defects at all times from one map on trajectories (Section~\ref{sec:sharp_accumulated_dissipation}, with the construction and its consequences in Supplement~\ref{supp:initial_nonsplitting}). An example common-reservoir model is followed through these steps.

\section{Partial access, operator weights, and relations}\label{sec:setup}
We first specify the relations whose evolution we will follow. Sector projections resolve the identity: projective measurements read the sector, and the projections are the vertices of the graph. An edge maps one sector into another, and its products specify the input subspace, output subspace, and compositions with other edges. Examining the operators themselves will let their changing amplitudes carry information about environmental records. We then derive the reduced dynamics from partial access, define the operator weights, and introduce the model that is followed through the later sections.

\subsection{Sectors and edge products}\label{sec:sectors}
Let \(\HH\) be a \(d\)-dimensional complex Hilbert space and \(\A=B(\HH)=M_d(\CZ)\), the algebra of bounded operators on the Hilbert space $\HH$. A state is a positive normalized linear functional \(\omega\) on \(\A\), of  the form \(\omega(X)=\Tr(\rho X)\), with \(\rho\geq0\) and \(\Tr\rho=1\); it is faithful precisely when \(\rho\) is invertible.  Supplement~\ref{supp:states_polar} gives the coordinate derivation and the polar decomposition used in the information proof.

Choose a partition of unity by nonzero orthogonal projections \(p_v\), with
\begin{equation}
p_vp_w=\delta_{vw}p_v,\qquad \sum_vp_v=I,\qquad
\HH_v=p_v\HH. \label{eq:sector_resolution}
\end{equation}
The subspaces \(\HH_v\) are the sectors, the vertices of the quantum graph, with sector (vertex) projections \(p_v\). For an input \(y\), the operator \(p_vXp_w\) first takes the component of \(y\) in \(\HH_w\), applies \(X\), and keeps the component of the result in \(\HH_v\). Its restriction is an arbitrary linear map \(\HH_w\to\HH_v\). 
Conversely every such map becomes an element of \(p_v\A p_w\) by extending it by zero on \(\HH_w^\perp\). We call \(p_v\A p_w\) the corner space from sector \(w\) to sector \(v\). Inserting the resolution of identity on both sides and then multiplying two blocks gives
\begin{align}
X&=\Big(\sum_vp_v\Big)X\Big(\sum_wp_w\Big)=\sum_{v,w}p_vXp_w,\notag\\
(p_uXp_v)(p_zYp_w)&=p_uX(p_vp_z)Yp_w =\delta_{vz}p_uXp_vYp_w,\notag\\
(p_vXp_w)^*&=p_wX^*p_v. \label{eq:corner_operations}
\end{align}
The middle indices give the composition rule. The adjoint reverses the direction of the operation.

For a map \(S\in p_v\A p_w\), suppose \(p=S^*S\) is a projection. If \(y\in p\HH\), then \(\norm{Sy}^2=\langle
y,py\rangle=\norm y^2\). If \(y\perp p\HH\), then \(\norm{Sy}^2=0\). Thus \(S=Sp\) and \(q=SS^*\) satisfies \(q^2=S(S^*S)S^*=SS^*=q\). Such an \(S\) is a partial isometry, and an edge from sector \(w\) to sector \(v\) is a partial isometry in \(p_v\A p_w\). Its source and range projections \(p,q\) satisfy \(p\leq p_w\), \(q\leq p_v\).  For example, the matrix
\begin{equation}
S=\begin{pmatrix}1&0&0\\0&1&0\end{pmatrix}:\CZ^3\to\CZ^2
\end{equation}
has \(S^*S=\operatorname{diag}(1,1,0)\) and \(SS^*=I_2\).   We specify the vertices and the two support projections separately for this reason.

 Every \(X:\HH_w\to\HH_v\) has a polar decomposition \(X=U|X|\), with \(|X|=(X^*X)^{1/2}\) and \(U\) a partial isometry from the support of \(X^*X\) onto the support of \(XX^*\). The eigenvalues \(s_j\) of \(|X|\) are the singular values of \(X\), and the operator, trace and Hilbert--Schmidt norms used throughout are \(\norm X=\max_js_j\), \(\norm X_1=\sum_js_j\) and \(\norm X_2^2=\sum_js_j^2\) (Supplement~\ref{supp:states_polar}). The polar decomposition is needed because an evolved edge is in general no longer a partial isometry: its positive part \(|X|\) is then not a projection. An edge can thus lose amplitude while its support and its polar partial isometry stay fixed.

\subsection{The two endpoint actions and the two inner products}\label{supp:edges}
The corner between two sectors is acted on from both sides.  Put \(A_v=p_v\A p_v\), \(A_w=p_w\A p_w\), and \(\mathcal X_{vw}=p_v\A p_w\). Given \(a\in A_v\), \(b\in A_w\),
the operator \(aXb\) has the same source and target sectors as \(X\). The right action changes the input fed to \(X\), and the left action changes the output produced by \(X\). If \(X,Y\) have the same endpoints, the products \(X^*Y\) and \(XY^*\) return operators to the input and output
sectors, respectively. Define
\begin{equation}
\langle X,Y\rangle_{A_w}:=X^*Y\in A_w,\qquad
{}_{A_v}\langle X,Y\rangle:=XY^*\in A_v.
\label{eq:corner_inner_products}
\end{equation}

Choose partial-isometry edges \(S_e\in p_{r(e)}\A p_{s(e)}\), where \(s(e)\) and \(r(e)\) label the source and range sectors of \(S_e\). Their products encode particular paths, e.g., for matrix units \(e_{ij}e_{kl}=\delta_{jk}e_{il}\), so \(e_{32}e_{21}=e_{34}e_{41}=e_{31}\). We can ask separately about the operator \(e_{31}\) and the equality of the two products implementing it. This distinction is physical even if the ranges of some edges coincide, for example, \(\ket{11}\bra{10}\) and \(\ket{11}\bra{01}\) have the same final projection and distinct source projections.

\subsection{The observer's conditional expectation}\label{sec:observer_expectation}
Let \(\A\subset\mathcal M\) be a finite-dimensional unital inclusion, so the accessible algebra and the full algebra have the same identity. A conditional expectation \(P:\mathcal M\to\A\) fixes \(\A\), is unital and completely positive (UCP), and obeys \(P(aZb)=aP(Z)b\) for \(a,b\in\A\).

We use $P$ to define the accessible system: it splits a joint observable \(Z\) into its accessible prediction \(P(Z)\),  and the orthogonal components  \(Z-P(Z)\in\ker P\) define the environment.   
In the tensor-product realization \(\mathcal M=\A\otimes B(\HH_E)\), the environment $E$ is the factor \(\HH_E\). For the system and environment, with \(\mathcal M=\A\otimes B(\HH_E)\), we identify \(a\in\A\) with \(a\otimes I_E\). With an environment density
\begin{equation}
\tau=\sum_j\pi_j\ket j\bra j,
\label{eq:env_tau}
\end{equation}
define \(V_jy=y\otimes\ket j\). The accessible
prediction for a joint observable \(Z\) is
\begin{equation}
P(Z)=\sum_j\pi_j V_j^*ZV_j .
\end{equation}
On simple tensors \(P(a\otimes b)=a\Tr(\tau b)\), so \(\Tr(\eta P(Z))=\Tr((\eta\otimes\tau)Z)\) for every system density \(\eta\): \(P\) is the partial trace against \(\tau\). The map \(P\) is a conditional expectation, and it is faithful, meaning \(P(Z^*Z)=0\) only for \(Z=0\), exactly when every \(\pi_j>0\) (Supplement~\ref{supp:partial_trace_expectation}). 

\subsection{The selected quantum relation}\label{sec:quantum_relation}
The sectors, as vertices, and the selected edges \(S_e\) between them form the quantum graph of the system. The selected operators determine a quantum relation in the sense of Weaver \cite{Weaver}, that is, a bimodule \(\mathcal V\subseteq B(\HH)\) over the commutant \(\B'\) of an algebra \(\B\subseteq B(\HH)\). The algebra relative to which we take the relation and the space of selected edges are determined by a projection basis
\begin{equation}
\B=\bigoplus_vB(\HH_v),\quad
\mathcal V_{vw}=\Span\{S_e:r(e)=v,\ s(e)=w\},\quad
\mathcal V=\bigoplus_{v,w}\mathcal V_{vw}. \label{eq:sector_relation}
\end{equation}
The space \(\mathcal V\) is a \(\B'\)-bimodule, since \(\B'\) consists of the operators of the form \(\sum_vt_vp_v\) with scalars \(t_v\), which act on each \(\mathcal V_{vw}\) by scalars.  The full rectangular corner \(\mathcal X_{vw}\) has the larger endpoint actions of Section~\ref{supp:edges}. Supplement~\ref{supp:selected_relation} checks these assertions explicitly and distinguishes a selected edge space from its full corner.

The span preserves the existence of a connection. A selected edge \(S\) also fixes its amplitude, since \(\Span\{\lambda S\}=\Span\{S\}\) for every nonzero scalar \(\lambda\), while \((\lambda S)^*(\lambda S)=|\lambda|^2S^*S\) records \(|\lambda|\). For \(|\lambda|\neq1\) the operator \(\lambda S\) is not a partial isometry; neither, in general, is an evolved edge. Other graph constructions associate an operator system to confusability of a quantum channel \cite{DuanSeveriniWinter2013}, classify adjacency data on small matrix algebras
\cite{doi:10.1007/s11005-022-01603-5,Matsuda2022,KieferSchaefer2026}, or study specific relations of dephasing graphs \cite{Ismert2026}. Here the selected concrete operators, their products, and their changes in amplitude specify the quantum graph, whose edges represent coherences between sectors.

\subsection{Reversible evolution and partial access}\label{sec:compression}
A \(*\)-automorphism is a bijective linear map preserving products and adjoints. Let \(\beta_t\) be the full automorphism group, e.g., for a unitary \(U_t\), the map \(\beta_t(X)=U_t^*XU_t\) has these properties,
\begin{equation}
\beta_t(S)^*\beta_t(S)=U_t^*S^*U_tU_t^*SU_t
=U_t^*(S^*S)U_t=\beta_t(p).
\end{equation}
Its range product is \(\beta_t(q)\) by the same multiplication. If \(BA=DC\), then \(\beta_t(B)\beta_t(A)=\beta_t(BA)=\beta_t(DC)=\beta_t(D)\beta_t(C)\). Thus even when a unitary moves the sectors, it carries their specified relations along with them.

The modular automorphism group \(\sigma_s^\rho(X)=\rho^{is}X\rho^{-is}\) of a faithful state with density \(\rho\) also preserves these products \cite{Takesaki2003}.
It describes structure associated with that state. 
The modular flow can itself be the reversible evolution implemented in an automorphism group, take \(\beta_s=\sigma_s^\Omega\) for a faithful state \(\Omega\) on the joint algebra \(\M\), with density \(\rho_\Omega\) and modular group \(\sigma_s^\Omega(Z)=\rho_\Omega^{is}Z\rho_\Omega^{-is}\). The observer reaches a subalgebra \(\A\) through a conditional expectation \(P\), which is faithful  and is fixed by the preparation of the environment, independently of \(\Omega\) (Section~\ref{sec:observer_expectation}).  
The reduced maps \(P\sigma_s^\Omega|_\A\) are in general not automorphisms: by Theorem~\ref{thm:partial_access} their defects vanish for every \(s\) exactly when \(\sigma_s^\Omega(\A)=\A\), which is Takesaki's condition \cite{Takesaki1972} for an \(\Omega\)-preserving expectation onto \(\A\). 
 The starting point for decoherence is thus a modular flow seen with partial access. 
 Supplement~\ref{app:modular} constructs the modular group explicitly, and carries out this construction.

\subsection{The reduced evolution and its multiplication defect}\label{sec:reduced_evolution}
We first compute what happens when the observer restricts access after a physical automorphism.
The accessible observable and its remainder are defined as
\begin{equation}
\Phi_t(X):= P\beta_t(X),\qquad R_t(X):=\beta_t(X)-\Phi_t(X),\quad X\in\A.
\label{eq:access_setup}
\end{equation}
Because \(P^2=P\), \(P(R_t(X))=0\). The finite-time multiplication defect is the sesquilinear expression
\begin{equation}
D_t(X,Y):=\Phi_t(X^*Y)-\Phi_t(X)^*\Phi_t(Y).
\label{eq:defect_kernel}
\end{equation}

\begin{theorem}[Partial access and multiplication]\label{thm:partial_access}
For the reduction in \eqref{eq:access_setup},
\begin{equation}
D_t(X,Y)=P(R_t(X)^*R_t(Y)). \label{eq:access_defect}
\end{equation}
For any \(B\in\A\) the conditional squared prediction error, the expectation \(P\) of the squared difference between \(\beta_t(X)\) and \(B\), is
\begin{align}
P((\beta_t(X)-B)^*(\beta_t(X)-B))
&=D_t(X,X)+(\Phi_t(X)-B)^*(\Phi_t(X)-B).
\label{eq:access_prediction}
\end{align}
The two-time discrepancy is
\begin{equation}
\Phi_{t+s}-\Phi_t\Phi_s=P\beta_t(\id-P)\beta_s|_\A .
\label{eq:access_memory}
\end{equation}
If \(P\) is faithful, \(D_t(X,X)=0\) exactly when \(\beta_t(X)\in\A\), and \(\Phi_t\) is an automorphism exactly
when \(\beta_t(\A)=\A\).
\end{theorem}
\begin{proof}
Write \(A_X=\Phi_t(X)\), \(R_X=R_t(X)\), and likewise for \(Y\). Multiplication in the larger algebra and
bimodularity of \(P\) give
\begin{align}
\Phi_t(X^*Y)
&=P((A_X+R_X)^*(A_Y+R_Y))\notag\\
&=P(A_X^*A_Y)+P(A_X^*R_Y)+P(R_X^*A_Y)+P(R_X^*R_Y)\notag\\
&=A_X^*A_Y+A_X^*P(R_Y)+P(R_X)^*A_Y+P(R_X^*R_Y)\notag\\
&=A_X^*A_Y+P(R_X^*R_Y).
\end{align}
To compare with an arbitrary prediction \(B\in\A\), set \( a =A_X-B\in\A\) and write \(\beta_t(X)-B=R_X+ a\). The mixed terms are \(P(R_X^* a)=P(R_X)^* a=0\) and \(P(a^*R_X)= a^*P(R_X)=0\). Hence \(P((R_X+a)^*(R_X+a))=P(R_X^*R_X)+a^*a\), which is \eqref{eq:access_prediction}. Subtracting the conditional error of \(A_X\) from that of \(B\) therefore leaves \( a^*a\). On a unit input \(y\) the excess error is \(\norm{(A_X-B)y}^2\geq0\). The comparison holds on every input, so \(D_t(X,X)\) is the least conditional error in the operator order. Equality as operators requires \((A_X-B)y=0\) for every \(y\), and thus the unique minimizing prediction is \(B=A_X\).

For \eqref{eq:access_memory}, inserting \(\id=P+(\id-P)\) between \(\beta_t\) and \(\beta_s\) gives
\begin{equation}
P\beta_{t+s}(X)=P\beta_tP\beta_s(X)+P\beta_t(\id-P)\beta_s(X).
\end{equation}
The term \(P\beta_tP\beta_s(X)\) is \(\Phi_t(\Phi_s(X))\), and the term \(P\beta_t(\id-P)\beta_s(X)\) is what becomes accessible at time \(t+s\) from the remainder at time \(s\). Finally, faithfulness of \(P\) means \(P(Z^*Z)=0\) implies \(Z=0\). Applying it to
\(R_X\) gives \(D_t(X,X)=0\) iff \(R_X=0\). If \(\Phi_t\) is an automorphism, this holds for every \(X\), hence \(\beta_t(\A)\subseteq\A\). The restriction of \(\beta_t\) is injective and its image has the same finite dimension as \(\A\), so the inclusion \(\beta_t(\A)\subseteq\A\), between spaces of equal dimension, is equality. If the image is \(\A\), \(P\) fixes every evolved observable and \(\Phi_t=\beta_t|_\A\) is an automorphism.
\end{proof}

Thus \(D_t(X,X)\) is the least conditional error of an accessible prediction of \(\beta_t(X)\), and partial access alone does not give open quantum system dynamics realized as a semigroup \eqref{eq:access_memory}. The two-qubit example in Supplement~\ref{supp:repeated_interactions}
computes a remainder that returns to the system and evaluates \eqref{eq:access_memory}.

\subsection{A semigroup from repeated interactions}\label{sec:fresh_collisions}
Equation \eqref{eq:access_memory} identifies a contribution from the remainder \(R_s(X)\) of the joint observable if the same environment acts twice. In a repeated interaction, the system meets one environment after another, each for a time \(h\), and the environment entering each interaction is independently prepared in \(\tau\) \cite{AttalPautrat}.
Take \(G=G^*\in\M\) with \(P(G)=0\)
and \(H=H^*\in\A\), and define
\begin{equation}
A_h:= hH+\sqrt h\,G,\qquad U_h:= e^{-iA_h},\qquad
\Psi_h(X):= P(U_h^*XU_h). \label{eq:access_collision}
\end{equation}
The reversible evolution of one interaction \(Z\mapsto U_h^*ZU_h\) is a modular automorphism. Let \(\Omega_h\) be the faithful state on \(\M\) with density \(e^{-A_h}/\Tr(e^{-A_h})\). Its modular group is \(\sigma_s^{\Omega_h}(Z)=e^{-isA_h}Ze^{isA_h}\), so setting \(s=-1\) gives \(U_h^*ZU_h=\sigma_{-1}^{\Omega_h}(Z)\) and \(\Psi_h=P\sigma_{-1}^{\Omega_h}|_\A\). The state \(\Omega_h\) is a reference state whose modular flow is the reversible evolution, it is in general correlated between system and environment, and it is not the preparation \(\tau\), \eqref{eq:env_tau}, which determines \(P\) (Supplement~\ref{supp:modular_compatible}).
If \(\eta_1\) is the system density leaving  the first interaction, which began in the state \(\eta_0\otimes\tau\), then the state entering the second interaction is \(\eta_1\otimes\tau\). By the defining trace identity for \(P\), the final expectation is \(\Tr(\eta_1\Psi_h(X))=\Tr(\eta_0\Psi_h(\Psi_h(X)))\). Repeating this argument gives \(\Psi_h^n\) after \(n\) interactions. To obtain a generator per unit of physical time we hold \(t=nh\) fixed. Expanding  one interaction through order \(h\) and using \(P(G)=0\), which removes the term of order \(\sqrt h\), gives
\begin{align}
\Psi_h(X)&=X+h\LL(X)+O(h^{3/2})\norm X,\notag\\
\LL(X)&=i[H,X]+P(GXG)-\tfrac12\bigl(P(G^2)X+XP(G^2)\bigr).
\label{eq:access_generator}
\end{align}
The completely positive map \(X\mapsto P(GXG)\) has a Kraus form
\(P(GXG)=\sum_kL_k^*XL_k\), with \(P(G^2)=\sum_kL_k^*L_k\). Consequently
\begin{equation}
\LL(X)=i[H,X]+\sum_k\left(L_k^*XL_k
-\tfrac12L_k^*L_kX-\tfrac12XL_k^*L_k\right).
\label{eq:gksl}
\end{equation}
This is the observable form of the Gorini--Kossakowski--Sudarshan--Lindblad (GKSL)
generator \cite{GKS1976,doi:10.1007/BF01608499}, with jump operators \(L_k\): each \(L_k\) is one way in which  the coupling \(G\) acts on the system in one interaction. The norm limit
\(\Psi_{t/n}^{\,n}\to e^{t\LL}\) establishes the quantum Markov semigroup (QMS)
\(\Phi_t=e^{t\LL}\) used from here on: repeated partial access to independently prepared environments, each interaction a modular automorphism \(\sigma_{-1}^{\Omega_h}\) seen with partial access, gives GKSL dynamics on the accessible algebra.  Supplement~\ref{supp:collision_derivation} gives the expansion, the construction of the jumps, the norm convergence, and the realization of a given generator by such a coupling. 

\subsection{Compression, positivity, and operator weights}\label{supp:weights}
We now write \(\Phi_t\) as a compression  of observables on a larger space by an isometry, the form from which positivity and the weights are read. Fix a physical time \(t\). Apply the positive-matrix factorization, derived in  Supplement~\ref{supp:collision_derivation}, to \(\Phi_t\).
It gives Kraus operators \(K_\mu:\HH\to\HH\) such that \(\Phi_t(X)=\sum_\mu K_\mu^*XK_\mu\), setting \(X=I\) fixes
their normalization. To obtain the action on states, enforce equality of the two ways of computing an expectation, 
\begin{align}
\Tr(\Phi_{t*}(\eta)X)=\Tr(\eta\Phi_t(X)),
\end{align}
such that,
\begin{align}
\sum_\mu K_\mu^*K_\mu&=I,\notag\\
\Phi_t(X)&=\sum_\mu K_\mu^*XK_\mu, \nonumber \\ 
\Phi_{t*}(\eta)&=\sum_\mu K_\mu\eta K_\mu^*.
\label{eq:duality}
\end{align}
Indeed \(\Tr(\eta K_\mu^*XK_\mu)=\Tr(K_\mu\eta K_\mu^*X)\).  To compare \(\Phi_t(X^*Y)\) with \(\Phi_t(X)^*\Phi_t(Y)\), place all Kraus indices in one 
Hilbert space \(\HH_E\) with orthonormal basis \(\{\ket\mu\}\). Define
\begin{equation}
V_t y:=\sum_\mu K_\mu y\otimes\ket\mu,\quad
\Phi_t(X)=V_t^* (X\otimes I_E) V_t. \label{eq:dilation}
\end{equation}
The map \(V_t\) is an isometry, because \(\sum_\mu K_\mu^*K_\mu=I\), and \eqref{eq:dilation} is the Stinespring form of the channel \cite{doi:10.1090/S0002-9939-1955-0069403-4}.

 The product of two compressed operators places a projection between the two original operators. Define \(P_t:=V_tV_t^*\), the projection onto \(\ran V_t\), \(Q_t:=I-P_t\), and \(C_t(X):=Q_t (X\otimes I_E)V_t\).  Splitting the identity  gives
\begin{align}
\Phi_t(X^*Y)&=V_t^*(X\otimes I_E)^*(Y\otimes I_E)V_t\notag\\
 &=V_t^* (X\otimes I_E) ^*(P_t+Q_t) (Y\otimes I_E) V_t\notag\\
 &=(V_t^* (X\otimes I_E) V_t)^*(V_t^* (Y\otimes I_E)V_t)  +(Q_t (X\otimes I_E)V_t)^*(Q_t (Y\otimes I_E) V_t)\notag\\
 &=\Phi_t(X)^*\Phi_t(Y)+C_t(X)^*C_t(Y). \label{eq:compression_identity}
\end{align}
It follows,
\begin{equation}
D_t(X,Y)=C_t(X)^*C_t(Y). \label{eq:defect_factorization}
\end{equation}
For \(X=Y\), \(\langle y,D_t(X,X)y\rangle=\norm{C_t(X)y}^2\). This gives the Schwarz inequality
\(\Phi_t(X)^*\Phi_t(X)\leq\Phi_t(X^*X)\) for \(\Phi_t\). The dynamics are therefore contractive, \(\norm{\Phi_t(X)}\leq\norm X\), since \(0\leq X^*X\leq\norm X^2I\) implies \(0\leq\Phi_t(X)^*\Phi_t(X)\leq\Phi_t(X^*X)\leq\norm X^2 I\). More generally, given a finite collection \(X_i\) and matrices \(a_i\),
\begin{equation}
\sum_{i,j}a_i^*D_t(X_i,X_j)a_j
=\Big(\sum_i C_t(X_i)a_i\Big)^*
 \Big(\sum_j C_t(X_j)a_j\Big)\geq0. \label{eq:kernel_positivity}
\end{equation}

The same factorization has an explicit channel-bimodule form,  \(D_t(X,Y)=\langle\delta_t(X),\delta_t(Y)\rangle_\A\); Supplement~\ref{supp:channel_bimodule} derives the tuple remainder \(\delta_t(X)\), with components \(XK_\mu-K_\mu\Phi_t(X)\), and relates this form to the endpoint inner products of Section~\ref{supp:edges}. The multiplication defect is thus a positive kernel, built from the parts \(C_t(X)\) of the evolved observables that leave the range of \(V_t\).

\subsection{Endpoint weights}
For an initial edge \(S\), let \(S(t)=\Phi_t(S)\) and define
\begin{align}
\Ws_S(t)&:= S(t)^*S(t),&\Wr_S(t)&:= S(t)S(t)^*,\label{eq:operator_weights}\\
\Ds_t(X)&:= D_t(X,X),&\Dr_t(X)&:= D_t(X^*,X^*).\label{eq:source_range_defects}
\end{align}
The operators $\Ws_S(t)$ and $\Wr_S(t)$ are edge weights, the source and range weights of the edge, respectively, which give the squared amplitude on the source and range sides.  Together with the sectors as vertices and the selected edges \(S_e\), these weights form the coherence-weighted quantum graph; the evolved operators \(\Phi_t(S_e)\) carry, beyond the weights, the phases that enter products of edges. If \(S^*S=p\), \(SS^*=q\), subtracting \eqref{eq:source_range_defects} from the transported original projections gives
\begin{align}
\Ws_S(t)&=\Phi_t(p)-\Ds_t(S),&
\Wr_S(t)&=\Phi_t(q)-\Dr_t(S). \label{eq:edge_relations}
\end{align}
This choice of comparison matters if \(p\) or \(q\) evolves. For instance, a fixed partial isometry has the same intrinsic products at all times, yet the transported original products can change. The four-state model in Supplement~\ref{sec:square} demonstrates this effect.

\subsection{Evolving vertices and compositions}\label{sec:relations}
Vertex projections evolve to \(\Phi_t(p_v)\). Positivity, unitality, and \(p_v^2=p_v\) imply
\begin{equation}
0\leq  \Phi_t(p_v) \leq I,\qquad\sum_v \Phi_t(p_v)=I,\qquad
 \Phi_t(p_v)-\Phi_t(p_v)^2 =D_t(p_v,p_v). \label{eq:vertex_defect}
\end{equation}
The positive \(\Phi_t(p_v)\) sum to \(I\), so they form a positive operator-valued measurement. 
An individual element is a projection exactly when its square equals itself. By \eqref{eq:vertex_defect}
and \eqref{eq:defect_factorization}, this occurs for \( \Phi_t(p_v)\) precisely when \(C_t(p_v)^*C_t(p_v)=0\), that is, when \(C_t(p_v)=0\). For any \(Y\), the factorization \eqref{eq:defect_factorization} then gives
\begin{equation}
0=D_t(p_v,Y)=\Phi_t(p_vY)- \Phi_t(p_v) \Phi_t(Y),\qquad
0=D_t(Y^*,p_v)=\Phi_t(Yp_v)-\Phi_t(Y) \Phi_t(p_v). \label{eq:vertex_multiplicative}
\end{equation}
When all vertex defects vanish,  \eqref{eq:vertex_multiplicative} with \(Y=p_w\), and then with \(Y=Xp_w\) for the vertex \(p_v\) and \(Y=X\) for the vertex \(p_w\), gives
\begin{align}
 \Phi_t(p_v)\Phi_t(p_w)&=\Phi_t(p_vp_w)=\delta_{vw} \Phi_t(p_v),\notag\\
\Phi_t(p_vXp_w)&= \Phi_t(p_v)\Phi_t(X) \Phi_t(p_w). \label{eq:moving_corners}
\end{align}
The \(\Phi_t(p_v)\) are then orthogonal projections, and the spaces \(\Phi_t(p_v)\A\,\Phi_t(p_w)\) are corners between transported sectors. If \(\Phi_t(p_v) =p_v\), the corresponding original sectors are fixed. For general \( \Phi_t(p_v)\), one can always expand \(\Phi_t(S)=\sum_{v,w}p_v\Phi_t(S)p_w\) using the original resolution. When \eqref{eq:moving_corners} holds, the dynamics themselves transport a single corner to a single corner, \(\Phi_t(p_v\A p_w)\subseteq\Phi_t(p_v)\A\,\Phi_t(p_w)\).

For a product \(BA\), inserting \(X=B^*\), \(Y=A\) into the defect definition gives
\begin{equation}
\Phi_t(BA)=\Phi_t(B)\Phi_t(A)+D_t(B^*,A). \label{eq:path_compression}
\end{equation}
If \(BA=DC\) initially, evolving that equality and using \eqref{eq:path_compression} on each side gives
\begin{equation}
\Phi_t(B)\Phi_t(A)-\Phi_t(D)\Phi_t(C) =-D_t(B^*,A)+D_t(D^*,C). 
\label{eq:square_defect}
\end{equation}
The factorization in \eqref{eq:defect_factorization} gives the bound
\begin{align}
\norm{D_t(B^*,A)}
 &\leq\norm{C_t(B^*)}\norm{C_t(A)}  =\sqrt{\norm{\Dr_t(B)}\norm{\Ds_t(A)}}. 
 \label{eq:path_cs}
\end{align}
Here \(\norm C^2=\norm{C^*C}\) supplies both square roots. Applying the triangle inequality in
\eqref{eq:square_defect} yields
\begin{align}
\norm{\Phi_t(B)\Phi_t(A)-\Phi_t(D)\Phi_t(C)}
 &\leq\sqrt{\norm{\Dr_t(B)}\norm{\Ds_t(A)}}
       +\sqrt{\norm{\Dr_t(D)}\norm{\Ds_t(C)}}. \label{eq:path_bound}
\end{align}

These products compare the transported observables at the same physical time. An implementation as successive interventions would additionally require the corresponding instruments and their order in time. Thus, a relation \(BA=DC\) between compositions of edges can thus fail only by the amount \eqref{eq:path_bound} that the defects of its individual four edges allow.

\subsection{A common reservoir with internal relaxation}\label{sec:selective_relaxation}
We will follow the edge, its records, and a composition defect in one model, the common-reservoir model, which returns in Sections~\ref{sec:representations}, \ref{sec:asymptotic_algebra} and~\ref{sec:sharp_loss}.

\begin{example}[The common-reservoir model]\label{ex:common_reservoir}
Two sector labels couple identically to a common reservoir, while a third sector has a different coupling strength. The sector label that the dynamics conserves, specifies the corner, and an internal two-level system exchanges excitations with the reservoir. Different internal preparations produce different fidelities between the records that two sectors leave in the reservoir, which appear as different eigenvalues of the same operator weight.

Take \(\HH=\CZ^3\otimes\CZ^2\), the sector factor \(\CZ^3\), which carries the sector label, times the internal factor \(\CZ^2\), which carries the two internal levels. Use \(e_{ij}=\ket i\bra j\) for the sector matrices and \(q_a=\ket a\bra a\) for
the internal levels. The sector projections are \(p_i:= e_{ii}\otimes I_2\), and the edges between whole sectors are \(S_{ij}:= e_{ij}\otimes I_2\), from sector \(j\) to sector \(i\). Multiplication gives \(S_{ij}^*S_{ij}=e_{ji}e_{ij}\otimes I_2=p_j\) and \(S_{ij}S_{ij}^*=p_i\).
The generator \eqref{eq:gksl} has a Hamiltonian \(H\) and two jump operators \(L_\pm\), which act on the sector label through one coupling matrix \(F\):
\begin{align}
F&=e_{00}+e_{11}+\kappa e_{22},\qquad H=\Omega I_3\otimes q_1,\notag\\
L_-&=F\otimes\ell_-,\quad \ell_-=\sqrt{\gamma_-}\ket0\bra1,\qquad
L_+=F\otimes\ell_+,\quad \ell_+=\sqrt{\gamma_+}\ket1\bra0 ,
\label{eq:selective_generator}
\end{align}
here $\Omega$  is the splitting of the two internal levels, \(\ell_-\) lowers the level and \(\ell_+\) raises it.  The rates \(\gamma_\pm\) of excitation and decay at unit coupling are positive and \(\kappa>0\). We first take \(\kappa\ne1\) and \(\gamma_+\ne\gamma_-\). The excitation and decay channels are common to all three sectors. The coupling amplitudes in sectors \(0,1,2\) are the diagonal entries of \(F\),
\(g_0=g_1=1\), \(g_2=\kappa\), respectively. If the system starts at level \(a\) of sector \(i\), its physical jump rate is \(g_i^2r_a\), with \(r_0=\gamma_+\) and \(r_1=\gamma_-\).

The operators \(H\) and \(L_\pm\) commute with the sector projections, which gives \(\LL(p_i)=0\). To compute the evolution of the excited-level observable \(I_3\otimes q_1\), first evaluate
\begin{equation}
\ell_-^*q_1\ell_-=0,\quad
\ell_+^*q_1\ell_+=\gamma_+q_0,\quad
\ell_-^*\ell_-=\gamma_-q_1,\quad\ell_+^*\ell_+=\gamma_+q_0 .
\end{equation}
The term \(\sum_\pm L_\pm^*(I_3\otimes q_1)L_\pm\) of the generator is \(F^2\otimes\gamma_+q_0\), and the anticommutators give  \(-F^2\otimes\gamma_-q_1\). Consequently
\begin{equation}
\LL(p_i)=0,\qquad
\LL(I_3\otimes q_1)=F^2\otimes(\gamma_+q_0-\gamma_-q_1).
\label{eq:selective_populations}
\end{equation}
The sector label is conserved while its internal levels relax. The relaxation rates are the same in sectors \(0\) and \(1\) and are multiplied by \(\kappa^2\) in sector \(2\); this difference is what the reservoir can learn about the sector. Section~\ref{sec:reservoir_weights} computes the edge amplitudes; Section~\ref{sec:classical_jump_records} obtains the same amplitudes from the reservoir records.
\end{example}

\section{Information in the edge weights}\label{sec:representations}
\subsection{Conditional records in a common dilation}
The preceding sections defined the weights entirely through system observables. We now identify the information about the sector that is carried by the  environment. The general result optimizes the  corner operator for a given pair of preparations.

Here we  assume that the sector projections are fixed by \(\Phi_t\), \(\Phi_t(p_v)=p_v\); this fixes the sector labels while allowing arbitrary channel dynamics within each sector.    With fixed sectors an edge keeps its weight in one corner at all times \eqref{eq:moving_corners}. Moving projections appear in Section~\ref{sec:relations}, in \eqref{eq:transported_corner_information} when their images stay projections, and in Section~\ref{sec:sharp_loss} for endpoints in the retained algebra; a single pair of preparations needs no sector at all \eqref{eq:unrestricted_pair_information}. 

For rank-one fixed sector labels  this assumption is a dephasing condition; sectors with internal degrees of freedom  allow relaxation and population transfer. Only one time enters, so \(\Phi_t\) may be any UCP map that fixes the sector projections. 
By \eqref{eq:vertex_defect},
\(D_t(p_v,p_v)=0\), and by the factorization \eqref{eq:defect_factorization} in the dilation \eqref{eq:dilation} this says \(Q_t(p_v\otimes I)V_t=0\). Thus
\begin{equation}
(p_v\otimes I)V_t=P_t(p_v\otimes I)V_t =V_tV_t^*(p_v\otimes I)V_t=V_t\Phi_t(p_v)=V_tp_v .
\end{equation}
For \(x\in\HH_v\), we have \(p_vx=x\), so \(V_tx\) lies in \(\HH_v\otimes\HH_E\),  denote the restriction of \(V_t\) to \(\HH_v\) by \(V_{v,t}\).  The environment must be identified across the sectors to calculate a coherence, since a coherence \(\langle V_{v,t}x,(X\otimes I)V_{w,t}y\rangle\) pairs the environment parts of the two sectors. For example, rotating the two environment factors separately by \(U_v\) and \(U_w\) would replace \(\langle V_{v,t}x,(X\otimes I)V_{w,t}y\rangle\) by \(\langle V_{v,t}x,(X\otimes U_v^*U_w)V_{w,t}y\rangle\); a common rotation cancels from the expression. A mixed physical environment can be handled by preparing a purification of its density, 
applying the system--environment unitary with the identity on the purifying space, and including both environment and purifying space in \(\HH_E\). Tracing out both recovers the original system channel.  The result below concerns this complete complementary record.

For a unit input \(x\in\HH_v\), trace out the output system to obtain
\begin{equation}
\sigma_v^E(x,t):=\Tr_{\HH_v}\ket{V_{v,t}x}\bra{V_{v,t}x}.
\label{eq:conditional_environment}
\end{equation}
To compute this partial trace, choose an output basis \(u_j\) in \(\HH_v\) and expand \(V_{v,t}x=\sum_j u_j\otimes\epsilon_j\). Orthogonality of the \(u_j\) gives \(\sigma_v^E=\sum_j\ket{\epsilon_j}\bra{\epsilon_j}\), a positive matrix of trace \(\sum_j\|\epsilon_j\|^2=\|V_{v,t}x\|^2=1\). This density on \(\HH_E\) is the record of the preparation \(x\) that the environment holds at time \(t\). Different preparations inside one sector can give different records.
For positive densities define root fidelity by \(\ff(\sigma,\tau)=\norm{\sqrt\sigma\sqrt\tau}_1\), where \(\norm C_1=\Tr\sqrt{C^*C}\). If both densities are pure, \(\sqrt\sigma=\sigma=\ket e\bra e\) and \(\sqrt\tau=\ket f\bra f\), so their product is \(\langle e,f\rangle\ket e\bra f\). It has one possible nonzero singular value, \(|\langle e,f\rangle|\).   For densities diagonal in a common basis, with eigenvalues \(\lambda_j,\lambda'_j\), it is \(\sum_j\sqrt{\lambda_j\lambda'_j}\). The root fidelity of two records is called their record fidelity; it measures the overlap of the two records. For a pair of sectors define
\begin{equation}
\ff_{vw}(t):=
\max_{\substack{x\in\HH_v,\ y\in\HH_w\\\norm x=\norm y=1}}
\ff(\sigma_v^E(x,t),\sigma_w^E(y,t)).
\label{eq:sector_record_overlap}
\end{equation}
The maximum selects the pair of preparations whose complete records have largest fidelity.
The records are computed in a dilation, and their fidelity does not depend on which one. By uniqueness of the minimal Stinespring dilation \cite{KretschmannSchlingemannWerner} (the form \eqref{eq:dilation} with the fewest Kraus operators),  every other dilation is obtained from the minimal one by an environment isometry \(J\). If \(\sigma\) changes to \(J\sigma J^*\), its positive square root is
\(J\sqrt\sigma J^*\), as seen by squaring and using \(J^*J=I\). Thus the product of the two square roots changes to \(J\sqrt\sigma\sqrt\tau J^*\). On the range of \(J\) this has the old singular values, and on its orthogonal complement it vanishes, so its trace norm is unchanged. In particular \(\ff_{vw}(t)\) depends only on the channel \(\Phi_t\): the dilation is a way to compute the records and assumes nothing beyond \(\Phi_t\).

\subsection{From a matrix element to a record fidelity}\label{supp:information}
The central result of this section equates the largest matrix element of the reduced dynamics, \(|\langle x,\Phi_t(X)y\rangle|\) between two preparations, over contractions \(X\) in the corner, with the fidelity of their environment records.

\begin{theorem}[Information and the edge weights]\label{thm:information_weights}
Let \(v\ne w\) be fixed sectors. For unit \(x\in\HH_v\), \(y\in\HH_w\),
\begin{equation}
\max_{\substack{X\in p_v\A p_w\\\norm X\leq1}}
|\langle x,\Phi_t(X)y\rangle|
=\ff(\sigma_v^E(x,t),\sigma_w^E(y,t)).
\label{eq:pair_information}
\end{equation}
Let \(\norm{\Phi_t}_{vw}\) denote the induced norm of the map on the rectangular corner \(p_v\A p_w\). Then
\begin{equation}
\max_{\substack{S\in p_v\A p_w\\ S\ \mathrm{partial\ isometry}}}
\norm{\Ws_S(t)} =\norm{\Phi_t}_{vw}^{\,2}=\ff_{vw}(t)^2 .
\label{eq:corner_information}
\end{equation}
The same maximum holds for the range weight.
\end{theorem}
\begin{proof}
For fixed preparations the corner operator \(X\) acts on the system alone. We will first trace out the environment to isolate the system operator paired with \(X\). We can then optimize that pairing and identify its value by tracing the same two joint vectors over the system. Set \(\psi=V_{v,t}x\), \(\chi=V_{w,t}y\). In a common environment basis \(\{\ket\mu\}\), write
\begin{equation}
\psi=\sum_\mu a_\mu\otimes\ket\mu,\qquad
\chi=\sum_\mu b_\mu\otimes\ket\mu,\qquad
a_\mu\in\HH_v,\quad b_\mu\in\HH_w.
\end{equation}
The cross operator needed below goes from \(\HH_v\) to \(\HH_w\); the partial trace on a tensor product gives it explicitly,
\begin{equation}
C=\Tr_E\ket\chi\bra\psi 
=\sum_{\mu,\nu}\ket{b_\mu}\bra{a_\nu}\Tr(\ket\mu\bra\nu)
=\sum_\mu\ket{b_\mu}\bra{a_\mu}.
\end{equation}
For \(X:\HH_w\to\HH_v\), \(XC\) is square on \(\HH_v\). The matrix element being optimized has the form
\begin{align}
\langle x,\Phi_t(X)y\rangle
&=\langle\psi,(X\otimes I)\chi\rangle
=\sum_{\mu,\nu}\langle a_\nu,Xb_\mu\rangle\langle\nu,\mu\rangle\notag\\
&=\sum_\mu\langle a_\mu,Xb_\mu\rangle=\Tr_{\HH_v}(XC).
\end{align}
Write \(C=U|C|\) in its polar decomposition (Section~\ref{sec:sectors}). If \(|C|u_j=s_ju_j\) with \(s_j>0\), the vectors
\(Uu_j\) are orthonormal. For a contraction \(X\),
\begin{equation}
|\Tr(XC)|=\left|\sum_js_j\langle u_j,XUu_j\rangle\right|
\leq\sum_js_j=\norm C_1.
\end{equation}
Taking \(X=U^*\) attains equality since \(U^*U\) fixes every \(u_j\) with positive singular value. Thus the maximizing operator \(X=U^*\) is already a partial isometry in the required corner. If \(C=0\), every value of the pairing is zero
and the maximum is again \(\norm C_1\).

It remains to identify the trace norm of this system cross operator with the fidelity of the environment densities. Following \cite[Lemma~20]{HaydenWinter2012}, the cross-operator form of Uhlmann's purification formula \cite{doi:10.1016/0034-4877(76)90060-4}, we obtain
\begin{equation}
\norm C_1=\ff(\sigma_v^E(x,t),\sigma_w^E(y,t)).
\label{eq:cross_trace_norm}
\end{equation}
Supplement~\ref{supp:cross_trace_norm} gives our calculation that does not use the purification formula. It writes \(C=BA^*\), where \(A:\HH_E\to\HH_v\) and \(B:\HH_E\to\HH_w\) are the linear maps with \(A\ket\mu=a_\mu\), \(B\ket\mu=b_\mu\), and uses \(\sigma_v^E=(A^*A)^{\mathsf T}\), \(\sigma_w^E=(B^*B)^{\mathsf T}\).  In the convention of \cite[Lemma 20]{HaydenWinter2012} the fidelity is squared, the trace norm \eqref{eq:cross_trace_norm} is its square root, our root fidelity \(\ff\).

To obtain \eqref{eq:corner_information}, use the variational expression
\(\norm Y=\max_{\norm x=\norm y=1}|\langle x,Yy\rangle|\). 
For fixed \(y\) with \(Yy\ne0\), the choice \(x=Yy/\norm{Yy}\) attains the inner maximum. Maximizing also over \(y\) gives its largest singular value. Since a fixed corner is invariant, \(\Phi_t(X)\in p_v\A p_w\) by \eqref{eq:moving_corners}, the maximum for \(Y=\Phi_t(X)\) is attained with \(x\in\HH_v\), \(y\in\HH_w\). The unit balls and spheres are compact in finite dimension, so each maximum is attained and we can take the joint maximum in either order to obtain
\begin{equation}
\norm{\Phi_t}_{vw} =\max_{x,y}\max_{\norm X\leq1}|\langle x,\Phi_t(X)y\rangle|
=\max_{x,y}\ff(\sigma_v^E(x,t),\sigma_w^E(y,t)).
\end{equation}
Choose a pair \(x,y\) attaining this last maximum. For its polar optimizer \(S=U^*\), the inequalities in both
directions can be read from
\begin{equation}
\ff_{vw}(t)=|\langle x,\Phi_t(S)y\rangle|
       \leq \|\Phi_t(S)\|\leq\|\Phi_t\|_{vw}=\ff_{vw}(t).
\end{equation}
The optimizer may depend on the chosen pair, but for a given edge this argument still gives an upper bound. If \(Y=\Phi_t(S)\) and \(Yv_j=s_j u_j\) is a singular-vector pair with \(s_j>0\), then
\(Y^*Yv_j=s_j^2v_j\) and \(YY^*u_j=s_j^2u_j\). Both endpoint norms are therefore \(\|Y\|^2\).
\end{proof}

As a check, if \(V\ket v=\ket v\otimes\ket{e_v}\) and \(V\ket w=\ket w\otimes\ket{e_w}\) on rank-one sectors, then \(\Phi(\ket v\bra w)=\langle e_v,e_w\rangle\ket v\bra w\). Its source weight is \(|\langle e_v,e_w\rangle|^2p_w\). In higher dimensions \eqref{eq:corner_information} shows that orthogonality of {every} pair of records is equivalent to extinction of the entire corner. For a finite-dimensional QMS the linear operator \(e^{t\LL}\) has inverse \(e^{-t\LL}\) on the vector space of observables. Thus a nonzero invariant corner cannot be annihilated at finite \(t\), although its norm can approach zero.

Note, the proof extends to both arbitrary pairs of preparations and sectors that move. The proof of \eqref{eq:pair_information} never uses the sectors. For any unital completely positive map \(\Phi_t\) on \(B(\HH)\) with dilation \eqref{eq:dilation}, and any unit vectors \(x,y\in\HH\) with records \(\sigma^E(x,t)=\Tr_{\HH}\ket{V_tx}\bra{V_tx}\),
\begin{equation}
\max_{\norm X\leq1}|\langle x,\Phi_t(X)y\rangle|
=\norm{\Phi_{t*}(\ket y\bra x)}_1
=\ff(\sigma^E(x,t),\sigma^E(y,t)).
\label{eq:unrestricted_pair_information}
\end{equation}
The sectors may also move, as long as their images stay projections: \(\Phi_t(p_v)\) is then a projection that can differ from \(p_v\), as it does under a unitary evolution that rotates the sectors. Suppose \(q_v=\Phi_t(p_v)\) and \(q_w=\Phi_t(p_w)\) are nonzero projections. Then \(p_v,p_w\in\MD(\Phi_t)\) and \eqref{eq:vertex_multiplicative} holds for \(p_v\) and for \(p_w\), so \(\Phi_t\) maps the corner \(p_v\A p_w\) into \(q_v\A q_w\), and
\begin{equation}
\norm{\Phi_t|_{p_v\A p_w}}
=\max_{\substack{x\in q_v\HH,\ y\in q_w\HH\\\norm x=\norm y=1}}
\ff(\sigma^E(x,t),\sigma^E(y,t)).
\label{eq:transported_corner_information}
\end{equation}
The edges are taken in the original corner and the preparations in the image sectors. Fixed sectors are the case \(q_v=p_v\), \(q_w=p_w\). Supplement~\ref{supp:cross_trace_norm} gives both arguments.

\subsection{Environmental fragments and classical records}\label{sec:measured_records}
In this section, we relate \eqref{eq:pair_information} to how measurements on the environment or on a fragment of it give classical records. Suppose only a fragment \(F\) of the environment \(E=F\otimes G\) is read. Write  the joint vectors \(\psi=V_{v,t}x\) and \(\chi=V_{w,t}y\) in the order \((\HH_i\otimes G)\otimes F\). In the proof of \eqref{eq:pair_information}, tracing over \(G\) as well as the output system allows every contraction \(Y:\HH_w\otimes G\to\HH_v\otimes G\) in the variational formula; the operators \(Y=X\otimes I_G\) of \eqref{eq:pair_information} are among the available contractions. Maximizing over all \(Y\) can only raise the maximum, and by the same proof the larger maximum is the fidelity of the marginals on \(F\), which gives
\begin{equation}
\ff(\sigma_v^E,\sigma_w^E)  \leq\ff(\Tr_G\sigma_v^E,\Tr_G\sigma_w^E).
\label{eq:fragment_record_bound}
\end{equation}
Thus two preparations distinguishable on \(F\) already bound the matrix element for those preparations.

 A measurement of the environment replaces each record by a classical distribution of outcomes. Let \(\{M_\xi\}\) be a positive operator-valued measurement (POVM) on the environment, with \(\xi\) labeling the outcomes  and \(M_\xi\geq0\) and \(\sum_\xi M_\xi=I_E\). For the joint vectors \(\psi=V_{v,t}x\), \(\chi=V_{w,t}y\), define the normalized outcome distributions:
\begin{equation}
P_x(\xi):=\langle\psi,(I\otimes M_\xi)\psi\rangle,\qquad
P_y(\xi):=\langle\chi,(I\otimes M_\xi)\chi\rangle.
\end{equation}
These are the records evaluated on the measurement, \(P_x(\xi)=\Tr(\sigma_v^E(x,t)M_\xi)\) and \(P_y(\xi)=\Tr(\sigma_w^E(y,t)M_\xi)\); in this sense the measurement replaces each record by a distribution of outcomes. Set \(\psi_\xi=(I\otimes M_\xi^{1/2})\psi\) and \(\chi_\xi=(I\otimes M_\xi^{1/2})\chi\). The operator \(X\otimes I\) commutes with each \(I\otimes M_\xi^{1/2}\), so for a corner contraction \(X\),
\begin{align}
|\langle x,\Phi_t(X)y\rangle|
&=\left|\sum_\xi\langle\psi_\xi,(X\otimes I)\chi_\xi\rangle\right|\notag\\
&\leq\sum_\xi\|\psi_\xi\|\,\|\chi_\xi\|
=\sum_\xi\sqrt{P_x(\xi)P_y(\xi)}.
\end{align}
Optimizing the left side by Theorem~\ref{thm:information_weights} gives
\begin{equation}
\ff(\sigma_v^E(x,t),\sigma_w^E(y,t)) \leq\sum_\xi\sqrt{P_x(\xi)P_y(\xi)}.
\label{eq:classical_record_bound}
\end{equation}
The right side is the classical root fidelity of the measured records \cite{FuchsCaves1996}. A measurement on a fragment \(F\) is included by taking \(M_\xi=M_\xi^F\otimes I_G\) for a POVM \(\{M_\xi^F\}\) on \(F\). Continuous outcomes replace the sum in \eqref{eq:classical_record_bound} by an integral. Distinct measured records therefore limit the coherence of the given preparations.

\subsection{Shared jumps and the eigenvalues of one edge}
Here we  identify a family for which the given physical edge realizes fidelity separately for each internal basis preparation. Take \(\HH=\CZ^r\otimes\CZ^m\), \(p_v=e_{vv}\otimes I_m\), \(q_a=\ket a\bra a\) on the internal factor  \(\CZ^m\)  of the system, with sector factor \(\CZ^r\), and define coupling to the environment via the jump operators
\begin{equation}
L_k:=\sum_v g_{v,k} e_{vv}\otimes\ell_k,\qquad
\ell_k:=\sqrt{r_k}\ket{b_k}\bra{a_k},\qquad
H:= I_r\otimes H_0 . \label{eq:shared_jumps}
\end{equation}
Here \(g_{v,k},r_k\geq0\) and \(H_0\) is diagonal in the basis \(\ket a\). A given jump index \(k\) is shared across all sectors, its amplitude may depend on \(v\). Each \(p_v\) commutes with \(H,L_k,L_k^*\), so \(\LL(p_v)=0\) and \(\Phi_t(p_v)=p_v\). Let \(S_{vw}=e_{vw}\otimes I_m\), and let \(\sigma_{v,a}^E(t)\) denote the record of \(\ket v\otimes\ket a\).

\begin{corollary}[Shared jumps and an individual edge]\label{prop:shared_jump_weight}
For the generator \eqref{eq:shared_jumps}, with \(g_{v,k},r_k\geq0\) and \(H_0\) diagonal, put \(f_a(t)=\ff(\sigma^E_{v,a}(t),\sigma^E_{w,a}(t))\). The evolved edge \(S_{vw}\), its source weight and its source defect are
\begin{align}
\Phi_t(S_{vw})&=e_{vw}\otimes\sum_a f_a(t)q_a,\notag\\
\Ws_{S_{vw}}(t)&=e_{ww}\otimes\sum_a f_a(t)^2q_a,\nonumber \\
\Ds_t(S_{vw})&=e_{ww}\otimes\sum_a(1-f_a(t)^2)q_a .
\label{eq:shared_jump_weight}
\end{align}
\end{corollary}
\begin{proof}
Start with the off-diagonal operator \(e_{wv}\otimes q_a\), the cross input, and evolve it on densities. Trace cyclicity sends \(\Tr(\eta L_k^*XL_k)\) to \(\Tr((L_k\eta L_k^*)X)\) and \(\Tr(\eta\,i[H,X])\) to \(\Tr((-i[H,\eta])X)\). In \(L_k\eta L_k^*\) with \(\eta=e_{wv}\otimes C\), for an internal matrix \(C\), multiplication of the sector units gives
\(e_{ii}e_{wv}e_{jj}=\delta_{iw}\delta_{jv}e_{wv}\). Hence the two surviving coupling factors are \(g_{w,k}\) and \(g_{v,k}\). In \(L_k^*L_k\eta\) there is only \(e_{ii}e_{wv}\), so the factor is \(g_{w,k}^2\). In \(\eta L_k^*L_k\) it is \(g_{v,k}^2\). The operators \(e_{wv}\otimes C\) are therefore mapped among themselves; write the evolution of the cross input as \(e_{wv}\otimes C_a(t)\), where $C_a(t)$ is the matrix on the internal factor that starts at the level projection \(q_a\). Then
\begin{equation}
\dot C_a=-i[H_0,C_a] +\sum_k\left(g_{w,k}g_{v,k}\ell_k C_a\ell_k^*
-\tfrac12 g_{w,k}^2\ell_k^*\ell_k C_a -\tfrac12 g_{v,k}^2 C_a\ell_k^*\ell_k\right),\qquad C_a(0)=q_a .
\label{eq:shared_cross_generator}
\end{equation}
For \(C_a=\sum_j c_jq_j\),  diagonal like its initial value \(q_a\), \(\ell_k C_a\ell_k^*=r_kc_{a_k}q_{b_k}\), and \(\ell_k^*\ell_k C_a=C_a\ell_k^*\ell_k=r_kc_{a_k}q_{a_k}\), and the commutator with \(H_0\) vanishes. The coefficients consequently solve the real linear equation
\begin{equation}
\dot c_j=\sum_k r_k c_{a_k}
\left(g_{w,k}g_{v,k}\delta_{j,b_k} -\tfrac12(g_{w,k}^2+g_{v,k}^2)\delta_{j,a_k}\right).
\end{equation}
Jump \(k\) takes the level \(a_k\) to the level \(b_k\) and contributes \(r_kg_{w,k}g_{v,k}\geq0\) to the coefficient of \(c_{a_k}\) in \(\dot c_{b_k}\), so every off-diagonal coefficient is nonnegative. If \(B\) is  the matrix of this linear equation, \(\dot c_j=\sum_iB_{ji}c_i\), choose \(c\) so \(B+cI\) is entrywise nonnegative. Each power of \(B+cI\) is then entrywise nonnegative, and \(e^{tB}=e^{-ct}e^{t(B+cI)}\) has the same property. Starting with \(c_j(0)=\delta_{ja}\) gives \(C_a(t)\geq0\), and it follows that the trace norm equals its trace.

To relate \(C_a\) to the records, denote the two dilated preparations by \(\psi_v=V_t(\ket v\otimes\ket a)\) and \(\psi_w=V_t(\ket w\otimes\ket a)\). Expanding \(\Tr_E(V_t\eta V_t^*)\) in a Kraus basis gives \(\sum_\mu K_\mu\eta K_\mu^*=\Phi_{t*}(\eta)\) even for the nonpositive cross input. Hence \(\Tr_E\ket{\psi_w}\bra{\psi_v}=e_{wv}\otimes C_a(t)\). Factoring off the sector matrix unit leaves its nonzero singular values unchanged. By the polar calculation of Theorem~\ref{thm:information_weights},
\(\|C_a\|_1=f_a\). On observables, \(\ell_k^*(\sum_j u_jq_j)\ell_k=r_ku_{b_k}q_{a_k}\), so the internal diagonal space is invariant and \(\Phi_t(S_{vw})=e_{vw}\otimes\sum_j u_jq_j\). The trace against the initial cross input singles out \(u_a\), giving
\begin{align}
u_a &=\Tr\bigl(\Phi_t(S_{vw})(e_{wv}\otimes q_a)\bigr)
 =\Tr\bigl(S_{vw}\Phi_{t*}(e_{wv}\otimes q_a)\bigr)\notag\\
&=\Tr\bigl((e_{vw}e_{wv})\otimes C_a\bigr ) =\Tr C_a=\norm{C_a}_1=f_a.
\end{align}
Finally, the source product is
\((e_{wv}e_{vw})\otimes\sum_{a,b}f_a^{*} f_b q_aq_b =e_{ww}\otimes\sum_a f_a^2q_a\), since the \(f_a\) are real and nonnegative.
Since \(\Phi_t(S_{vw}^*S_{vw})=p_w=e_{ww}\otimes\sum_aq_a\), subtracting gives \(\Ds_t(S_{vw})=e_{ww}\otimes\sum_a(1-f_a^2)q_a\).
\end{proof}

A time-resolved measurement of the jumps attains these fidelities, for every generator of the  family \eqref{eq:shared_jumps} (Supplement~\ref{supp:shared_classical_records}) and explicitly in the model (Section~\ref{sec:classical_jump_records}).

The  sharing of the jump index \(k\) across sectors enters in the product \(g_{w,k}g_{v,k}\). To see its physical meaning, give the
environment an additional record of the sector at each jump by replacing \(L_k\) with the separate
\(L_{j,k}=g_{j,k}e_{jj}\otimes\ell_k\). For \(v\ne w\) the term \(L_{j,k}\eta L_{j,k}^*\) then contains \(e_{jj}e_{wv}e_{jj}=0\) for every \(j\), so the gain term \(g_{w,k}g_{v,k}\ell_kC_a\ell_k^*\) of \eqref{eq:shared_cross_generator} is absent and the coherence between the sectors survives only while no jump occurs. The diagonal-sector population rates can agree in both descriptions, although the dynamics of the coherence between the sectors differ. In the terms of~\cite{Aberg2004}, the two descriptions are different gluings of the same sector channels.  It is therefore essential to specify which jump operators are shared between sectors: the sector channels alone do not determine the eigenvalues of an individual edge.

\subsection{Reservoir amplitudes and their internal preparation}\label{sec:reservoir_weights}
We now compute the evolved edge \(\Phi_t(S_{21})\) in the common-reservoir model of Example~\ref{ex:common_reservoir}.

\begin{example}[Common-reservoir model: edge amplitudes]\label{ex:reservoir_amplitudes}
We continue with the common-reservoir model of Example~\ref{ex:common_reservoir}. Put \(G_\gamma:=\gamma_++\gamma_-\),  the relaxation rate of the internal populations at unit coupling. To compute \(\Phi_t(S_{21})\) for the edge \(S_{21}=e_{21}\otimes I_2\) from sector \(1\) to sector \(2\), first act with the generator on an arbitrary \(e_{ij}\otimes B\), with \(B\) a matrix on the internal factor \(\CZ^2\). Define \(R:=\ell_-^*\ell_-+\ell_+^*\ell_+=\gamma_+q_0+\gamma_-q_1\). The matrix identities \(Fe_{ij}F=g_ig_je_{ij}\),
\(F^2e_{ij}=g_i^2e_{ij}\), and \(e_{ij}F^2=g_j^2e_{ij}\) give
\begin{align}
\LL(e_{ij}\otimes B) 
=e_{ij}\otimes\left(i[\Omega q_1,B]  +g_ig_j\sum_{\nu=\pm}\ell_\nu^*B\ell_\nu
 -\tfrac12(g_i^2RB+g_j^2BR)\right).
\end{align}
For example the term \(L_\nu^*(e_{ij}\otimes B)L_\nu\) is \((F\otimes\ell_\nu^*)(e_{ij}\otimes B)(F\otimes\ell_\nu)  =g_i g_j e_{ij}\otimes\ell_\nu^*B\ell_\nu\).
In the anticommutator one factor \(F^2\) acts entirely on row \(i\) or on column \(j\), which gives the two separate squares \(g_i^2,g_j^2\).  At \(i=2,j=1,B=I_2\), the jump term has coefficient \(\kappa\) and the two anticommutator terms have mean coefficient
\(m:=(1+\kappa^2)/2\). Therefore
\begin{align}
\LL(S_{21})
&=(\kappa-m)e_{21}\otimes R\notag\\
&=-\tfrac12(\kappa-1)^2 e_{21}\otimes(\gamma_+q_0+\gamma_-q_1).
\label{eq:selective_initial_weight}
\end{align}
The right side of \eqref{eq:selective_initial_weight} is \(e_{21}\) times a diagonal matrix, so we try a diagonal one: set \(B=f_0q_0+f_1q_1\). The Hamiltonian commutator vanishes, and
\(RB=BR=\gamma_+f_0q_0+\gamma_-f_1q_1\). The jump terms are \(\ell_+^*B\ell_+=\gamma_+f_1q_0\) and \(\ell_-^*B\ell_-=\gamma_-f_0q_1\).  For \(\Phi_t(S_{21})=e_{21}\otimes B_t\), the coefficient at \(q_0\) changes by \(\gamma_+(\kappa f_1-mf_0)\) per unit time. At \(q_1\) it changes by \(\gamma_-(\kappa f_0-mf_1)\). Since \(B_0=I_2\), the diagonal matrix with \(f_0=f_1=1\), this gives
\begin{equation}
\frac d{dt}\begin{pmatrix}f_0\\f_1\end{pmatrix}
=M_\kappa\begin{pmatrix}f_0\\f_1\end{pmatrix},\qquad
M_\kappa=
\begin{pmatrix}
 -m\gamma_+&\kappa\gamma_+\\
 \kappa\gamma_-&-m\gamma_-
\end{pmatrix},\qquad
\begin{pmatrix}f_0(0)\\f_1(0)\end{pmatrix}=\begin{pmatrix}1\\1\end{pmatrix}.
\label{eq:selective_weight_ode}
\end{equation}
In particular \(f'_0(0)=-(\kappa-1)^2\gamma_+/2\) and \(f'_1(0)=-(\kappa-1)^2\gamma_-/2\). Thus the two eigenvalues of the source weight, \(f_0^2,f_1^2\), have different initial slopes when \(\gamma_+\ne\gamma_-\).
For example, \(\frac{d}{dt}|_{t=0}\langle1,a|\Ws_{S_{21}}(t)|1,a\rangle
=2f_a(0)f'_a(0)=-(\kappa-1)^2r_a\). This derivative concerns a definite internal preparation.

Write \(\boldsymbol 1=(1,1)^{\mathsf T}\). As in the proof of Corollary~\ref{prop:shared_jump_weight}, \(e^{tM_\kappa}\) is entrywise nonnegative. Its exponential series after a positive diagonal shift includes a strictly positive diagonal contribution, so both components of \(f(t)=e^{tM_\kappa}\boldsymbol1\) are positive for finite \(t\). Also \(M_\kappa\boldsymbol1=(\kappa-m)(\gamma_+,\gamma_-)^{\mathsf T}\) is strictly negative. Thus \(f'(t)=e^{tM_\kappa}M_\kappa\boldsymbol1\leq0\) componentwise, giving \(f_a(t)\leq1\). To determine the limits of \(f_a(t)\), compute
\begin{equation}
\tr M_\kappa=-mG_\gamma<0,\qquad
\det M_\kappa=(m^2-\kappa^2)\gamma_+\gamma_-
=\tfrac14(\kappa^2-1)^2\gamma_+\gamma_->0 .
\end{equation}
The discriminant is \(m^2(\gamma_+-\gamma_-)^2+4\kappa^2\gamma_+\gamma_->0\). Both eigenvalues are real and negative, since their sum is negative and their product positive. Thus \(B_t\to0\). At every finite time \(B_t\) is positive and invertible. In its polar decomposition the partial isometry of \(e_{21}\otimes B_t\) remains \(S_{21}\). The decay in squared amplitude is visible in the eigenvalues before it is visible in the support projection.
\end{example}

\subsection{\texorpdfstring{Recovering the amplitudes from the reservoir}{Recovering the amplitudes from the reservoir}}\label{sec:environmental_weight}
The  amplitudes \(f_0\), \(f_1\) of Example~\ref{ex:reservoir_amplitudes} can be read from the records that the reservoir holds.

\begin{example}[Common-reservoir model: environment records]\label{ex:reservoir_records}
  In the dilation \(V_t\) of \eqref{eq:dilation}, which preserves the sector label, an input of the system at internal level \(a\), in sector \(1\) or in sector \(2\), has joint outputs
\begin{align}
V_t(\ket1\otimes\ket a)&=\ket1\otimes\ket{\Psi_{1,a}(t)},&
V_t(\ket2\otimes\ket a)&=\ket2\otimes\ket{\Psi_{\kappa,a}(t)}.
\label{eq:conditional_dilation}
\end{align}
These equations define the vectors \(\Psi_{g,a}(t)\). The first index on \(\Psi\) is the coupling strength. The vectors after factoring out the sector label lie in
\(\CZ^2\otimes\HH_E\). Tracing a diagonal outer product over the internal level gives a record, as in \eqref{eq:conditional_environment}. The cross outer product traced over the reservoir acts on the internal level; it is the cross operator of the proof of Theorem~\ref{thm:information_weights}. Explicitly,
\begin{align}
\sigma^E_{g,a}(t)&=\Tr_{\CZ^2}\ket{\Psi_{g,a}(t)}\bra{\Psi_{g,a}(t)},\notag\\
C_a(t)&=\Tr_E\ket{\Psi_{\kappa,a}(t)}\bra{\Psi_{1,a}(t)} .
\label{eq:environment_cross_operator}
\end{align}
\end{example}
\begin{proposition}[Reservoir record amplitudes]\label{prop:environment_fidelity}
For \(a=0,1\), the level amplitude from \eqref{eq:selective_weight_ode} obeys
\begin{equation}
f_a(t)=\ff(\sigma_{1,a}^E(t),\sigma_{\kappa,a}^E(t)). \label{eq:environment_fidelity_weight}
\end{equation}
\end{proposition}
\noindent The proof is given in Supplement~\ref{supp:reservoir_fidelity_proof}.

Substituting \eqref{eq:environment_fidelity_weight} into  the source defect of Corollary~\ref{prop:shared_jump_weight}, \eqref{eq:shared_jump_weight} with \(v=2\), \(w=1\), gives the source defect for each internal preparation,
\begin{equation}
\Ds_t(S_{21})= e_{11}\otimes\sum_{a=0}^1 \bigl[1-\ff(\sigma^E_{1,a}(t),\sigma^E_{\kappa,a}(t))^2\bigr]q_a .
\label{eq:environmental_source_defect}
\end{equation}
 The identity \eqref{eq:environment_fidelity_weight} resolves the preparation dependence of the source, range, and path products of this edge: each is given, level by level, by the fidelity of two records, i.e.,  for the level \(a\), \(\langle1,a|\Ws_{S_{21}}(t)|1,a\rangle=\langle2,a|\Wr_{S_{21}}(t)|2,a\rangle=\ff(\sigma^E_{1,a}(t),\sigma^E_{\kappa,a}(t))^2\).

\subsection{Classical jump records and the same edge weights}\label{sec:classical_jump_records}
A measurement of the reservoir turns its records into classical data, and this measurement attains the fidelities.

\begin{example}[Common-reservoir model: classical jump records]
A time-resolved measurement of the two jump channels of the common-reservoir model, i.e., excitation and decay of the internal level, gives a classical record. Starting at level \(a\),
the directions of the jumps alternate. For \(n\) events at times \(0<s_1<\cdots<s_n<t\), set \(s_0=0,\ s_{n+1}=t\), and let \(a_j\) equal \(a\) for even \(j\), \(1-a\) for odd \(j\). At coupling \(g\) the exit rate during interval \(j\) is \(g^2r_{a_j}\). An event following a waiting time \(u\) contributes \(g^2r_{a_j}e^{-g^2r_{a_j}u}\). The last interval has no event and contributes the probability \(e^{-g^2r_{a_n}(t-s_n)}\) that none occurs. Multiplication gives the probability density \(p_g(s_1,\ldots,s_n)\) of the record with \(n\) events at these times,
\begin{equation}
 p_g(s_1,\ldots,s_n)=
 \left(\prod_{j=0}^{n-1}g^2r_{a_j}\right)
 \exp\!\left[-g^2\sum_{j=0}^{n}r_{a_j}(s_{j+1}-s_j)\right].
\end{equation} 
For \(n=0\) the product is one and the probability is \(e^{-g^2r_a t}\).
To compare the records  of sector \(1\), with coupling \(1\), and of sector \(2\), with coupling \(\kappa\), we use one reference measure \(\mu\) on the space of records, with respect to which both have densities. A record is a number \(n\) of events with their times, and \(\mu\) gives unit mass to the empty record and is Lebesgue measure on the set \(\{0<s_1<\cdots<s_n<t\}\) of ordered event times, for each \(n\geq1\). Denote by \(P^{(g)}_{a,t}\) the probability distribution of the record up to time \(t\), for a system started at level \(a\) with coupling \(g\); it has density \(p_g\) with respect to \(\mu\). The classical root fidelity of \(P^{(1)}_{a,t}\) and \(P^{(\kappa)}_{a,t}\) is
\begin{equation}
h_a(t)=\int\sqrt{\frac{dP^{(1)}_{a,t}}{d\mu}
                  \frac{dP^{(\kappa)}_{a,t}}{d\mu}}\,d\mu
\end{equation}
for any measure with respect to which both distributions have densities. The no-event record has probabilities \(e^{-r_at}\) and \(e^{-\kappa^2r_at}\), giving a contribution \(e^{-mr_at}\). If the first event occurs at \(s\), the two first-event densities are \(r_ae^{-r_as}\) and \(\kappa^2r_ae^{-\kappa^2r_as}\). Their geometric mean is
\(\kappa r_ae^{-mr_as}\), and the rest of the records start in level \(1-a\). Given the first event at time \(s\), what remains on \([s,t]\) is a record of length \(t-s\) started in level \(1-a\), and the two densities factorize, so the integral over the remaining events is \(h_{1-a}(t-s)\). Adding the empty record to the integral over the time \(s\) of the first event gives
\begin{equation}
h_a(t)=e^{-mr_at}+\int_0^t\kappa r_a e^{-mr_a s}h_{1-a}(t-s)\,ds.
\end{equation}
To turn this integral equation into a differential equation, substitute \(u=t-s\) in the integral and multiply by \(e^{mr_at}\). This gives \(e^{mr_at}h_a(t)=1+\int_0^t\kappa r_ae^{mr_au}h_{1-a}(u)\,du\). Differentiating gives
\begin{equation}
\frac{d}{dt}(e^{mr_at}h_a(t))=\kappa r_a e^{mr_at}h_{1-a}(t),
\quad\hbox{so}\quad
\dot h_a=-mr_a h_a+\kappa r_a h_{1-a}.
\end{equation}
Initially the record is empty, \(h_a(0)=1\). Uniqueness of the solution of \eqref{eq:selective_weight_ode} now gives \(h_a=f_a\), recovering the quantum fidelity from this environment measurement.  Consequently the corresponding eigenvalue of the source weight is
\begin{align}
\langle1,a|\Ws_{S_{21}}(t)|1,a\rangle
&=\ff(\sigma^E_{1,a}(t),\sigma^E_{\kappa,a}(t))^2\notag\\
&=\left(\int
\sqrt{\frac{dP^{(1)}_{a,t}}{d\mu}\,
      \frac{dP^{(\kappa)}_{a,t}}{d\mu}}\,d\mu\right)^2.
\label{eq:classical_record_weight}
\end{align}
The same number enters the source and range products and the path product.  This demonstrates that a  more distinguishable classical record decreases the corresponding edge weight and increases the same component of the path defect.
\end{example}

\subsection{The same internal matrix appears in a path relation}\label{sec:selective_path}
We now follow a relation between two paths.

\begin{example}[Common-reservoir model: a path relation]\label{ex:reservoir_path}
In the common-reservoir model of Example~\ref{ex:common_reservoir}, the equal coupling on sectors \(0,1\) makes \(S_{10}\) commute with \(H\), both jumps, and their adjoints. 
Substitution into \eqref{eq:gksl} gives
\begin{equation}
\Phi_t(S_{10})=S_{10}. \label{eq:selective_surviving_edge}
\end{equation}
Its adjoint \(S_{01}\) is fixed as well. The vertices are fixed, so \eqref{eq:edge_relations} and the Hermiticity of \(B_t\) give
\begin{align}
\Ds_t(S_{21})&=p_1-(e_{12}\otimes B_t)(e_{21}\otimes B_t)
=e_{11}\otimes(I_2-B_t^2),\notag\\
\Dr_t(S_{21})&=p_2-(e_{21}\otimes B_t)(e_{12}\otimes B_t)
=e_{22}\otimes(I_2-B_t^2). \label{eq:selective_edge_defects}
\end{align}
The equation for \(S_{02}\) on the diagonal internal space has the same jump coefficient \(g_0g_2=\kappa\) and the same anticommutator coefficient \(m\) as the equation for \(S_{21}\) in Example~\ref{ex:reservoir_amplitudes}, so \(\Phi_t(S_{02})=e_{02}\otimes B_t\). Initially \(S_{02}S_{21}=S_{01}\). Evolving this relation in the two possible orders now gives
\begin{align}
\Phi_t(S_{02}S_{21})&=\Phi_t(S_{01})=S_{01},\notag\\
\Phi_t(S_{02})\Phi_t(S_{21})&=(e_{02}e_{21})\otimes B_t^2=e_{01}\otimes B_t^2,\notag\\
D_t(S_{02}^*,S_{21})&=e_{01}\otimes(I_2-B_t^2).
\label{eq:selective_path_defect}
\end{align}
The source defect of \(S_{21}\), the range defect of \(S_{02}\), and this mixed defect all have norm \(\norm{I_2-B_t^2}\),  hence \eqref{eq:path_cs} is an equality. The operation \(S_{01}\) remains fixed, while implementing it by a path through sector \(2\) loses the specified internal amplitudes leaving only \(e_{01}\otimes B_t^2\).

The same model separates the weight of a given edge from the maximum over edges in \eqref{eq:corner_information} particularly simply. In the \(01\) corner, \(S_{01}=e_{01}\otimes I_2\) is fixed, so the corner norm is one. Choose another edge there, \(T=e_{01}\otimes\ket0\bra1\).
For \(B=\ket0\bra1\), each jump term \(\ell_\nu^*B\ell_\nu\) is zero, while \(RB+BR=(\gamma_++\gamma_-)B\) and \(i[\Omega q_1,B]=-i\Omega B\). Since \(g_0=g_1=1\), the formula for \(\LL(e_{ij}\otimes B)\) in Example~\ref{ex:reservoir_amplitudes} now gives
\begin{equation}
 \Phi_t(T)=e^{-(G_\gamma/2+i\Omega)t}T,\qquad
 \Ws_T(t)=e^{-G_\gamma t}\,e_{11}\otimes q_1 .
\end{equation}
 An observation of the edge \(T\), of interference between  level \(1\) and level \(0\)   (Supplement~\ref{supp:chosen_edge}), sees its amplitude decay as \(e^{-G_\gamma t/2}\), even though another edge between the same sectors, \(S_{01}\), retains full amplitude. 

Thus \eqref{eq:classical_record_weight} and \eqref{eq:selective_path_defect} identify the same record overlap fidelity in an individual weight and in a composition defect.
\end{example}

Theorem~\ref{thm:information_weights} concerns the largest singular value of an evolved edge. Supplement~\ref{sec:supp_spectral} extends it to the sum of the \(k\) largest, optimized over edges, by preparing \(k\) orthogonal inputs together with a reference system.

\section{Surviving relations and pointer algebras}\label{sec:asymptotic_algebra}
The identities of Section~\ref{sec:representations} describe the amplitude retained by an edge at a given time. We now ask for which operations \(X\), the reduced evolution \(\Phi_t\) preserves the products \(XY\) and \(YX\) with every observable \(Y\) for all times.
These operations form the retained algebra \(\N\). Its structure will distinguish two situations: an edge that decays while another edge with the same endpoints is retained, as \(T\) and \(S_{01}\) in Example~\ref{ex:reservoir_path}, and  a pair of endpoints between which no edge is retained.  We first derive the product rule that produces these defects in a QMS, which  supplies the finite-time integral used in the survival and obstruction arguments.

\subsection{The product rule for a GKSL generator}\label{sec:dissipation}
For small positive \(h\), the exponential series of \(\Phi_h=e^{h\LL}\) gives \(\Phi_h(Z)=Z+h\LL(Z)+O(h^2)\) in operator norm. Since the channel preserves adjoints, \(\Phi_h(X)^*=\Phi_h(X^*)\) for a positive map, so does its derivative, \(\LL(X)^*=\LL(X^*)\). 
Multiplying the two expansions, of \(\Phi_h(X)^*=\Phi_h(X^*)\) and of \(\Phi_h(Y)\), gives
\begin{equation}
\Phi_h(X)^*\Phi_h(Y)  =X^*Y+h\bigl(\LL(X^*)Y+X^*\LL(Y)\bigr)+O(h^2).
\end{equation}
In contrast, \(\Phi_h(X^*Y)=X^*Y+h\LL(X^*Y)+O(h^2)\). Subtracting the first expansion from the second gives the defect \(D_h(X,Y)=h\Gamma(X,Y)+O(h^2)\), with
\begin{align}
\Gamma(X,Y)&:=\LL(X^*Y)-\LL(X^*)Y-X^*\LL(Y)\notag\\
&=\sum_k[L_k,X]^*[L_k,Y]. \label{eq:gamma}
\end{align}
The commutator form in \eqref{eq:gamma} is derived in Supplement~\ref{supp:gamma_expansion}, the map \(\Gamma\) is Lindblad's dissipation function \cite{doi:10.1007/BF01608499}. The same function can be read directly from one interaction  \eqref{eq:access_collision}. The interaction evolves \(X\in\A\) to \(U_h^*XU_h=X+i\sqrt h\,[G,X]+O(h)\). Since \(P(G)=0\), bimodularity gives \(P([G,X])=0\), so the term \(i\sqrt h\,[G,X]\) is not accessible: to leading order this term is the remainder \eqref{eq:access_setup} of \(U_h^*XU_h\). By Theorem~\ref{thm:partial_access} the defect of the  interaction is the accessible part of the product of two remainders, \(hP([G,X]^*[G,Y])+O(h^{3/2})\). By \eqref{eq:access_generator} the same defect is \(h\Gamma(X,Y)+O(h^{3/2})\). Therefore
\begin{equation}
\Gamma(X,Y)=P\bigl([G,X]^*[G,Y]\bigr).
\label{eq:gamma_access}
\end{equation}
For \(X=Y\) the dissipation is the conditional mean square of the commutator \([G,X]\) with the coupling. In particular \(\Gamma(X,X)\geq0\). Taking its expectation on any \(y\) gives \(\sum_k\norm{[L_k,X]y}^2\), so \(\Gamma(X,X)=0\) exactly when every \([L_k,X]\) vanishes. The same calculation with \(X^*\) applies to the adjoints of the jumps.

The dissipation function is also the rate at which an edge weight falls behind its own transport. With \(S(t)=\Phi_t(S)\), differentiation of the weight \(\Ws_S(t)=S(t)^*S(t)\) gives
\begin{align}
\frac{d}{dt}\Ws_S(t)
 &=\LL( S(t)^*)S(t) + S(t)^*\LL( S(t) )\notag\\
 &=\LL( S(t)^* S(t))-\Gamma(S(t),S(t))
 =\LL(\Ws_S(t))-\Gamma(S(t),S(t)). \label{eq:weight_evolution}
\end{align}
By \eqref{eq:weight_evolution} the weight changes by its transport under \(\LL\) less the dissipation. The transport term, \(\LL(\Ws_S(t))\), has no definite sign, so the weight need not decrease. What holds at every time is the bound of \eqref{eq:edge_relations}, \(\Ws_S(t)\leq\Phi_t(p)\).

\subsection{Accumulation between times zero and \(t\)}\label{sec:accumulation}
To express a finite-time defect using \(\Gamma\), we need to compare evolving the product with multiplying the evolved factors. We introduce an intermediate time \(u\), evolve the factors separately until \(u\), multiply them,
and evolve the product for the remaining time \(t-u\). This defines
\begin{equation}
F(u):=\Phi_{t-u}\bigl((\Phi_uX)^*(\Phi_uY)\bigr),\qquad 0\leq u\leq t.
\end{equation}
At the endpoints, \(F(0)=\Phi_t(X^*Y)\) and \(F(t)=\Phi_t(X)^*\Phi_t(Y)\).
Set \(X_u=\Phi_uX,\ Y_u=\Phi_uY\). When \(u\) increases by small \(h\), the outer map \(\Phi_{t-u}\) changes to \(e^{(t-u)\LL}(I-h\LL+O(h^2))\). Meanwhile the product \(X_u^*Y_u\) inside it changes by \(h(\LL(X_u^*)Y_u+X_u^*\LL(Y_u))+O(h^2)\). Collecting the terms of order \(h\) gives 
\begin{align}
F'(u)
 &=-\Phi_{t-u}\LL(X_u^*Y_u)   +\Phi_{t-u}\bigl(\LL(X_u^*)Y_u+X_u^*\LL(Y_u)\bigr)\notag\\
 &=-\Phi_{t-u}\bigl(\Gamma(X_u,Y_u)\bigr).
\end{align}
The endpoint difference \(F(0)-F(t)\) is therefore
\begin{equation}
D_t(X,Y)=\int_0^t\Phi_{t-u}\bigl(\Gamma(\Phi_uX,\Phi_uY)\bigr)\,du .
\label{eq:integrated_kernel}
\end{equation}
This integral transports each instantaneous contribution from its time of production to the observation time. 

To compose two time intervals, add and subtract \(\Phi_t\bigl((\Phi_sX)^*\Phi_sY\bigr)\) in the definition \eqref{eq:defect_kernel} of \(D_{s+t}(X,Y)\). With \(\Phi_{s+t}=\Phi_t\Phi_s\) this gives
\begin{equation}
D_{s+t}(X,Y)=\Phi_t(D_s(X,Y))+D_t(\Phi_sX,\Phi_sY).
\label{eq:kernel_cocycle}
\end{equation}
Using positivity of the second defect kernel, \eqref{eq:kernel_positivity}, for a finite family \(X_1,\ldots,X_n\) gives
\begin{equation}
[D_{s+t}(X_i,X_j)]_{ij}\ \geq\
[\Phi_t(D_s(X_i,X_j))]_{ij}. \label{eq:transported_order}
\end{equation}
The defect at a later time is thus the transported earlier defect plus a positive contribution from the later interval.

We now ask on which operations the defects never appear. An observable \(X\) might preserve one particular product \(XY\) without preserving its other products. To obtain an algebra of operations that can be composed with arbitrary accessible observables, impose both multiplication identities
\(\Phi(XY)=\Phi(X)\Phi(Y)\) and \(\Phi(YX)=\Phi(Y)\Phi(X)\) for every \(Y\). For a UCP map the multiplicative domain \cite{Choi1974} is
\begin{equation}
\MD(\Phi):=\{X:D(X,X)=D(X^*,X^*)=0\}. \label{eq:md}
\end{equation}
Indeed, as for \(p_v\) in \eqref{eq:vertex_multiplicative}, the right side gives \(C(X)=C(X^*)=0\) in \eqref{eq:defect_factorization}. Thus \(D(X^*,Y)=0\) proves the
first product identity, and \(D(Y^*,X)=0\) proves the second. Conversely, take \(Y=X^*\) in the two identities. 
Supplement~\ref{supp:exact_survival_proof} shows that \(\MD(\Phi)\) is a unital \(C^*\)-algebra.

\begin{proposition}[Exact survival at positive times]\label{prop:exact-survival}
For a norm-continuous QMS on a finite-dimensional algebra \(\A\),
\begin{equation}
\MD(\Phi_t)=\N\defeq\bigcap_{s\geq0}\MD(\Phi_s)\quad(t>0).
\label{eq:constant_domain}
\end{equation}
The algebra \(\N\) is invariant and the restriction \(\alpha_t=\Phi_t|_\N\) extends to a group of \(*\)-automorphisms.
\end{proposition}
\noindent The proof is given in Supplement~\ref{supp:exact_survival_proof}.

\subsection{Computing the retained algebra from the jumps}\label{supp:pointer_details}
For a generator \eqref{eq:gksl} on \(\A=M_d\), define \(\operatorname{ad}_H(X)=[H,X]\) and let
\begin{equation}
\RR=C^*\!\left(I,\,(\operatorname{ad}_H)^j(L_k),\,
  (\operatorname{ad}_H)^j(L_k^*):j\geq0,\ k\right),
\label{eq:pointer_noise_algebra}
\end{equation}
be the $C^*$-algebra generated by the identity, the jump operators, their adjoints, and their repeated commutators with \(H\). 
The retained algebra is the commutant of this algebra, \(\N=\RR'\), the standard commutant description \cite{doi:10.1142/S0129055X16500033}. In words, an observable is retained exactly when it commutes with every jump operator, every adjoint of a jump operator, and all their repeated commutators with \(H\); a retained observable then evolves by \(\Phi_t(X)=e^{itH}Xe^{-itH}\). Supplement~\ref{supp:exact_survival_proof} gives the derivation. 

Use the fact that a finite-dimensional \(C^*\)-algebra is a direct sum of full matrix algebras, here \(\RR\cong\bigoplus_aM_{n_a}\) with central projections \(z_a\).  Each summand \(M_{n_a}\) acts on \(z_a\HH\) as \(n_a\times n_a\) matrices repeated \(m_a\) times, which gives
\begin{equation}
\HH=\bigoplus_a(\CZ^{m_a}\otimes\CZ^{n_a}),\qquad
\RR=\bigoplus_a(I_{m_a}\otimes M_{n_a}) .
\label{eq:pointer_noise_decomposition}
\end{equation} 
 This shows that the commutant of a summand \(I_{m_a}\otimes M_{n_a}\) is \(M_{m_a}\otimes I_{n_a}\), which gives
\begin{equation}
\N=\RR'=\bigoplus_a(M_{m_a}\otimes I_{n_a}) .
\label{eq:pointer_commutant_supp}
\end{equation}
Thus \(m_a\) counts the matrix size that survives in the observable algebra, and \(n_a\) is the physical multiplicity of a retained matrix unit; each \(M_{m_a}\otimes I_{n_a}\) is a central summand of \(\N\), with unit \(z_a\), and is one matrix block of \(\N\). The same decomposition appears in quantum error correction, where the commutant of the algebra generated by the noise operators carries the noiseless subsystems \cite{KnillLaflammeViola2000,HolbrookKribsLaflamme2003,DauphinaisKribsVasmer2024}; here they are the matrix factors \(M_{m_a}\).

\subsection{Faithful stationarity and the asymptotic expectation}\label{supp:survival}
We now assume there is a faithful stationary state \(\omega\), one satisfying
\begin{equation}
\omega\Phi_t=\omega\quad(t\geq0),\qquad  X \geq0,\ \omega( X)=0\ \Longrightarrow\ X=0 .
\label{eq:faithful_stationary}
\end{equation}
\begin{proposition}[Asymptotic expectation]\label{prop:asymptotic_expectation}
There is an \(\omega\)-preserving conditional expectation \(\EE:\A\to\N\) such that
\begin{equation}
\EE\Phi_t=\Phi_t \EE=\alpha_t \EE,\qquad
\norm{\Phi_t-\alpha_t \EE}\longrightarrow0 .
\label{eq:asymptotic_decomposition}
\end{equation}
\end{proposition}
\noindent The proof is given in Supplement~\ref{supp:asymptotic_proof}. By Takesaki's theorem \cite{Takesaki1972}, the existence of this \(\omega\)-preserving expectation means that \(\N\) is invariant under the modular group \(\sigma_s^\rho\) of \(\omega\), where \(\rho\) is the density of \(\omega\), and that \(\EE\sigma_s^\rho=\sigma_s^\rho \EE\) (Supplement~\ref{supp:modular_compatible}).

Proposition \ref{prop:asymptotic_expectation} establishes the finite-dimensional asymptotic decomposition in the form used in the decoherence theory of
\cite{doi:10.1007/s00023-012-0199-3,doi:10.1063/1.5030954,SassoUmanita2021}. It also identifies the distinction between a retained observable and a fixed observable; retained observables evolve reversibly, by \(\alpha_t(X)=e^{itH}Xe^{-itH}\) (Section~\ref{supp:pointer_details}), while the fixed ones satisfy \(\alpha_t(X)=X\).

There is a scalar way to measure how much of an observable \(X\) is outside the retained algebra \(\N\). Write \(\norm X_\omega^2=\omega(X^*X)\). This is the norm of \(X\) in the Hilbert space of the stationary state. We apply \(\omega\) first to the definition \eqref{eq:defect_kernel}, then to its integral \eqref{eq:integrated_kernel}, stationarity removes the outer channel from both expressions, giving
\begin{align}
\omega(D_t(X,X))
 &=\norm X_\omega^2-\norm{\Phi_t X}_\omega^2\notag\\
 &=\int_0^t\omega(\Gamma(\Phi_uX,\Phi_uX))\,du .
\label{eq:stationary_defect_balance}
\end{align}
The scalar integrand is nonnegative for every \(X\), so \(\omega(D_t(X,X))\) is nondecreasing, with no assumption on the sectors.  This number is the loss of \(X\) under the dynamics, \(\norm X_\omega^2-\norm{\Phi_tX}_\omega^2\). The loss vanishes for every \(t\) exactly when \(X\in\N\), as shown below. The loss therefore quantifies, for a single observable, the splitting of Blanchard and Olkiewicz \cite{doi:10.1142/S0129055X03001631} into a part of the algebra that evolves reversibly and a part that vanishes in time. 
To identify its 
limit, note that
\begin{equation}
\omega(\EE(X)^*(X- \EE(X)))
=\omega\bigl( \EE(\EE(X)^*(X- \EE(X)))\bigr)
=\omega(\EE(X)^*(\EE(X)-\EE^2(X)))=0 .
\end{equation}
Expanding the square of \(X=\EE(X)+(X-\EE(X))\) gives \(\norm X_\omega^2=\norm{\EE(X)}_\omega^2+\norm{X-\EE(X)}_\omega^2\). Also \(\alpha_t\) preserves \(\omega\) on \(\N\), hence \(\norm{\alpha_t \EE(X)}_\omega=\norm{\EE(X)}_\omega\). Since
\eqref{eq:asymptotic_decomposition} implies \(\norm{\Phi_tX-\alpha_t \EE(X)}_\omega\to0\), we obtain
\begin{equation}
\lim_{t\to\infty}\omega(D_t(X,X)) =\int_0^\infty\omega(\Gamma(\Phi_tX,\Phi_tX))\,dt
=\norm{X-\EE(X)}_\omega^2 .
\label{eq:defect_distance}
\end{equation}
Replacing \(X\) by \(X^*\) gives the range statement. This reasoning nowhere uses the trace, so the stationary state need not be tracial. Thus the defect accumulated over all time, measured in \(\omega\), is the squared distance  in the norm \(\norm{\cdot}_\omega\)  from \(X\) to the retained algebra.

\subsection{The stationary state of the reservoir}\label{sec:reservoir_stationarity}
Section~\ref{supp:survival} assumes a faithful stationary state. The common-reservoir model has one, and it can be checked directly.

\begin{example}[Common-reservoir model: the stationary state]\label{ex:reservoir_stationary}
For the common-reservoir model \eqref{eq:selective_generator}, define
\begin{equation}
\tau:=\frac{\gamma_-q_0+\gamma_+q_1}{G_\gamma},\qquad
\rho:=\frac{I_3}{3}\otimes\tau . \label{eq:selective_stationary}
\end{equation}
By \eqref{eq:selective_populations} the internal populations relax to \(\tau\) in every sector, at the rate \(g_i^2G_\gamma\) in sector \(i\) (Supplement~\ref{supp:reservoir_retained_proofs}). 
The diagonal state  \(\rho\) is stationary as a matrix, since \(L_\nu\eta L_\nu^*\) and the anticommutator terms preserve diagonality there, the commutator vanishes, and the two population currents cancel in each sector. 
All six diagonal entries are strictly positive, so \(\rho\) is faithful.
\end{example}

\subsection{Commutative pointer observables}\label{sec:pointer_algebra}
We now specialize to a commutative retained algebra, whose minimal projections are the pointer observables of the decoherence literature \cite{Zurek2003}.

\begin{proposition}[Commutative pointer algebra]\label{prop:commutative_pointer}
Under  the assumptions \eqref{eq:faithful_stationary} of a faithful stationary state, write
\begin{equation}
\N=\bigoplus_a(M_{m_a}\otimes I_{n_a}). \label{eq:pointer_commutant}
\end{equation}
It is commutative exactly when every \(m_a=1\). In this case its minimal projections \(z_a\) are fixed and
\begin{align}
\mathcal P=\N&=\Span\{z_1,\ldots,z_r\}\cong C(\{1,\ldots,r\}),
\qquad \Phi_t(z_a)=z_a,\label{eq:pointer_algebra}\\ \intertext{and the dynamics converges to the expectation onto it,}
\Phi_t&\longrightarrow  \EE,\qquad
  \EE(X)=\sum_a\frac{\omega(z_aXz_a)}{\omega(z_a)}z_a .
\label{eq:pointer_expectation}
\end{align} Here \(C(\{1,\ldots,r\})\) is the algebra of complex functions on the labels, and \(\mathcal P\) is the pointer algebra.
\end{proposition}
\begin{proof}
A  central summand \(M_m\otimes I_n\) with \(m\geq2\) contains  \(e_{12}\otimes I_n\) and \(e_{21}\otimes I_n\), whose products in the two orders are \(e_{11}\otimes I_n\) and \(e_{22}\otimes I_n\), so it is noncommutative. When all \(m_a=1\), each block is scalar and multiplication is
pointwise in the label \(a\). An automorphism  preserves products and adjoints, so it must permute the minimal projections. Distinct \(z_a,z_b\) have norm distance one, as seen by testing \(z_a-z_b\) on their ranges.  The path \(t\mapsto\alpha_t(z_a)\) is continuous, starts at \(z_a\) and stays in the finite set \(\{z_b\}\), whose points are at distance one from each other, so it is constant. Thus \(\alpha_t\) fixes each \(z_a\), hence their span, and \eqref{eq:asymptotic_decomposition} gives \(\Phi_t\to \EE\). Since \( \EE(X)\in\N=\Span\{z_a\}\), write \( \EE(X)=\sum_a c_a(X)z_a\) with complex coefficients \(c_a(X)\); then \(z_a \EE(X)z_a=c_a(X)z_a\). Bimodularity and stationarity of \( \EE\) show
\begin{equation}
c_a(X)\omega(z_a)=\omega(z_a \EE(X)z_a)=\omega( \EE(z_aXz_a))=\omega(z_aXz_a).
\end{equation}
Faithfulness ensures \(\omega(z_a)>0\), which proves \eqref{eq:pointer_expectation}.
\end{proof}
For \(a\ne b\), bimodularity gives \( \EE(z_aXz_b)=z_a\EE(X)z_b=0\), hence
\begin{equation}
\Phi_t(z_aXz_b)\longrightarrow0 . \label{eq:pointer_corner_decay}
\end{equation}
For an edge with full supports \(S^*S=z_b\) and \(SS^*=z_a\), \(\Phi_t\) fixes its support projections while \(\Phi_t(S)\) decays. Equations \eqref{eq:edge_relations} and \eqref{eq:defect_distance} consequently give
\begin{align}
\Ds_t(S)&\longrightarrow z_b,\qquad \Dr_t(S)\longrightarrow z_a,\notag\\
\int_0^\infty\omega(\Gamma(\Phi_tS,\Phi_tS))\,dt&=\norm S_\omega^2=\omega(z_b).
\label{eq:pointer_edge_extinction}
\end{align}

 The corresponding state limit, conditional pointer densities, and characters are derived in Supplement~\ref{supp:pointer_states}. When the retained algebra is commutative, every edge between two distinct pointer sectors loses all of its weight. What survives is a classical description, the pointer observables and their probabilities, with no relation left between different labels.

\subsection{The retained algebra in the common-reservoir family}\label{sec:distinct_pointer_couplings}
We compute the retained algebra and the asymptotic expectation of the common-reservoir model, for any number of sectors.

\begin{example}[Common-reservoir model: the retained algebra]\label{ex:reservoir_retained}
Generalize the common-reservoir model \eqref{eq:selective_generator} to \(r\) sectors of the system by setting
\begin{equation}
F=\sum_{i=0}^{r-1}g_i e_{ii},\quad g_i>0,\qquad
H=\Omega I_r\otimes q_1,\qquad L_\pm=F\otimes\ell_\pm .
\label{eq:pointer_family_generator}
\end{equation}
The calculation of Example~\ref{ex:reservoir_stationary}  gives a faithful stationary state \(I_r/r\otimes\tau\). Let \(P_\lambda\) be the spectral projection of \(F\) for the coupling value \(\lambda\). The algebra \(\RR\) of Section~\ref{supp:pointer_details} and its commutant are now (Supplement~\ref{supp:reservoir_retained_proofs})
\begin{align}
\RR&=\bigoplus_{\lambda\in\spec F}(\CZ P_\lambda\otimes M_2),\notag\\
\N&=\RR'=\bigoplus_{\lambda\in\spec F}
\bigl(B(P_\lambda\CZ^r)\otimes I_2\bigr)
\cong\bigoplus_\lambda M_{m_\lambda},\qquad m_\lambda=\rank P_\lambda.
\label{eq:pointer_family_algebra}
\end{align}
Here \(B(P_\lambda\CZ^r)\) is the algebra of all operators on the range of \(P_\lambda\). Every member of this commutant also commutes with \(H\), so \(\alpha_t=\id_\N\).

The asymptotic expectation keeps the part of \(X=\sum_{i,j}e_{ij}\otimes X_{ij}\) between sectors of equal coupling and replaces each internal matrix \(X_{ij}\) by its expectation in \(\tau\) (Supplement~\ref{supp:reservoir_retained_proofs}):
\begin{equation}
\EE(X)=\sum_{g_i=g_j}\Tr(\tau X_{ij})e_{ij}\otimes I_2 . \label{eq:pointer_family_expectation}
\end{equation}
For all distinct couplings this reduces to
\begin{equation}
\N=\Span\{p_i\}\cong\CZ^r,\qquad \EE(X)=\sum_i\Tr(\tau X_{ii})p_i .
\label{eq:pointer_family_classical}
\end{equation}
For the couplings \((1,1,\kappa)\) of Example~\ref{ex:common_reservoir}, the spectral projection of \(F\) for the coupling value \(1\) is \(P_1=e_{00}+e_{11}\). 
Equations \eqref{eq:pointer_family_algebra} and \eqref{eq:pointer_family_expectation} give
\begin{align}
\RR&=(\CZ P_1\oplus\CZ e_{22})\otimes M_2,\qquad
\N=(M_2\otimes I_2)\oplus\CZ I_2\cong M_2\oplus\CZ,\label{eq:selective_surviving_algebra}\\
\EE(X)&=\left(\sum_{i,j=0}^1\Tr(\tau X_{ij})e_{ij} +\Tr(\tau X_{22})e_{22}\right)\otimes I_2 .
\label{eq:selective_expectation}
\end{align}
All \(p_i\) have rank two in \(M_6\). The edge \(S_{10}\) belongs to \(\N\) and implements \(p_0\sim_\N p_1\), where equivalence \(\sim_\N\) means the existence of a partial isometry in \(\N\) with the indicated endpoints. In contrast,
\begin{equation}
p_2\N p_1=\{0\},\qquad p_1\not\sim_\N p_2 .
\label{eq:selective_equivalence_loss}
\end{equation}
Thus the loss of this equivalence concerns every choice of edge with  these endpoints, source \(p_1\) and range \(p_2\). At \(\kappa=1\) all couplings agree, so \(F=I_3\) has the single spectral projection \(I_3\) and \eqref{eq:pointer_family_algebra} gives \(\N=M_3\otimes I_2\): thus the full label matrix algebra survives. The  stationary \(\rho\) from \eqref{eq:selective_stationary} does not depend on \(\kappa\) and commutes with every \(S_{ij}\). Its modular group therefore fixes every \(S_{ij}\) for every \(\kappa\), while the retained algebra changes at \(\kappa=1\):
\begin{equation}
\sigma_s^\rho(S_{ij})=S_{ij},\qquad
\N_1=M_3\otimes I_2,\qquad
\N_\kappa=(M_2\otimes I_2)\oplus\CZ I_2\ \ (\kappa\ne1).
\label{eq:modular_endpoint_comparison}
\end{equation}
The modular action \(\sigma_s^\rho\) is defined in Supplement~\ref{app:modular}. The modular data are the same in the two cases, while the coupling determines the retained algebra. 
\end{example}

\section{Sharp loss of relations and its record dependence}\label{sec:sharp_loss}
An edge in \(\N\) acts inside each central summand of \(\N\) separately. The retained algebra therefore restricts how many source directions can be matched to target directions
without  crossing its central  summands: a direction is matched when an edge in \(\N\) pairs it with a target direction in the same central summand, and unmatched otherwise. We first compute the least loss allowed by  the restriction to central summands  and construct an edge attaining it;  the lower bound on the defects is sharp in this sense. Theorem~\ref{thm:finite_record_obstruction} then uses the record fidelities to bound the remaining weights at finite times on the unmatched directions.

\subsection{Three ways to count endpoint dimensions}\label{sec:sharp_retained_dimensions}
Let \(p,q\in\N\) be projections equivalent in the full matrix algebra \(\A=M_d\), the ambient algebra, in the sense of Murray and von Neumann: the notation \(p\sim_\A q\)
means there is an edge \(S\in\A\) with \(S^*S=p\) and \(SS^*=q\). Such an edge identifies orthonormal bases of \(p\HH\) and \(q\HH\). Conversely, if these bases have the same number of vectors, mapping the \(j\)th vector of the first basis to the \(j\)th vector of the second and extending the map by zero gives the desired partial isometry, thus ambient equivalence is exactly equality of ordinary ranks.

Inside \(\N=\bigoplus_a(M_{m_a}\otimes I_{n_a})\) of \eqref{eq:pointer_commutant_supp}, an implementing edge, meaning an edge with \(S^*S=p\) and \(SS^*=q\), respects each central summand \(M_{m_a}\otimes I_{n_a}\). Write \(pz_a=p_a\otimes I_{n_a}\) for the part of \(p\) in the \(a\)th central summand, whose unit is \(z_a\), and \(k_a(p)=\rank_{M_{m_a}}p_a\). An edge in \(\N\) can match \(p\) to \(q\) exactly when \(k_a(p)=k_a(q)\) for every \(a\). The ambient rank of \(p\) is instead \(\sum_a n_a k_a(p)\), since a rank-one vector of the retained matrix factor occurs in \(n_a\) physical copies.  We can define one more count, the center-valued trace, used to state the sharp bound of Theorem~\ref{thm:sharp_endpoint_defects}, it normalizes each retained rank by the size \(m_a\) of its matrix factor. For \(X=\bigoplus_a(X_a\otimes I_{n_a})\) define
\begin{equation}
\vartheta_\N(X):=\sum_a\frac{\Tr_{m_a}X_a}{m_a}z_a,\qquad
\vartheta_\N(p):=\sum_a\frac{k_a(p)}{m_a}z_a, \label{eq:center_trace}
\end{equation}
these are  the center-valued trace of \(\N\) \cite{doi:10.1007/978-1-4612-6188-9}, the standard dimension function of a finite von Neumann algebra, written out for matrices. The center \(Z(\N)=\{z\in\N: zn=nz\ \text{for every }n\in\N\}\) consists here of scalar multiples of the identity in each matrix summand, with a possibly different scalar in each summand. 
The normalization is the tracial state of each block, \(X_a\mapsto\Tr_{m_a}(X_a)/m_a\). Multiplying the normalized traces by the central projections \(z_a\), instead of adding them, keeps the dimensions in distinct summands separate (Supplement~\ref{supp:block_trace}).

For two self-adjoint elements \(a,b\) of the center \(Z(\N)\), lists of real numbers indexed by the central summands, their meet \(a\wedge b\) is their componentwise minimum. The positive and negative parts of such an element \(c\) are \(c_+ = (|c|+c)/2\) and \(c_- =(|c|-c)/2\), so \(c = c_+- c_-\). Checking the two cases \(a_j\geq b_j\) and \(a_j<b_j\) on scalar components gives
\begin{equation}
a-a\wedge b=(a-b)_+,\qquad b-a\wedge b=(a-b)_- .
\end{equation}
For \(a=(1,0)\), \(b=(1/2,1)\), the meet is \((1/2,0)\), leaving excess source dimension \((1/2,0)\) and excess target dimension \((0,1)\). The order of operations here is essential. In \(\N=M_2\), take \(p=\ket0\bra0\) and \(q=\ket+\bra+\), with \(\ket+=(\ket0+\ket1)/\sqrt2\). Their ranges have zero intersection, yet \(\vartheta_\N(p)\wedge\vartheta_\N(q)=1/2\). The map \(\ket+\bra0\) matches their entire dimensions within the same matrix algebra. The meet of the two traces, and not the trace of \(p\wedge q\), which is zero here, records this possible matching.

\subsection{Sharp bounds on the endpoint defects}\label{supp:sharp}
The count by the center-valued trace  \eqref{eq:center_trace}  also applies, beyond matrices, to a unital inclusion of finite von Neumann algebras \(\N\subseteq \A\),  where \(\vartheta_\N\) is the canonical center-valued trace of \(\N\), which is faithful and normal \cite{doi:10.1007/978-1-4612-6188-9}.  A von Neumann algebra is finite when its unit is not equivalent to a proper subprojection of itself. Write \(p\precsim_\N q\) if \(p\) is equivalent in \(\N\) to a
subprojection of \(q\).  In a finite von Neumann algebra \cite{doi:10.1007/978-1-4612-6188-9} \(p\precsim_\N q\) iff \(\vartheta_\N(p)\leq\vartheta_\N(q)\), and equality of these traces  characterizes equivalence. The elements \(a\wedge b\) and \((a-b)_\pm\) of Section~\ref{sec:sharp_retained_dimensions} are defined as before: the center is abelian, so  its self-adjoint elements are real functions on its spectrum, and the minimum \(a\wedge b=(a+b-|a-b|)/2\) and the positive  and negative parts \((a-b)_\pm=(|a-b|\pm(a-b))/2\) are taken pointwise on its spectrum; for matrices they are the componentwise ones.

Fix \(p,q\in\N\) with \(p\sim_{\A} q\), and a normal conditional expectation \(\EE: \A\to\N\). The set of implementing edges in \( \A\)  is
\begin{equation}
\mathcal I_{\A}(p,q):=\{S\in \A :S^*S=p,\ SS^*=q\}.
\end{equation}
Put \(a=\vartheta_\N(p)\), \(b=\vartheta_\N(q)\), \(d_{p,q}=a-b\), and \(\mu_{p,q}=a\wedge b\). For \(S\) in this set, define the endpoint defects by
\begin{equation}
\Ds_{\EE}(S):=p-\EE(S)^*\EE(S),\qquad
\Dr_{\EE}(S):=q-\EE(S)\EE(S)^* . \label{eq:endpoint_defects}
\end{equation}

\begin{theorem}[Sharp endpoint defects]\label{thm:sharp_endpoint_defects}
Define \(\mathcal R_S:=\mu_{p,q}-\vartheta_\N(\EE(S)^*\EE(S))\). Then \(\mathcal R_S\geq0\) and
\begin{align}
\vartheta_\N(\Ds_{\EE}(S))&=(d_{p,q})_++\mathcal R_S,\notag\\
\vartheta_\N(\Dr_{\EE}(S))&=(d_{p,q})_-+\mathcal R_S .
\label{eq:sharp_central_defects}
\end{align}
In particular \(\vartheta_\N(\Ds_{\EE}(S))\geq(d_{p,q})_+\) and \(\vartheta_\N(\Dr_{\EE}(S))\geq(d_{p,q})_-\) for every implementing edge. 
There exists one edge \(S_0\in\mathcal I_{\A}(p,q)\), depending only on the inclusion and endpoints, which attains both lower bounds for every normal conditional expectation onto \(\N\). A given \(S\) attains this lower bound for \(\EE\) iff \(\EE(S)\) is a partial isometry with initial center-valued dimension \(\mu_{p,q}\).
\end{theorem}
\begin{proof}
First we bound what the expectation \(\EE(S)\) of any implementing edge can carry. We then attain the bound by matching as many directions as possible within the retained algebra and completing the edge in the ambient algebra using only the unmatched directions. 
Since \(S=qSp\) and \(p,q\in\N\), bimodularity gives \( \EE(S)=q \EE(S) p\). The Schwarz inequality after \eqref{eq:defect_factorization}, applied to the UCP map \(\EE\), gives
\begin{equation}
0\leq  \EE(S)^* \EE(S)\leq \EE(S^*S)=p,\qquad
0\leq \EE(S)\EE(S)^*\leq \EE(SS^*)=q .
\end{equation}
Set \(c=\vartheta_\N(\EE(S)^* \EE(S))=\vartheta_\N( \EE(S) \EE(S)^*)\). Applying \(\vartheta_\N\) to these inequalities gives \(c\leq a =\vartheta_\N(p) \) and \(c\leq b =\vartheta_\N(q) \). Therefore \(c\leq\mu_{p,q}\) and \(\mathcal R_S\geq0\). Subtracting \(c\) from the dimensions \(a\) of \(p\) and \(b\) of \(q\) gives
\begin{align}
a-c&=(a-\mu_{p,q})+(\mu_{p,q}-c)=(d_{p,q})_++\mathcal R_S, \label{eq:a_endpointdefect}\\
b-c&=(b-\mu_{p,q})+(\mu_{p,q}-c)=(d_{p,q})_-+\mathcal R_S \label{eq:b_endpointdefect}.
\end{align}
This proves \eqref{eq:sharp_central_defects}. The same residual $\mathcal{R}_S$ appears in both \eqref{eq:a_endpointdefect} and \eqref{eq:b_endpointdefect} because the two products of \(\EE(S)\) have the same center-valued trace.  The proof of \eqref{eq:sharp_central_defects} just given uses  only the center-valued trace on \(\N\), and no trace on \(\A\), so it applies to expectations \(\EE\) that preserve no trace.

We now construct the edge \(S_0\) of the statement, which attains both bounds. In the matrix case its part \(v\) in \(\N\), i.e., the operator   \(\EE(S_0)\), can be constructed directly. Choose orthonormal bases \(x_{a,1},\ldots,x_{a,k_a(p)}\) of \(\ran p_a\) and
\(y_{a,1},\ldots,y_{a,k_a(q)}\) of \(\ran q_a\). Set \(h_a=\min(k_a(p),k_a(q))\) and
\begin{equation}
 v_a=\sum_{j=1}^{h_a}\ket{y_{a,j}}\bra{x_{a,j}},\qquad
 v=\bigoplus_a(v_a\otimes I_{n_a}).
\end{equation}
The product \(v_a^*v_a\) is the sum of the first \(h_a\) source projections, and \(v_av_a^*\) the sum of the first \(h_a\) target projections. Both have normalized dimension \(h_a/m_a\). 
Whenever the source has unused directions in a summand, the target there has none, and conversely, since \(h_a\) equals \(k_a(p)\) or \(k_a(q)\).

Now we implement exactly this choice with the comparison of projections, \(p\precsim_\N q\) exactly when \(\vartheta_\N(p)\leq\vartheta_\N(q)\), so it also works in a finite von Neumann algebra without a finite basis. To show that the bounds are attained we must exhibit an implementing edge in \(\A\). We build it from a partial isometry \(v\in\N\) between the matched parts \(e\leq p\) and \(f\leq q\), the largest parts that an edge in \(\N\) can pair, and a partial isometry \(r\in\A\) between the unmatched parts \(p-e\) and \(q-f\). Let \(z\) be the largest central projection with \(az\geq bz\); for matrices it is the sum of the central projections of the summands where the retained rank of \(p\) is at least that of \(q\).  On \(z\), then \(\vartheta_\N(qz)=bz\leq az=\vartheta_\N(pz)\), so \(qz\precsim_\N pz\); choose \(v_+\in\N\) mapping a subprojection of \(pz\) onto \(qz\). On \(1-z\), in the same way \(p(1-z)\precsim_\N q(1-z)\); choose \(v_-\in\N\) mapping
\(p(1-z)\) onto a subprojection of \(q(1-z)\). Explicitly,
\begin{align}
v_+^*v_+&\leq pz,&v_+v_+^*&=qz,\notag\\
v_-^*v_-&=p(1-z),&v_-v_-^*&\leq q(1-z).
\end{align}
Orthogonality of the central supports makes \(v=v_++v_-\) a partial isometry. Write \(e=v^*v\leq p\) and \(f=vv^*\leq q\).  Both have center-valued trace \(bz+a(1-z)=\mu_{p,q}\): since \(\vartheta_\N(v^*v)=\vartheta_\N(vv^*)\), \(\vartheta_\N(e)=\vartheta_\N(v_+^*v_+)+\vartheta_\N(v_-^*v_-)=\vartheta_\N(qz)+\vartheta_\N(p(1-z))=bz+a(1-z)\), likewise for \(f\), and \(a\wedge b\) equals \(b\) on \(z\), where \(a\geq b\), and \(a\) on \(1-z\).  The unmatched projections \(p'=p-e\) and \(q'=q-f\) thus satisfy
\begin{align}
p'&=zp',&q'&=(1-z)q',\notag\\
\vartheta_\N(p')&=(d_{p,q})_+,&
\vartheta_\N(q')&=(d_{p,q})_- .
\label{eq:residual_central_supports}
\end{align}
In the ambient algebra \(p\sim_{ \A} q\) by hypothesis, and \(e\sim_{ \A} f\) via \(v\). Therefore
\begin{equation}
\vartheta_{ \A}(p')=\vartheta_{\A }(p)-\vartheta_{\A }(e) =\vartheta_{ \A }(q)-\vartheta_{\A }(f)=\vartheta_{ \A }(q').
\end{equation}
Projections of equal center-valued trace are equivalent \cite{doi:10.1007/978-1-4612-6188-9} via  the center-valued trace \cite[Theorem~8.4.3]{KadisonRingroseII}, so in \( \A\) there is \(r\in q' \A p'\) with \(r^*r=p'\), \(rr^*=q'\). For matrices one obtains \(r\) by  mapping an orthonormal basis of \(p'\HH\) onto one of \(q'\HH\); the two subspaces have the same dimension. 
The operator \(r\) can cross central components of \(\N\), i.e., maps the part of \(p\) under \(z\) to the part of \(q\) under \(1-z\), but its expectation, which lies in \(\N\), must vanish.
Indeed \(r=q'rp'\) and \(\EE\) is bimodular, so \(\EE(r)=q'\EE(r)p'\), where \(q'=(1-z)q'\) and \(p'=p'z\) by \eqref{eq:residual_central_supports}. Since \(z\) commutes with \(\EE(r)\in\N\),
\begin{equation}
\EE(r)=q'\EE(r)p'=(1-z)q'\EE(r)p'z=(1-z)zq'\EE(r)p'=0.
\end{equation}
This calculation works for every expectation onto \(\N\). Also \(v^*r=0\) since \(f q'=0\), and \(vr^*=0\) since
\(e p'=0\). Thus, all mixed products disappear for the sum \(S_0=v+r\), giving
\begin{align}
S_0^*S_0&=e+p'=p,& S_0S_0^*&=f+q'=q,\notag\\
\EE(S_0)&=v,& \Ds_{\EE}(S_0)&=p',\qquad \Dr_{\EE}(S_0)=q'.
\label{eq:sharp_edge_construction}
\end{align}
Both lower bounds are attained by the same \(S_0\).

For the equality criterion let \( \EE(S) =u| \EE(S)|\) be the polar decomposition (Section~\ref{sec:sectors}) and \(e_{\EE}=\supp |\EE(S)|=u^*u,\ f_{\EE}=uu^*\).
The identity \(\EE(S)=q\EE(S)p\) gives \(e_{\EE}\leq p,\ f_{\EE}\leq q\). The contraction \(\EE(S)\) satisfies
\(\EE(S)^*\EE(S)=e_{\EE} \EE(S)^*\EE(S) e_{\EE}\leq e_{\EE}\), since it vanishes on \(1-e_{\EE}\) and its spectrum on \(e_{\EE}\) is in \([0,1]\). When \(\mathcal R_S=0\), the relations \(\EE(S)^*\EE(S)\leq e_{\EE}\sim_\N f_{\EE}\), \(e_{\EE}\leq p\) and \(f_{\EE}\leq q\) give
\begin{equation}
\mu_{p,q}=\vartheta_\N(\EE(S)^*\EE(S))\leq\vartheta_\N(e_{\EE}) =\vartheta_\N(f_{\EE})\leq a\wedge b=\mu_{p,q},
\end{equation}
so each inequality is an equality. Faithfulness of the center-valued trace, applied to \(e_{\EE}-\EE(S)^*\EE(S)\geq0\), gives \(\EE(S)^*\EE(S)=e_{\EE}\). Hence \(|\EE(S)|=(\EE(S)^*\EE(S))^{1/2}=e_{\EE}\) and \(\EE(S)=u e_{\EE}=u\),
a partial isometry whose initial dimension is \(\mu_{p,q}\).
Conversely for a partial isometry with that initial dimension, \(\vartheta_\N(\EE(S)^*\EE(S))=\mu_{p,q}\) directly, giving \(\mathcal R_S=0\).  Thus we have shown the two bounds, their attainment by \(S_0\), and the criterion for equality.
\end{proof}

For a commutative pointer algebra \(\N=\mathcal P\) of \eqref{eq:pointer_algebra}, the center-valued trace is the identity, \(\vartheta_\mathcal P=\id\). Projections in \(\mathcal P\) are sums of the \(z_a\) over sets of outcome labels \(a\), each naming one outcome \(z_a\) of a measurement of the pointer observables, and their product is the sum over the intersection. At each outcome label \(a\), \(p\wedge q=pq\),
\((p-q)_+=p(1-q)\), and \((p-q)_-=q(1-p)\). Equation \eqref{eq:sharp_central_defects} therefore gives
\begin{align}
\Ds_{\EE}(S)&=p(1-q)+\mathcal R_S,\qquad
\Dr_{\EE}(S)=q(1-p)+\mathcal R_S,\notag\\
\mathcal R_S&=pq-\EE(S)^*\EE(S)\geq0 . \label{eq:sharp_pointer_defects}
\end{align}
The retained partial isometry \(v\) in the proof of Theorem~\ref{thm:sharp_endpoint_defects} can be \(v=pq\), giving
\begin{equation}
S_0=pq+r,\qquad r^*r=p(1-q),\quad rr^*=q(1-p),\quad \EE(r)=0 .
\label{eq:sharp_pointer_optimizer}
\end{equation}
For example, take the diagonal algebra of \(M_3\), \(p=e_{00}+e_{11}\), and \(q=e_{00}+e_{22}\). Then
\(pq=e_{00}\), \(p(1-q)=e_{11}\), and \(q(1-p)=e_{22}\). The edge \(S_0=e_{00}+e_{21}\) implements \(p\sim_\A q\) and its diagonal expectation is \(e_{00}\). Its source and range defects are exactly \(e_{11}\) and \(e_{22}\).
For disjoint pointer projections, every implementing edge has \(\EE(S)=q\EE(S)p=0\).

For a noncommutative example, in the retained algebra \eqref{eq:selective_surviving_algebra} of the common-reservoir model  take \(p=p_0+p_1\), \(q=p_0+p_2\), \(S_0=p_0+S_{21}\). Orthogonality of the two source supports, \(p_0\) and \(p_1\), and the two range supports, \(p_0\) and \(p_2\), gives 
\(S_0^*S_0=p\), \(S_0S_0^*=q\). Equation \eqref{eq:selective_expectation} gives \(\EE(S_0)=p_0\). In the central summand \(M_2\otimes I_2\) the ranks of \(p,q\) are two and one, while in the  scalar central retained block  \(\CZ I_2\) they are zero and one.
Hence
\begin{equation}
\vartheta_\N(p)=(1,0),\qquad \vartheta_\N(q)=(1/2,1),\qquad
(d_{p,q})_+=(1/2,0),\quad(d_{p,q})_-=(0,1).
\end{equation}
The endpoint defects of \(S_0\) are \(p_1\) and \(p_2\), with \(\vartheta_\N(p_1)=(1/2,0)\) and \(\vartheta_\N(p_2)=(0,1)\), so \(S_0\) attains both bounds. At finite time their unmatched components are given by \eqref{eq:selective_edge_defects}. The common \(p_0\) component contributes zero defect for all \(t\). 

\subsection{Finite-time loss forced by the records}
 Return to a matrix QMS, with \(\N\) and \(\alpha_t\) from Proposition~\ref{prop:exact-survival}; no stationary state is assumed yet. Its central block projections \(z_a\) are fixed even if
\(\alpha_t\) rotates projections inside a central summand. Apply \eqref{eq:sector_record_overlap} to the two fixed sectors \(I-z_a\) and \(z_a\), writing \(\ff_{\bar aa}(t)\) for the record overlap fidelity between them. By Theorem~\ref{thm:information_weights},
\begin{equation}
\ff_{\bar aa}(t) = \norm{\Phi_t|_{(I-z_a)\A z_a}} ,
\label{eq:complement_overlap}
\end{equation}
 the norm of \(\Phi_t\) on the corner between them.  If \(z_a=I\), set \(\ff_{\bar aa}=0\). Set \(d_a=k_a(p)-k_a(q)\) and \(d_a^+=\max(d_a,0)\); then \(d_{p,q}=\sum_a(d_a/m_a)z_a\), by the formula for \(\vartheta_\N(p)\) in \eqref{eq:center_trace}.

\begin{theorem}[Finite-time loss forced by the records]\label{thm:finite_record_obstruction}
Let \(S^*S=p\) and \(SS^*=q\), with \(p,q\in\N\). For every \(a,t\) there is a projection
\(e_a(t)\leq\alpha_t(pz_a)\) of ambient rank at least \(n_ad_a^+\) satisfying
\begin{equation}
e_a(t)\Ws_S(t)e_a(t)\leq  \ff_{\bar aa}(t)^2 \,e_a(t).
\label{eq:finite_record_obstruction}
\end{equation}
Replacing \(S\) by \(S^*\) gives the range bound.
\end{theorem}
\begin{proof}
Write \(S(t)=\Phi_t(S)\). Since \(S=qSp\) and \(p,q,z_a\in\MD(\Phi_t)\),
\( S(t)=\alpha_t(q) S(t)\alpha_t(p)\). The central projection \(z_a\) is fixed and commutes with both evolved endpoints, thus the compression of \(S(t)\) to the \(a\)th central summand has the following domain and target,
\begin{equation}
 B_a=z_a S(t) z_a:\alpha_t(pz_a)\HH\longrightarrow\alpha_t(qz_a)\HH .
\end{equation}
In a retained central summand, \(\alpha_t\) is unitary conjugation on the \(m_a\)-dimensional matrix factor. It preserves the retained ranks \(k_a(p),k_a(q)\).  Tensoring with \(I_{n_a}\) makes the two physical dimensions \(n_ak_a(p)\) and \(n_ak_a(q)\). Rank--nullity now gives
\begin{equation}
 \dim\ker B_a=n_ak_a(p)-\rank B_a
 \geq n_ak_a(p)-\min(n_ak_a(p),n_ak_a(q))=n_a d_a^+ .
\end{equation}
Let \(e_a(t)\) be the projection onto \(\ker B_a\subseteq\alpha_t(pz_a)\HH\). It is a projection on the physical Hilbert space, and may depend on \(S\) and on \(t\). 
For \(y=e_a(t)y\) we have \(y=z_ay\), since \(e_a(t)\leq\alpha_t(pz_a)\leq z_a\), and \(B_ay=0\) reads \(z_aS(t)y=0\). Hence \(S(t)y=(I-z_a)S(t)z_ay\). The projections \(z_a\) and \(I-z_a\) are fixed, so \eqref{eq:moving_corners} gives \((I-z_a)S(t)z_a=\Phi_t((I-z_a)Sz_a)\). Therefore
\begin{align}
\langle y,\Ws_S(t)y\rangle  & = \norm{(I-z_a)S(t)z_ay}^2 =\norm{\Phi_t((I-z_a)Sz_a)y}^2 \notag\\
& \leq\ff_{\bar aa}(t)^2\,\norm{(I-z_a)Sz_a}^2\,\norm y^2
\leq\ff_{\bar aa}(t)^2\norm y^2.
\label{eq:complement_vector_bound}
\end{align}
The first inequality holds by  \eqref{eq:complement_overlap} and the second by \(\norm{(I-z_a)Sz_a}\leq\norm S\leq1\),  and Theorem~\ref{thm:information_weights} gives \(\norm{\Phi_t|_{(I-z_a)\A z_a}}=\ff_{\bar aa}(t)\). Apply  \eqref{eq:complement_vector_bound} to \(y=e_a(t)z\) for arbitrary \(z \in\HH \). Then \(\|y\|^2=\langle z,e_a(t)z\rangle\), so  \(\langle z,e_a(t)\Ws_S(t)e_a(t)z\rangle\leq\ff_{\bar aa}(t)^2\langle z,e_a(t)z\rangle\) for every \(z\), which is \eqref{eq:finite_record_obstruction}.
\end{proof}
The record fidelity with the complement is bounded by the record fidelities \(\ff_{ba}(t)\) of the  separate central sectors \(z_b\) and \(z_a\)  for \(b\ne a\). For a contraction \(X\) in the corner \((I-z_a)\A z_a\) and a unit vector \(y=z_ay\), \eqref{eq:moving_corners} gives \(z_b\Phi_t(X)y=\Phi_t(z_bXz_a)y\). These vectors are orthogonal and sum to \(\Phi_t(X)y\), each has norm at most \(\ff_{ba}(t)\) by Theorem~\ref{thm:information_weights}, and \(\norm{\Phi_t(X)y}\leq1\). Hence
\begin{equation}
\ff_{\bar aa}(t)^2\leq\min\Bigl\{1,\ \sum_{b\ne a}\ff_{ba}(t)^2\Bigr\}.
\label{eq:complement_vs_pairs}
\end{equation}
 Under faithful stationarity \eqref{eq:faithful_stationary}, with \(\EE\) from Proposition~\ref{prop:asymptotic_expectation}, \(\EE((I-z_a)Xz_a)=(I-z_a)\EE(X)z_a=0\). Therefore
\(\ff_{\bar aa}(t)\leq\norm{\Phi_t-\alpha_t\EE}\to0\), and by \eqref{eq:finite_record_obstruction} the weight on the range of \(e_a(t)\) is lost completely as \(t\to\infty\).

\begin{corollary}[Norm-one defects when the retained ranks differ]\label{cor:asymptotic_loss}
 Assume faithful stationarity \eqref{eq:faithful_stationary}. Let \(p,q\in\N\) with \(p\sim_\A q\), and suppose \(k_a(p)\ne k_a(q)\) for some \(a\). Then every \(S\in\A\) with \(S^*S=p\), \(SS^*=q\) satisfies
\begin{equation}
\lim_{t\to\infty}\norm{D_t(S,S)}
=\lim_{t\to\infty}\norm{D_t(S^*,S^*)}=1 ,
\label{eq:forced_asymptotic_defect}
\end{equation}
and \(\norm{D_t(S,S)}\geq1- \ff_{\bar aa}(t)^2\) at every time, for each \(a\) with \(d_a>0\). If \(k_a(p)=k_a(q)\) for every \(a\), there exists an implementing edge \(S_0\in\N\) with both defects identically zero.
\end{corollary}
\begin{proof}
Equal ambient ranks give \(\sum_an_ad_a=0\), so if the retained ranks differ, some \(d_a\) is positive and some \(d_b\) is negative. Let \(d_a>0\). By \eqref{eq:edge_relations}, \(D_t(S,S)=\alpha_t(p)-\Ws_S(t)\), and the projection \(e_a(t)\leq\alpha_t(p)\) is nonzero, so \eqref{eq:finite_record_obstruction} gives \(e_a(t)D_t(S,S)e_a(t)\geq\bigl(1- \ff_{\bar aa}(t)^2\bigr)e_a(t)\). For a unit vector \(y=e_a(t)y\) this reads \(\langle y,D_t(S,S)y\rangle\geq1-\ff_{\bar aa}(t)^2\). Since \(0\leq D_t(S,S)\leq I\), this proves the inequality and, as the record fidelities tend to zero, the source limit: \(1\geq\norm{D_t(S,S)}\geq1-\ff_{\bar aa}(t)^2\to1\). The range limit is the same statement for \(S^*\) and a block central summand with \(d_b<0\). If every \(d_a\) vanishes, Section~\ref{sec:sharp_retained_dimensions} gives an implementing edge in \(\N\), and \(\N=\MD(\Phi_t)\) by Proposition~\ref{prop:exact-survival}.
\end{proof}
The rank count of Theorem~\ref{thm:finite_record_obstruction} identifies directions whose remaining amplitude must reach another central  summand. For the edge \(S_0=v+r\) of \eqref{eq:sharp_edge_construction} the weight separates into a matched and an unmatched part. Since \(v\in\N\), \(\Phi_t(S_0)=\alpha_t(v)+\Phi_t(r)\), and the mixed products of \(\Ws_{S_0}(t)\), \(\alpha_t(v)^*\Phi_t(r)\) and its adjoint, vanish because \(\alpha_t(v)^*\Phi_t(r)=\Phi_t(v^*r)=0\). Also \(r=rp'\) implies
\(\Phi_t(r)=\Phi_t(r)\alpha_t(p')\). Thus
\begin{equation}
\Ws_{S_0}(t)=\alpha_t(v^*v)+\Phi_t(r)^*\Phi_t(r).
\end{equation}
The first term is a projection: the matched directions lose nothing at any time. The second vanishes on them and is supported on the unmatched directions \(\alpha_t(p')\), where \eqref{eq:finite_record_obstruction} bounds it.
Indeed \(p'e=0\) for \(e=v^*v\), so \(\Phi_t(r)\alpha_t(e)=\Phi_t(r)\alpha_t(p'e)=0\).

\subsection{Record separation rates and finite-time loss}\label{sec:record_rates}
Assume now that \(\Phi_t=e^{t\LL}\) fixes the sectors for all \(t\), e.g., dephasing or the common-reservoir model. For a fixed projection \(p_v\), \(\LL(p_v)=\LL(p_v^2)=0\), hence \(\Gamma(p_v,p_v)=0\), and equation \eqref{eq:gamma} gives \([L_k,p_v]=0\) for
every \(k\). Taking adjoints gives \([L_k^*,p_v]=0\), the dissipator on \(p_v\) vanishes, and
\(i[H,p_v]=\LL(p_v)=0\). All jumps and the Hamiltonian therefore preserve the sector subspaces. Let \(\LL_{vw}\) denote the restriction of the generator to the corner \(p_v\A p_w\), which \(\LL\) maps into itself: \(\LL_{vw}(X)=\LL(X)\) for \(X=p_vXp_w\).

For any unit vector \(x\in\HH\) define the jump means and covariance matrix, 
\begin{align}
m_k(x)&:=\langle x,L_kx\rangle,\notag\\
\cov_x(k,l)&:=\langle L_kx,L_lx\rangle-m_k(x)^{*}m_l(x).
\label{eq:jump_moments}
\end{align}
Set \(a_k=L_kx-m_k(x)x\), expanding its inner products gives \(\cov_x(k,l)=\langle a_k,a_l\rangle\), so \(z^*\cov_x z=\|\sum_k z_k a_k\|^2\geq0\). 
For positive \(C,D\), define the Bures distance by \(d_{\mathrm B}(C,D)^2=\Tr C+\Tr D-2\|\sqrt C\sqrt D\|_1\). We apply it to the covariance matrices \(\cov_x\) and \(\cov_y\), whose traces need not be one. This expression is nonnegative even when \(C\) and \(D\) do not have unit trace, so  the two are compared without normalization. Indeed the polar trace formula \(\norm X_1=\max_{\norm U\leq1}|\Tr(UX)|\) and the
Hilbert--Schmidt Cauchy--Schwarz inequality give, for any contraction \(U\),
\begin{equation}
 |\Tr(U\sqrt C\sqrt D)|\leq\|U\sqrt C\|_2\|\sqrt D\|_2  \leq\sqrt{\Tr C\,\Tr D}.
\end{equation}
Taking the maximum over \(U\) shows \(d_{\mathrm B}(C,D)^2\geq(\sqrt{\Tr C}-\sqrt{\Tr D})^2\).
We minimize the  squared distance between the jump means of \(x\) and \(y\)  plus the squared Bures distance between their covariance matrices by setting
\begin{equation}
\mathfrak r_{vw}= \min_{\substack{x\in\HH_v,\ y\in\HH_w\\\norm x=\norm y=1}}
\left(\sum_k|m_k(x)-m_k(y)|^2+ d_{\mathrm B}(\cov_x,\cov_y)^2\right).
\label{eq:record_rate}
\end{equation}
This minimum is a rate: by Proposition~\ref{thm:record_rate} it is the initial decay rate of \(\ff_{vw}(t)^2\).

\begin{proposition}[Initial separation of the records]\label{thm:record_rate}
If the QMS \(\Phi_t=e^{t\LL}\) fixes the sector projections,
\begin{equation}
\ff_{vw}(t)^2=1-\mathfrak r_{vw}t+o(t),\qquad
\ff_{vw}(t)^2\leq e^{-\mathfrak r_{vw}t},\quad t\geq0.
\label{eq:record_rate_bound}
\end{equation}
If \(d_v\geq d_w\), where \(d_v=\dim\HH_v\), then
\begin{equation}
\mathfrak r_{vw}=\min_{\substack{S\in p_v\A p_w,\ S^*S=p_w\\
 y\in\HH_w,\ \norm y=1}}\langle y,\Gamma(S,S)y\rangle.
\label{eq:record_rate_dissipation}
\end{equation}
If \(d_v<d_w\), then \(\mathfrak r_{vw}\) is the minimum of \(\langle x,\Gamma(S,S)x\rangle\) over \(S\in p_w\A p_v\) with \(S^*S=p_v\) and unit \(x\in\HH_v\). The rate is independent of \(H\) for the given invariant sectors.
\end{proposition}
\noindent The proof is given in Supplement~\ref{supp:record_rate_proof}.

The sum  minimized in \eqref{eq:record_rate} is the initial separation rate of the records of the pair \(x,y\), for any two unit vectors (Supplement~\ref{supp:record_rate_proof}). A comparison, derived in Supplement~\ref{supp:record_rate_proof}, bounds the rate: it is at most twice the spectral decay rate of the corner generator \(\LL_{vw}\), with equality exactly when \(\ff_{vw}(t)^2\) is a pure exponential, as it is for rank-one sectors, i.e.,  the case of dephasing.

 The sectors \(I-z_a\) and \(z_a\) of Theorem~\ref{thm:finite_record_obstruction} are fixed, so \eqref{eq:record_rate_bound}  applies to them: \(\ff_{\bar aa}(t)^2\leq e^{-\mathfrak r_{\bar aa}t}\), with \(\mathfrak r_{\bar aa}\) the rate \eqref{eq:record_rate} of this pair, and \eqref{eq:finite_record_obstruction} reduces to \(e_a(t)\Ws_S(t)e_a(t)\leq e^{-\mathfrak r_{\bar aa}t}e_a(t)\). 

\subsection{Retained defects as integrals of dissipation}\label{sec:sharp_accumulated_dissipation}
To relate the endpoint defects \eqref{eq:endpoint_defects} to a time integral, let \(\Phi_t=e^{t\LL}\) be a uniformly continuous semigroup of normal UCP maps on a finite von Neumann algebra \( \A\) with
\begin{equation}
\Phi_t|_\N=\alpha_t,\qquad \EE\Phi_t=\alpha_t \EE,\qquad
\norm{\Phi_t(X)-\alpha_t\EE(X)}\longrightarrow0\quad(X\in \A).
\label{eq:retained_dynamical_assumptions}
\end{equation}
Here \(\EE:\A \to\N\) is a normal conditional expectation and \(\alpha_t\) a group of automorphisms. 
Proposition~\ref{prop:asymptotic_expectation} gives these hypotheses in the matrix setting. Define
\begin{equation}
H_t^{\mathrm s}(S):=\alpha_{-t}\EE(D_t(S,S)),\qquad
H_t^{\mathrm r}(S):=\alpha_{-t}\EE(D_t(S^*,S^*)).
\label{eq:retained_defects}
\end{equation}
The conditional expectation \(\EE\) takes the part  of the physical defect in \(\N\). The inverse automorphism expresses it in the original retained coordinates.

\begin{proposition}[Accumulated retained dissipation]\label{prop:retained_accumulation}
For an implementing edge \(S\in\mathcal I_\A(p,q)\) with endpoints \(p,q\) in \(\N\), \(H_t^{\mathrm s},H_t^{\mathrm r}\) increase in operator order from zero and converge  in operator norm, as \(t\to\infty\), to \(\Ds_{\EE}(S),\Dr_{\EE}(S)\), respectively. In particular,
\begin{equation}
H_t^{\mathrm s}(S)=\int_0^t\alpha_{-u}\EE \bigl(\Gamma(\Phi_uS,\Phi_uS)\bigr)\,du . \label{eq:retained_dissipation_integral}
\end{equation}
\end{proposition}
\noindent The proof is given in Supplement~\ref{supp:retained_dissipation_proof}.

Letting \(t\to\infty\) in \eqref{eq:retained_dissipation_integral} and then applying Theorem~\ref{thm:sharp_endpoint_defects} gives
\begin{align}
\vartheta_\N\left(\int_0^\infty\alpha_{-t}\EE\Gamma(\Phi_tS,\Phi_tS)\,dt\right)
&=(d_{p,q})_++\mathcal R_S,\label{eq:sharp_accumulated_dissipation_1}\\
\vartheta_\N\left(\int_0^\infty\alpha_{-t}\EE\Gamma(\Phi_t(S^*),\Phi_t(S^*))\,dt\right)
&=(d_{p,q})_-+\mathcal R_S .
\label{eq:sharp_accumulated_dissipation_2}
\end{align}
By Proposition~\ref{prop:retained_accumulation} the integral in \eqref{eq:sharp_accumulated_dissipation_1} up to time \(t\), which realizes  \(H_t^{\mathrm s}(S)\), tends to \(\Ds_{\EE}(S)\) in operator norm, and likewise for the range. Since \(\mathcal R_S\geq0\), the edge  \(S_0\) of Theorem~\ref{thm:sharp_endpoint_defects},  constructed in \eqref{eq:sharp_edge_construction}, for which \(\mathcal R_{S_0}=0\), attains both  minima, the least values \((d_{p,q})_+\) and \((d_{p,q})_-\) of the left sides of \eqref{eq:sharp_accumulated_dissipation_1} and \eqref{eq:sharp_accumulated_dissipation_2} over implementing edges: no implementing edge accumulates less dissipation, as measured by \(\vartheta_\N\).   The increase in operator order is a property of \(H_t^{\mathrm s}\) and \(H_t^{\mathrm r}\); the weight \(\Ws_S(t)\) need not decrease, because of the transport term in \eqref{eq:weight_evolution}.

These bounds concern one edge at a time. There is also a way to test preservation of a  given family of operations that does not require a dimension count; it is given in Supplement~\ref{supp:do-proof}.

The defects at all times  can be read from one map: under the hypotheses of Proposition~\ref{prop:asymptotic_expectation}, a QMS on \(M_d\) with a faithful stationary state \eqref{eq:faithful_stationary}, send an observable \(X\) to its trajectory, the function \(j(X):t\mapsto\Phi_t(X)\). The map \(j\) is unital and completely positive  but, unless \(\N=\A\), not multiplicative: the value at time \(t\) of \(j(X^*Y)-j(X)^*j(Y)\) is the defect \(D_t(X,Y)\). For the common-reservoir inclusion \eqref{eq:selective_surviving_algebra}, the trajectory map \(j\) cannot be replaced by a multiplicative one: no \(*\)-homomorphism sends each observable \(X\) to a function of time that starts at \(X\) and converges, as \(t\to\infty\), to an element of the retained algebra \(\N\) (Supplement~\ref{supp:initial_nonsplitting}).

\begin{example}[Common-reservoir model: accumulated dissipation]
In the common-reservoir model, \(\EE\) takes the \(\tau\)-weighted trace on the internal factor and
\(\alpha_t=\id\).
Applying it to the source defect of \(S_{21}\) gives the scalar coefficient explicitly,
\begin{equation}
 H_t^{\mathrm s}(S_{21})
  =\bigl[1-\tau_{00} f_0(t)^2-\tau_{11} f_1(t)^2\bigr]p_1 .
\end{equation}
Each \(f_a\) is positive and \(f'_a\leq0\) by \eqref{eq:selective_weight_ode}, so the derivative of this coefficient, \(-2\sum_a\tau_{aa}f_a f'_a\), is nonnegative.  The coefficient tends from zero to one when \(\kappa\ne1\).
Replacing \(S_{21}\) by its adjoint gives the same coefficient times \(p_2\). For \(\kappa\ne1\) the two integrals in \eqref{eq:sharp_accumulated_dissipation_1} and \eqref{eq:sharp_accumulated_dissipation_2} for \(S_{21}\) are therefore \(p_1\) and \(p_2\); since \(\vartheta_\N(p_1)=(d_{p_1,p_2})_+\) and \(\vartheta_\N(p_2)=(d_{p_1,p_2})_-\), the residual \(\mathcal R_{S_{21}}\) vanishes: \(S_{21}\) accumulates exactly the dissipation of the mismatch of the retained ranks.
\end{example}

\section{Discussion }\label{sec:discussion}
 We have described decoherence as a change in the products of operations. An observer with partial access predicts an evolved observable by a conditional expectation, and the multiplication defect of the reduced evolution is the conditional squared error of that prediction (Theorem~\ref{thm:partial_access}). Repeated partial access to independently prepared environments turns this defect into a rate, the dissipation function \eqref{eq:gamma_access}, and gives the Lindblad semigroup. No assumption beyond partial access to an observable algebra is needed to obtain open dynamics, and the reversible evolution may itself be a modular flow. 
The modular flow of the stationary state does not determine the retained algebra: in the common-reservoir model, equal and unequal couplings give the same stationary state, and so the same modular flow, but different retained algebras.

We introduced a quantum graph with edge weights that carry what the environment has learned. For fixed sectors, the largest weight that any edge between two sectors retains is the squared fidelity of the records that the two sectors leave in the environment (Theorem~\ref{thm:information_weights}). This is an equality, whereas information--disturbance relations give bounds \cite{KretschmannSchlingemannWerner,HaydenWinter2012}, and it holds for individual edges: when jumps are shared between sectors, each eigenvalue of the weight of one physical edge is the squared fidelity of two records, and a time-resolved record of the jumps attains this fidelity (Corollary~\ref{prop:shared_jump_weight}). Examples in the common-reservoir model demonstrate this, where a single record fidelity gives both the weight that an edge retains and the defect of a relation between two paths in the graph \eqref{eq:selective_path_defect}.

The retained algebra decides which relations between sectors can be preserved.  
When two endpoints have different retained ranks, Theorem~\ref{thm:sharp_endpoint_defects} gives the least defect that any edge between them must have. 
The dynamics of the loss are governed by the  records in the environment: on unmatched directions, the weight of every edge between them is at most a squared record fidelity at all times (Theorem~\ref{thm:finite_record_obstruction}), and under a faithful stationary state its defects reach norm one (Corollary~\ref{cor:asymptotic_loss}).  
When the retained ranks are equal, some edge between the two endpoints is preserved in full. A commutative retained algebra is the algebra of pointer observables, and no relation between distinct pointer sectors survives.

The information acquired by an environment, the algebra of observables that survives at long times, and the selection of pointer observables are often treated separately. Here they are statements about one object. The weights of the quantum graph are record fidelities at every time, under a faithful stationary state their limits are fixed by the retained algebra. Pointer selection is the case in which that algebra is commutative, and a relation between two sectors is either kept or lost. When the retained algebra is not commutative a relation can be preserved in part, and the retained ranks of its endpoints  quantify exactly how much is preserved.  
In summary, we have related quantum information to operator structure in decoherence: at every time, the products of operations fail exactly as far as the environment's records can be distinguished.

\section*{Acknowledgements} The author thanks Lara Ismert for helpful conversations. This work was carried out at Los Alamos National Laboratory, operated by Triad National Security, LLC, for the National Nuclear Security Administration of the U.S. Department of Energy under Contract No.~89233218CNA000001.

\paragraph{Use of language models.} Large language models (Claude, Anthropic; ChatGPT, OpenAI)  were used in revising this manuscript: to edit the text for clarity, to provide feedback on derivations, proofs, and cross-references, to run numerical checks of the model calculations, and to compile the source.  
Language models also contributed to the proof strategy in Supplement~\ref{supp:do-proof}, specifically to the use of an ultraweakly closed ideal to identify the largest reducing projection on which the prescribed family of operations is preserved.  Initial drafts, results, theorems, proofs, and examples were developed by the author with the assistance described above. The author is wholly responsible for the content.

\clearpage
\appendix
\renewcommand{\thesection}{S\arabic{section}}
\renewcommand{\theHsection}{supp.\arabic{section}}
\setcounter{equation}{0}
\renewcommand{\theequation}{S\arabic{equation}}
\renewcommand{\theHequation}{supp.\arabic{equation}}
\addtocontents{toc}{\protect\setcounter{tocdepth}{-1}}

\section{Operator coordinates and channel bimodules}\label{supp:operator_coordinates}
\subsection{States, faithfulness, and the polar decomposition}\label{supp:states_polar}
We supply the coordinate calculations for states used in Section~\ref{sec:sectors}.

Let \(\HH\) be a \(d\)-dimensional complex Hilbert space and \(\A=B(\HH)\)  the algebra of bounded operators on \(\HH\), the observable algebra of the system, identified with the algebra \(M_d(\CZ)\) of matrices in a chosen orthonormal basis. For a self-adjoint \(A\), the notation \(A\geq0\) means \(\langle y,Ay\rangle\geq0\) for every vector \(y\), and \(A\leq B\) means \(B-A\geq0\). A state is a linear map \(\omega:\A\to\CZ\) with \(\omega(I)=1\) and
\(\omega(A)\geq0\) for positive \(A\), to recover its density matrix  set \(e_{ij}=\ket i\bra j\) and \(\rho_{ij}=\omega(e_{ji})\). Since every \(X\in\A\) is \(X=\sum_{i,j}X_{ij}e_{ij}\) and \(\Tr(\rho e_{ij})=\rho_{ji}\), we get
\begin{equation}
\omega(X)=\sum_{i,j}X_{ij}\omega(e_{ij})
         =\sum_{i,j}X_{ij}\rho_{ji}=\Tr(\rho X).
\end{equation}
The choice \(X=\ket y\bra y\) gives \(\langle y,\rho y\rangle \geq 0\).
Taking \(X=I\) gives \(\Tr\rho=1\). Thus the matrix \(\rho\) with entries \(\rho_{ij}\) is a density, and the values on the matrix units determine it uniquely.

A state is faithful if \(\omega(A)>0\) for every nonzero positive \(A\). If its density has smallest eigenvalue \(\lambda_0>0\), then for \(A\geq0\),
\begin{equation}
\omega(A)-\lambda_0 \Tr A=\Tr\!\left(A^{1/2}(\rho-\lambda_0 I)A^{1/2}\right)\geq0 .
\end{equation}
Therefore \(\omega(A)=0\) forces \(\Tr A=0\), hence \(A=0\), since all eigenvalues of \(A\) are nonnegative. Conversely, a unit vector in \(\ker\rho\) gives a nonzero projection \(\ket y\bra y\) with expectation zero. This equivalence between faithfulness and invertibility will turn scalar equalities into operator equalities in Section~\ref{supp:survival}, under the assumption of a faithful stationary state made there.

There is a useful calculation for an operator \(X\) between two sectors, such as an edge after its amplitudes have changed: its polar decomposition, which separates a partial isometry from the amplitudes. For any 
rectangular \(X:\HH_w\to\HH_v\), diagonalize \(X^*X\) on the input space. Write
\(X^*Xy_j=s_j^2y_j\), with the \(y_j\) orthonormal. For \(s_j>0\) put \(x_j=Xy_j/s_j\). Then
\begin{equation}
\langle x_i,x_j\rangle
 =\frac{\langle y_i,X^*Xy_j\rangle}{s_i s_j}=\delta_{ij},\qquad
XX^*x_j=s_j^2x_j .
\end{equation}
If \(s_j=0\), then \(\norm{Xy_j}^2=0\). Decomposing the input in this basis now gives
\begin{align}
X&=\sum_{s_j>0}s_j\ket{x_j}\bra{y_j}=U|X|,\qquad
|X|=(X^*X)^{1/2},\notag\\
U&=\sum_{s_j>0}\ket{x_j}\bra{y_j},\qquad
U^*U=\sum_{s_j>0}\ket{y_j}\bra{y_j},\qquad
UU^*=\sum_{s_j>0}\ket{x_j}\bra{x_j}.\label{eq:polar}
\end{align}
The operator \(U\) is the partial isometry of the polar decomposition \(X=U|X|\), including its value zero on the unused input directions. The support of \(X^*X\) is the projection \(U^*U\), its nonzero eigenvalues \(s_j^2\) record the strengths on that support. The operator, trace, and Hilbert--Schmidt norms used throughout are given respectively by
\begin{equation}
\norm X=\max_j s_j,\qquad \norm X_1=\Tr|X|=\sum_j s_j,\qquad
\norm X_2^2=\Tr(X^*X)=\sum_j s_j^2.\label{eq:norms}
\end{equation}

\subsection{The partial trace against a state as a conditional expectation}\label{supp:partial_trace_expectation}
Section~\ref{sec:observer_expectation} uses the map \(P(Z)=\sum_j\pi_jV_j^*ZV_j\), with \(V_jy=y\otimes\ket j\) and the environment density \(\tau=\sum_j\pi_j\ket j\bra j\). Here we verify that \(P\) is the partial trace against \(\tau\) and a conditional expectation, and that it is faithful when every \(\pi_j>0\). In fact, on \(Z=a\otimes b\), the formula gives
\(P(Z)=a\sum_j\pi_j\bra j b\ket j=a\Tr(\tau b)\). Finite sums of simple tensors span the joint algebra, so for every system density \(\eta\), \(\Tr(\eta P(Z))=\Tr((\eta\otimes\tau)Z)\). The identities \((a\otimes I)V_j=V_j a\) and \(V_j^*(b\otimes I)=bV_j^*\) give the bimodule property \(P((a\otimes I)Z(b\otimes I))=aP(Z)b\). The map \(P\) is completely positive if for every \(n\) the map \(\id_n\otimes P\)  on \(n\times n\) matrices over the joint algebra is positive. It is a sum of compressions by \(I_n\otimes V_j\) with positive coefficients, \((\id_n\otimes P)(Z)=\sum_j\pi_j(I_n\otimes V_j)^*Z(I_n\otimes V_j)\), and each term is positive when \(Z\geq0\). To test whether \(P(Z^*Z)=0\) forces \(Z=0\), evaluate
\begin{equation}
\langle y,P(Z^*Z)y\rangle=\sum_j\pi_j\norm{Z(y\otimes\ket j)}^2.
\end{equation}
If every \(\pi_j>0\), vanishing of the left side for every \(y\) forces \(Z=0\), since the vectors \(y\otimes\ket j\) span the joint space, this is faithfulness of \(P\).

\subsection{The selected relation and its commutant}\label{supp:selected_relation}
For the algebra \(\B\) and selected spaces in \eqref{eq:sector_relation}, we verify the commutant action used in Section~\ref{sec:quantum_relation}.

The commutant \(\B'\) consists of the operators \(T\) satisfying \(Ta=aT\) for every
\(a\in\B\). In particular \(p_vTp_w=Tp_vp_w=0\) for \(v\ne w\). In one block, \(e_{ii}Te_{jj}=Te_{ii}e_{jj}=0\) for \(i\ne j\),  write the resulting diagonal as
\(\sum_i t_i e_{ii}\). Commutation with \(e_{ij}\) requires \(t_i e_{ij}=Te_{ij}=e_{ij}T=t_j e_{ij}\), thus all entries within that block are equal. There is no constraint between distinct blocks, so \(T=\sum_v t_vp_v\). For \(X_{vw}\in\mathcal V_{vw}\) and two elements \(\sum_ua_up_u\) and \(\sum_zb_zp_z\) of \(\B'\), with scalars \(a_u\), \(b_z\), this gives
\begin{equation}
\Big(\sum_u a_up_u\Big)X_{vw}\Big(\sum_z b_zp_z\Big)
=\sum_{u,z}a_ub_zp_up_vX_{vw}p_wp_z=a_vb_wX_{vw}.
\end{equation}
Thus \(\mathcal V\) is a \(\B'\)-bimodule. The full   corner is also invariant under the algebras \(A_v\), \(A_w\) of Section~\ref{supp:edges}. A selected subspace can be invariant under the scalars \(a_v\), \(b_w\), which only rescale each \(X_{vw}\), even if arbitrary matrices in \(A_v\) or \(A_w\) send it outside that subspace. Reflexivity is defined as \(\B'\subseteq\mathcal V\) \cite{Weaver}, which here says that every \(p_v\) lies in \(\mathcal V_{vv}\), and symmetry  as \(\mathcal V^*=\mathcal V\). For rank-one sectors the selected nonzero corners describe an ordinary 
directed relation on the set of sectors, the commutative case of Weaver's quantum relations \cite{Weaver}. A reflexive and symmetric \(\mathcal V\), for sectors of any rank, is an operator system, a self-adjoint subspace containing \(I\), as for the noncommutative graphs of Duan, Severini and Winter \cite{DuanSeveriniWinter2013}, which are the case of a single sector, \(\B=B(\HH)\). 

For instance, take two sectors of dimension two and select only the edge
\(X=\ket{v,1}\bra{w,1}\) between them. Multiplication by the commutant on either side gives a scalar
multiple of \(X\). But \(a=\ket{v,2}\bra{v,1}\in A_v\) sends \(X\) to \(aX=\ket{v,2}\bra{w,1}\), linearly independent of \(X\). Thus a selected edge space can be a \(\B'\)-bimodule, and so define a quantum relation, even when it does not contain the entire matrix corner. Quantum graphs on a matrix algebra are of this kind: for the dephasing graph on \(M_n\) \cite{Ismert2026}, whose adjacency map keeps the diagonal matrix units, there is one sector and the edge space is the diagonal matrices. The quantum Cuntz--Krieger algebras of quantum graphs \cite{BEVW2022,BHINW2023} are defined by relations that generalize those of a Cuntz--Krieger family: partial isometries \(S_i\) with mutually orthogonal ranges and \(S_i^*S_i=\sum_ja_{ij}S_jS_j^*\), for a matrix \((a_{ij})\) of zeros and ones. Our selected edges are partial isometries in one fixed matrix algebra, with no relation imposed beyond their products.

\subsection{The defect factorization in a channel bimodule}\label{supp:channel_bimodule}
There is an equivalent matrix-tuple description of \eqref{eq:defect_factorization} which will clarify its relation to the corner inner products of Section~\ref{supp:edges}.
Let \(m\) be the number of Kraus operators and write \(K=(K_1,\ldots,K_m)\in\A^m\). For tuples set
\begin{equation}
\langle x,y\rangle_\A=\sum_\mu x_\mu^*y_\mu,\qquad
(axb)_\mu=ax_\mu b,\qquad \langle K,K\rangle_\A=I ,
\label{eq:channel_module}
\end{equation}
for tuples \(x,y\in\A^m\) and \(a,b\in\A\). Right multiplication changes each component of \(y\) to \(y_\mu b\). Hence \(\langle x,yb\rangle_\A=\langle x,y\rangle_\A b\). Left multiplication obeys
\(\langle ax,y\rangle_\A=\sum_\mu x_\mu^*a^*y_\mu=\langle x,a^*y\rangle_\A\). To resolve \(y\) into a component of the form \(Kb\) and an orthogonal remainder, solve \(\langle K,y-Kb\rangle_\A=0\), normalization gives \(b=\langle K,y\rangle_\A\). 

For \(y=XK\) the coefficient is \(\langle K,XK\rangle_\A{}=\sum_\mu K_\mu^*XK_\mu=\Phi_t(X)\).
The remainder \(\delta_t(X)=XK-K\Phi_t(X)\) therefore has components \(XK_\mu-K_\mu\Phi_t(X)\).
Multiplying those components gives
\begin{align}
\langle\delta_t(X),\delta_t(Y)\rangle_\A
 &=\sum_\mu\bigl(K_\mu^*X^*-\Phi_t(X)^*K_\mu^*\bigr)
       \bigl(YK_\mu-K_\mu\Phi_t(Y)\bigr)\notag\\
 &=\Phi_t(X^*Y)-\Phi_t(X^*)\Phi_t(Y)-\Phi_t(X)^*\Phi_t(Y)  +\Phi_t(X)^*\Big(\sum_\mu K_\mu^*K_\mu\Big)\Phi_t(Y)\notag\\
 &=D_t(X,Y). \label{eq:module_defect}
\end{align}
To see  the same remainder acting on an ordinary system vector \(z\in\HH\), form the column
\begin{equation}
\sum_\mu (XK_\mu-K_\mu\Phi_t(X))z\otimes\ket\mu
 =(X\otimes I)V_tz-V_tV_t^*(X\otimes I)V_tz=C_t(X)z ,
\end{equation}
with \(\ket\mu\) the basis of \(\HH_E\) in \eqref{eq:dilation}. Taking inner products of two such columns gives \eqref{eq:module_defect} on vectors. In Section~\ref{supp:edges} the two factors of an inner product belong to a single corner and multiplication computes their endpoint products. Here the tuple \(\delta_t(X)\in\A^m\) keeps track of a separate component for each environment index.

\section{Modular structure and compatible interactions}\label{app:modular}
Section~\ref{sec:setup} uses two maps on the joint algebra: the automorphism \(\beta_t\) of the joint evolution and the conditional expectation \(P\) of partial access.  Here we give the state-dependent modular construction and derive the constraints it imposes on compatible interactions. This shows why the modular data alone do not determine the dissipative coupling or the retained algebra.

\subsection{The modular evolution of a faithful state}\label{sec:why_not_modular}
A faithful state  specifies an automorphism group, its modular group \cite{Takesaki2003,Landsman2026}. We
construct this group in matrices, keeping apart the operators acting on \(\HH\), which are the observables, and the operators acting on the Hilbert space of matrices. Write \(\mathcal K=L^2(\A) \) for the  matrices \(M_d(\CZ)\) themselves regarded as a Hilbert space, with \(\langle \xi,\eta\rangle_{\mathcal K}=\Tr(\xi^*\eta)\) for \(\xi,\eta\in\mathcal K\). Represent \(X\in\A\) on \(\mathcal K\) by \(L_X\xi=X\xi\). If \(\omega(X)=\Tr(\rho X)\), the vector \(\rho^{1/2}\) represents \(\omega\) as
\begin{equation}
\langle\rho^{1/2},L_X\rho^{1/2}\rangle_{\mathcal K} 
=\Tr(\rho^{1/2}X\rho^{1/2})=\Tr(\rho X).
\end{equation}
The inverse of \(\rho^{1/2}\) exists because \(\omega\) is faithful, that is, \(\rho\) has no zero eigenvalue, so every \(\xi\in\mathcal K\) is \(X\rho^{1/2}\) for the unique
\(X=\xi\rho^{-1/2}\). Hence the antilinear Tomita operator \(S_\omega\), defined by \(S_\omega(X\rho^{1/2})=X^*\rho^{1/2}\), acts on an arbitrary matrix as
\begin{equation}
S_\omega\xi=(\xi\rho^{-1/2})^*\rho^{1/2}
=\rho^{-1/2}\xi^*\rho^{1/2}.
\end{equation}
For our convention on inner products the adjoint of an antilinear map satisfies \(\langle S_\omega\xi,\eta\rangle=\langle S_\omega^*\eta,\xi\rangle\). Evaluating the left side with \(\langle\xi,\eta\rangle_{\mathcal K}=\Tr(\xi^*\eta)\) gives
\begin{align}
\langle S_\omega\xi,\eta\rangle
 &=\Tr(\rho^{1/2}\xi\rho^{-1/2}\eta)  =\Tr(\rho^{-1/2}\eta\rho^{1/2}\xi)\notag\\
 &=\langle\rho^{1/2}\eta^*\rho^{-1/2},\xi\rangle.
\end{align}
It follows that, for every vector \(\eta\in\mathcal K\), \(S_\omega^*\eta=\rho^{1/2}\eta^*\rho^{-1/2}\). Composing the two maps, \(S_\omega\) first,  gives
\begin{equation}
\Delta_\omega\xi=S_\omega^*S_\omega\xi =\rho^{1/2}(\rho^{-1/2}\xi^*\rho^{1/2})^*\rho^{-1/2}
=\rho\,\xi\,\rho^{-1}.
\end{equation}
This positive operator is the modular operator of \(\omega\) in Tomita--Takesaki theory \cite{Takesaki2003}, and conjugation by its imaginary powers \(\Delta_\omega^{is}\) is the modular evolution. To compute these powers, diagonalize \(\rho=\sum_i r_i e_{ii}\), the Hilbert--Schmidt orthonormal matrices \(e_{ij}\) obey \(\Delta_\omega e_{ij}=(r_i/r_j)e_{ij}\), thus
\(\Delta_\omega^{is}e_{ij}=r_i^{is}r_j^{-is}e_{ij}\). By linearity, for every \(\xi\in\mathcal K\), 
\(\Delta_\omega^{is}\xi=\rho^{is}\xi\rho^{-is}\). The modular evolution of an observable \(X\) is the conjugate \(\Delta_\omega^{is}L_X\Delta_\omega^{-is}\) of its left multiplication operator, evaluating it on \(\xi\) gives
\begin{equation}
\Delta_\omega^{is}L_X\Delta_\omega^{-is}\xi
 =\rho^{is}\bigl(X\rho^{-is}\xi\rho^{is}\bigr)\rho^{-is}
 =(\rho^{is}X\rho^{-is})\xi .
\end{equation}
Conjugation by \(\Delta_\omega^{is}\) thus takes left multiplication by \(X\) to left multiplication by another observable, \(\Delta_\omega^{is}L_X\Delta_\omega^{-is}=L_{\sigma_s^\rho(X)}\), with
\begin{equation}
\sigma_s^\rho(X):=\rho^{is}X\rho^{-is}, \label{eq:modular_group_intro}
\end{equation}
 \(\sigma_s^\rho\) is the modular automorphism group of \(\omega\), and \(s\) is modular time. We reserve \(t\) for physical evolution.

 We now ask which operators the modular group only  multiplies by a phase. Suppose \(S\in\A\) satisfies \(\sigma_s^\rho(S)=\lambda^{is}S\)  for all \(s\), with \(\lambda>0\). 
In the eigenbasis of \(\rho\) used above, \(\rho=\sum_ir_ie_{ii}\), the entries of \(\sigma_s^\rho(S)\) are \((r_i/r_j)^{is}S_{ij}\), so each nonzero entry
\(S_{ij}\) requires \(r_i/r_j=\lambda\). Therefore \((\rho S)_{ij}=r_iS_{ij}=\lambda S_{ij}r_j=\lambda(S\rho)_{ij}\) for all \(i,j\), and \(\rho S=\lambda S\rho\). Equivalently, differentiating \(\rho^{is}S\rho^{-is}=\lambda^{is}S\) at \(s=0\) gives \(i[\log\rho,S]=i(\log\lambda)S\), that is, \((\log\rho)S=S(\log\rho+(\log\lambda)I)\). Repeating it gives \((\log\rho)^nS=S(\log\rho+(\log\lambda)I)^n\), and summing the exponential series gives the same intertwining relation. For a partial isometry \(S\) with this property, for instance an edge, this determines the ratio of the state's weights on its  two supports \(SS^*\) and \(S^*S\) through
\begin{equation}
\omega(SS^*)=\Tr(S^*\rho S) 
=\lambda\Tr(S^*S\rho)=\lambda\,\omega(S^*S). \label{eq:modular_weight_constraint}
\end{equation}
For \(S=e_{ij}\) these numbers are \(r_i\) and \(r_j\).  Under the modular group such an edge acquires a phase and its two endpoint operators remain projections. This is the modular flow with full access: every observable of the algebra on which the state is defined is available, the flow is an automorphism group, and the two products of such an edge stay equal to its projections for every \(s\). Partial access enters when the modular flow of a state on a larger algebra is seen through a conditional expectation \(P\); the products can then fall below the projections (Theorem~\ref{thm:partial_access}).

\subsection{A density with a kernel}\label{supp:access}
When \(\rho\) has a kernel the  rule \(X\rho^{1/2}\mapsto X^*\rho^{1/2}\) that defined \(S_\omega\) is not well defined, because \(X\rho^{1/2}\) no longer uniquely determines \(X\).  For example, with \(\rho=\ket0\bra0\) and \(X=\ket0\bra1\), one has \(X\rho^{1/2}=0\) and \(X^*\rho^{1/2}=\ket1\bra0\), thus the  rule \(X\rho^{1/2}\mapsto X^*\rho^{1/2}\) would assign a nonzero vector to zero. The two indices on the following
relative Tomita map \(S_{ \rho,\rho_1}\) name two densities, \(\rho\) and an auxiliary density \(\rho_1\). Here, this \(S_{\rho,\rho_1}\) is an antilinear map on \(\mathcal K\). Let \(e_\omega\) be the projection onto \(\ran\rho\), and introduce a faithful auxiliary density \(\rho_1\), any invertible density matrix; the vectors \(X\rho_1^{1/2}\) then determine \(X\), and  the evolution of observables obtained at the end does not depend on \(\rho_1\). Now \(S_{\rho,\rho_1}(X\rho_1^{1/2})=X^*\rho^{1/2}\) is well defined on the full \(\mathcal K\). The computation of Section~\ref{sec:why_not_modular}, with \(\rho_1^{1/2}\) in place of \(\rho^{1/2}\) as the reference vector, gives
\begin{align}
S_{\rho,\rho_1}\xi&=\rho_1^{-1/2}\xi^*\rho^{1/2},&
S_{\rho,\rho_1}^*\eta&=\rho^{1/2}\eta^*\rho_1^{-1/2},\notag\\
\Delta_{\rho,\rho_1}\xi=S_{\rho,\rho_1}^*S_{\rho,\rho_1}\xi&=\rho\xi\rho_1^{-1},&
\Delta_{\rho,\rho_1}^{1/2}\xi&=\rho^{1/2}\xi\rho_1^{-1/2}.
\end{align}
The formula for \(\Delta_{\rho,\rho_1}^{1/2}\) follows by squaring the positive product of commuting left and right multiplication maps. Define the antilinear map \(J_{\rho,\rho_1}\) on \(\mathcal K\) by \(J_{\rho,\rho_1}\xi=\xi^*e_\omega\); it will be the antilinear factor in the polar decomposition of \(S_{\rho,\rho_1}\). Then
\begin{equation}
J_{\rho,\rho_1}\Delta_{\rho,\rho_1}^{1/2}(X\rho_1^{1/2})
=(\rho^{1/2}X)^*e_\omega=X^*\rho^{1/2}.
\end{equation}
This is the polar decomposition \(S_{\rho,\rho_1}=J_{\rho,\rho_1}\Delta_{\rho,\rho_1}^{1/2}\) that defines the relative modular operator \cite{Araki1977}, the antilinear analogue of \eqref{eq:polar}: \(J_{\rho,\rho_1}\) is antilinear, isometric on its initial subspace \(e_\omega\mathcal K\) and zero on the complement, which is the definition of a partial anti-isometry. If \(\xi=e_\omega\xi\), then 
\(\norm{J_{\rho,\rho_1}\xi}_2^2=\Tr(\xi\xi^*)=\norm{\xi}_2^2\). If \(e_\omega\xi=0\), then \(J_{\rho,\rho_1}\xi=0\). The image of
its initial subspace is \(\mathcal K e_\omega\), i.e., it acts on the space \(\mathcal K\) of matrices.

Choose eigenbases of \(\rho\) and \(\rho_1\). On matrix units built from a left eigenvector of \(\rho\) and a right eigenvector of \(\rho_1\), the eigenvalue of \(\Delta_{\rho,\rho_1}\) is the quotient of the corresponding eigenvalues: \(\rho\ket{x_i}\bra{y_j}\rho_1^{-1}=(r_i/r'_j)\ket{x_i}\bra{y_j}\) when \(\rho x_i=r_ix_i\) and \(\rho_1y_j=r'_jy_j\). It is zero exactly when the left vector lies in \(\ker\rho\).  Define the imaginary powers \(\Delta_{\rho,\rho_1}^{is}\) by the scalar power \((r_i/r'_j)^{is}\) on the positive eigenvalues and zero on the kernel, which is \((I-e_\omega)\mathcal K\). Then \(\Delta_{\rho,\rho_1}^{is}\xi=\rho^{is}\xi\rho_1^{-is}\), with \(\rho^{is}\) zero on \(\ker\rho\), and conjugating a left multiplication gives
\begin{equation}
\Delta_{\rho,\rho_1}^{is}L_X\Delta_{\rho,\rho_1}^{-is}=L_{\rho^{is}X\rho^{-is}},
\qquad \rho^{is}|_{\ker\rho}=0,\qquad
\left.\rho^{is}X\rho^{-is}\right|_{s=0}=e_\omega Xe_\omega . \label{eq:modular_compression}
\end{equation}
The factors involving \(\rho_1\) cancel because they act by right multiplication, which commutes with \(L_X\): \(\rho^{is}X(\rho^{-is}\xi\rho_1^{is})\rho_1^{-is}=(\rho^{is}X\rho^{-is})\xi\).  On \(e_\omega\A e_\omega\) the
density is invertible and  \(X\mapsto\rho^{is}X\rho^{-is}\) is a group of automorphisms. On the full algebra its value at  \(s=0\)  is the compression \(X\mapsto e_\omega Xe_\omega\), not the identity.  When a density has a kernel, the modular group is thus carried by the corner \(e_\omega\A e_\omega\) only. The stationary states of the four-state model of Section~\ref{sec:square} below are of this kind; the modular statements of the main text assume a faithful state.

\subsection{What a modularly compatible interaction fixes}\label{supp:modular_compatible}
The modular group and the conditional expectation \(P\) of partial access are linked by Takesaki's theorem on conditional expectations \cite{Takesaki1972}. Let \(\Omega\) be a faithful state on the joint algebra \(\M\) and suppose \(\Omega P=\Omega\). Takesaki's theorem  identifies invariance of \(\A\) under \(\sigma_s^\Omega\) as the criterion for a
normal \(\Omega\)-preserving conditional expectation: such a \(P\) onto \(\A\) exists exactly when \(\sigma_s^\Omega(\A)=\A\).  Writing \(\omega=\Omega|_\A\),  the restriction to \(\A\) of the modular group of \(\Omega\) is that of \(\omega\), and \(P\) intertwines the two:
\begin{equation}
P\sigma_s^\Omega=\sigma_s^\omega P,\qquad
P\sigma_s^\Omega|_\A=\sigma_s^\omega. \label{eq:modular_restriction}
\end{equation}
The equality \(P\sigma_s^\Omega|_\A=\sigma_s^\omega\) uses \(P|_\A=\id\). The modular evolution seen with partial access is thus again an automorphism group, and substituting this automorphism into \eqref{eq:defect_kernel} gives zero. Takesaki's theorem applies to the asymptotic expectation \(\EE\) of Proposition~\ref{prop:asymptotic_expectation}, which preserves the faithful stationary state: \(\N\) is invariant under \(\sigma_s^\rho\) and \(\EE\sigma_s^\rho=\sigma_s^\rho \EE\). The modular group therefore says which algebras admit a state-preserving expectation, though it does not say how fast anything decays: a compatible interaction can nevertheless produce dissipative dynamics in the limit of repeated interactions \eqref{eq:access_collision}.

The modular flow can also be taken as the reversible evolution in Theorem~\ref{thm:partial_access}. Let \(\Omega\) be a faithful state on \(\M\) and put \(\beta_s=\sigma_s^\Omega\). Let \(P\) be any faithful normal conditional expectation onto \(\A\), not assumed to preserve \(\Omega\). The reduced maps \(\Phi_s=P\sigma_s^\Omega|_\A\) are unital and completely positive, and Section~\ref{supp:faithful_partial_access} shows that their multiplicative domain is \(\A\cap\sigma_{-s}^\Omega(\A)\). Their multiplication defects therefore vanish for every \(s\) exactly when \(\sigma_s^\Omega(\A)=\A\) for every \(s\), which by Takesaki's theorem is the condition for an \(\Omega\)-preserving conditional expectation onto \(\A\). The multiplication defect of the modular flow, seen with partial access, is thus the obstruction to state-preserving access. The reduced flow need not be a semigroup, as \eqref{eq:access_memory} shows.

The semigroup of Section~\ref{sec:fresh_collisions} starts from modular flows in the same way. For the operator \(A_h=hH+\sqrt h\,G\) of \eqref{eq:access_collision}, let \(\Omega_h\) be the faithful state on \(\M\) whose density \(\rho_{\Omega_h}\) is proportional to \(e^{-A_h}\); this matrix is positive and invertible because \(A_h\) is self-adjoint, so the state is faithful. The normalization  \(\Tr(e^{-A_h})\) enters \(\rho_{\Omega_h}^{is}\) as a scalar phase and cancels from \(\rho_{\Omega_h}^{is}Z\rho_{\Omega_h}^{-is}\), so the density, the modular group and its value at modular time \(s=-1\), the time at which it is the evolution \(Z\mapsto U_h^*ZU_h\) of \eqref{eq:access_collision}, are
\begin{equation}
\rho_{\Omega_h}=\frac{e^{-A_h}}{\Tr(e^{-A_h})},\qquad
\sigma_s^{\Omega_h}(Z)=e^{-isA_h}Ze^{isA_h},\qquad
\sigma_{-1}^{\Omega_h}(Z)=U_h^*ZU_h .
\label{eq:modular_reference_state}
\end{equation}
The state \(\Omega_h\) is a reference state in one sense: its modular flow is the reversible evolution. The state \(\Omega_h\) is in general correlated between system and environment, because \(A_h\) contains the coupling \(G\), and it is not the state in which the environment is prepared. The preparation is \(\tau\), and \(\tau\) alone determines the access map \(P\) (Section~\ref{sec:observer_expectation}). One interaction  with partial access, the map \(\Psi_h\) of \eqref{eq:access_collision}, is therefore
\begin{equation}
\Psi_h=P\sigma_{-1}^{\Omega_h}|_\A .
\label{eq:modular_access_step}
\end{equation}
By Theorem~\ref{thm:partial_access}, for faithful \(P\) the  interaction \(\Psi_h\) is an automorphism exactly when \(\sigma_{-1}^{\Omega_h}(\A)=\A\); otherwise it has a multiplication defect. Repeating the  interaction, with an independently prepared environment each time, gives \(\Psi_{t/n}^{\,n}\to e^{t\LL}\) (Section~\ref{sec:fresh_collisions}), the Lindblad semigroup with generator \eqref{eq:gksl}. The multiplication defect, decoherence and the semigroup are thus obtained by starting from modular flows  of the joint algebra \(\M\) and seeing each flow with partial access.

A short example shows what is fixed when the interaction commutes with the modular flow of a product state \(\rho\otimes\tau\); we call such an interaction compatible. Let the system density \(\rho>0\) and an operator \(L \ne0 \) satisfy \(\rho L=\lambda L\rho\) with \(\lambda>0\), and let \([H,\rho]=0\). Prepare an environment qubit in \(\tau=\pi_0\ket0\bra0+\pi_1\ket1\bra1\), with \(\pi_0,\pi_1>0\) and \(\pi_0+\pi_1=1\), and for \(c>0\) take the coupling
\begin{equation}
G=\sqrt c\,\bigl(L\otimes\ket1\bra0+L^*\otimes\ket0\bra1\bigr)
\label{eq:modular_coupling}
\end{equation}
in the interaction  \eqref{eq:access_collision}. Since \(\tau\) is diagonal, \(P(G)=0\). The two products of \(\rho\otimes\tau\) with the summand \(L\otimes\ket1\bra0\) are \(\lambda\pi_1\,L\rho\otimes\ket1\bra0\) and \(\pi_0\,L\rho\otimes\ket1\bra0\), so \(G\) commutes with \(\rho\otimes\tau\) exactly when
\begin{equation}
\lambda\pi_1=\pi_0 .
\label{eq:modular_rate_ratio}
\end{equation}
Assume \eqref{eq:modular_rate_ratio}, then \(U_h\) commutes with \(\rho\otimes\tau\), which means  that the state with density \(\rho\otimes\tau\), which \(P\) preserves, is invariant under the interaction. Two consequences follow for each  interaction \(\Psi_h\), and both pass to the limit \(e^{t\LL}\). The  interaction preserves \(\rho\), since \(\Tr(\rho\Psi_h(X))=\Tr\bigl((\rho\otimes\tau)U_h^*(X\otimes I)U_h\bigr)=\Tr(\rho X)\).  The  interaction commutes with \(\sigma_s^\rho\), since conjugating \(U_h^*(X\otimes I)U_h\) by \((\rho\otimes\tau)^{is}\) gives \(U_h^*(\sigma_s^\rho(X)\otimes I)U_h\), and \(P\) intertwines the two modular groups by \eqref{eq:modular_restriction}. In the generator \eqref{eq:access_generator} 
the term \(P(G(X\otimes I)G)=c\pi_0L^*XL+c\pi_1LXL^*\)  has the form \(\sum_kL_k^*XL_k\), so the jump operators are \(\sqrt{c\pi_0}\,L\) and \(\sqrt{c\pi_1}\,L^*\). Compatibility thus fixes the ratio \(\pi_1/\pi_0=\lambda^{-1}\) of the two jump rates and leaves the coupling strength \(c\) of \eqref{eq:modular_coupling}, which multiplies both rates, free.

\section{Repeated interactions and changing products}\label{supp:repeated_interactions}
\subsection{A remainder that returns to the system}
The following example evaluates the two identities of
Theorem~\ref{thm:partial_access}, \eqref{eq:access_prediction} and \eqref{eq:access_memory}, before any repeated-interaction limit is taken.

\begin{example}[A remainder that returns to the system]
Let the system and the environment  each be one qubit and prepare the environment in \(I_2/2\). Write \(p_a=\ket a\bra a\), \(Z=p_0-p_1\), and take \(g>0\).   Split by the two system projections \(p_0\) and \(p_1\),  the coupling unitary is
\begin{equation}
U_t=e^{-igtZ\otimes Z/2}
 =p_0\otimes e^{-igtZ/2}+p_1\otimes e^{igtZ/2}.
\end{equation}
For \(S=\ket0\bra1\) the only nonzero block product in \(U_t^*(S\otimes I)U_t\)
has \(p_0\) on the left and \(p_1\) on the right. The two environment factors therefore multiply to \(e^{igtZ}\). Since \(Z^2=I\) and \(\Tr(Z)/2=0\), we obtain
\begin{align}
\beta_t(S\otimes I)&=S\otimes e^{igtZ}
=\cos(gt)S\otimes I+i\sin(gt)S\otimes Z,\notag\\
\Phi_t(S)&=\cos(gt)S,\qquad R_t(S)=i\sin(gt)S\otimes Z,\notag\\
P(R_t(S)^*R_t(S))&=\sin^2(gt)\,S^*S.
\end{align}
The source projection \(S^*S\) is fixed, whereas the squared amplitude of the edge is \(\cos^2(gt)\). Their difference agrees with \eqref{eq:access_defect}.  At two successive times the amplitude difference is
\begin{equation}
(\Phi_{t+s}-\Phi_t\Phi_s)(S) =\bigl(\cos(g(t+s))-\cos(gt)\cos(gs)\bigr)S
=-\sin(gt)\sin(gs)S.
\end{equation}
At \(t=s=\pi/(2g)\), two successive reductions send the edge to zero, while the uninterrupted evolution returns the edge with amplitude \(-1\). The two calculations use different access to the same environment between the two intervals.
\end{example}

\subsection{Expansion of one interaction, jump operators, and norm limit }\label{supp:collision_derivation}

We continue the construction of Section~\ref{sec:fresh_collisions}, starting with \(A_h,U_h,\Psi_h\) in \eqref{eq:access_collision}.

In the expansion of \(e^{iA_h}Xe^{-iA_h}\), multiplying the two exponential series through second order gives
\begin{align}
U_h^*XU_h
 &=X+iA_hX-iXA_h+A_hXA_h-\tfrac12 A_h^2X-\tfrac12 XA_h^2
 +O(\norm{A_h}^3)\norm X\notag\\
 &=X+i[A_h,X]-\tfrac12[A_h,[A_h,X]]+O(h^{3/2})\norm X .
\end{align}
 Here \(A_hXA_h=hGXG+h^{3/2}(HXG+GXH)+h^2HXH\), and likewise \(A_h^2=hG^2+O(h^{3/2})\). To average the remaining terms, use \(P(G)=0\) and \(P(aZb)=aP(Z)b\), which gives
\begin{equation}
P([G,X])=P(G)X-XP(G)=0,\quad
P(G^2X)=P(G^2)X,\quad P(XG^2)=XP(G^2).
\end{equation}
The remainder in the exponential expansion is \(O(h^{3/2})\) uniformly on \(\norm X\leq1\), since \(\norm{[A_h,X]}\leq2\norm{A_h}\norm X\) and \(\norm{A_h}=O(\sqrt h)\). Together these observations give \eqref{eq:access_generator}. In particular \(\LL(I)=0\).

We now construct the GKSL form of \(\LL\) and prove the limit. Specialize now to \(\A=B(\HH)\), the map \(F(X)=P(GXG)\) is completely positive. On an \(n\times n\) matrix \(X\) over \(\A\)  the map \(\id_n\otimes F\) first forms the congruence  \((I_n\otimes G)X(I_n\otimes G)\) by \(I_n\otimes G\) and then applies \(\id_n\otimes P\), and both steps preserve positivity. We can construct  the jump operators of \(\LL\) from the complete positivity of \(F\), through its Choi matrix.  
Let \(\Omega_d=\sum_i\ket i\otimes\ket i\) be an unnormalized vector in \(\HH\otimes\HH\), two copies of the system space (the first copy is auxiliary), and define
\begin{equation}
C_F=(\id_d\otimes F)(\ket{\Omega_d}\bra{\Omega_d})    =\sum_{ij}e_{ij}\otimes F(e_{ij})\geq 0.
\end{equation}
Factor this positive matrix using its eigendecomposition, \(C_F=\sum_k\ket{w_k}\bra{w_k}\), where the square root of each positive eigenvalue  is included in \(w_k\). Write \( \ket{w_k}=\sum_i\ket i\otimes \ket{w_{ki}}\) with \(\ket{w_{ki}}\in\HH\) and  compare the \((i,j)\)  blocks of the two expressions for \(C_F\).  We obtain \(F(e_{ij})=\sum_k\ket{w_{ki}}\bra{w_{kj}}\). 
Defining \(L_k\) by \(L_k^*\ket i:=  \ket{w_{ki}}\) gives \(F(e_{ij})=\sum_kL_k^*e_{ij}L_k\), because \(L_k^*e_{ij}L_k=\ket{L_k^*i}\bra{L_k^*j}\).
The \(e_{ij}\) span \(\A\), hence \(F(X)=\sum_kL_k^*XL_k\) for arbitrary \(X\).
At \(X=I\) this reads \(P(G^2)=\sum_kL_k^*L_k\). Substitution in \eqref{eq:access_generator} gives \eqref{eq:gksl}. 

 To verify the continuous-time limit, take norms of maps induced by the operator norm on \(\A\), namely \(\|F\|=\sup_{\|X\|\leq1}\|F(X)\|\).
A UCP map \(\Psi\) is contractive. Indeed its Kraus form, just constructed for any CP map, has
\(\sum_j K_j^*K_j=\Psi(I)=I\). For unit \(x,y\), Cauchy--Schwarz gives
\begin{equation}
|\langle x,\Psi(X)y\rangle|
 \leq\norm X\Big(\sum_j\norm{K_jx}^2\Big)^{1/2}
              \Big(\sum_j\norm{K_jy}^2\Big)^{1/2}=\norm X.
\end{equation}
In particular \(\norm{\Psi_h}\leq1\). We compare one interaction \(\Psi_h\) with the semigroup \(e^{h\LL}\) over the same time. The exponential series for the bounded generator gives \(e^{h\LL}=\id+h\LL+O(h^2)\), so \eqref{eq:access_generator} gives \(\norm{\Psi_h-e^{h\LL}}\leq C h^{3/2}\) for small \(h\). 
To  pass from one interaction to  \(n\) interactions,  we write \(\Psi_h^n-e^{nh\LL}\) as a sum in which each term contains the difference for one interaction once: put \(B=\Psi_h\), \(C_h=e^{h\LL}\). In the sum
\begin{equation}
\sum_{j=0}^{n-1}B^{n-1-j}(B-C_h)C_h^j
=\sum_{j=0}^{n-1}\bigl(B^{n-j}C_h^j-B^{n-1-j}C_h^{j+1}\bigr)
=B^n-C_h^n
\end{equation}
successive terms cancel in this order, without interchanging any factors; \(B\) and \(C_h\) need not commute. Each term has norm at most \(\norm{B-C_h}\norm{C_h^j}\), because \(\norm B\leq1\). Using
\(\norm{C_h^j}\leq e^{jh\norm\LL}\) for \(h=t/n\) now gives, for fixed \(t\) as \(n\to\infty\), 
\begin{equation}
\norm{\Psi_{t/n}^{\,n}-e^{t\LL}}\leq C e^{t\norm\LL} n(t/n)^{3/2}
=C e^{t\norm\LL}t^{3/2}n^{-1/2}\longrightarrow0 .
\end{equation}
Each fixed matrix amplification converges as well, since its finitely many entries converge. The positive cone, the set of positive matrices, is closed, so the limits are completely positive, and unitality, \(\Psi(I)=I\), passes to the limit by evaluating at \(I\). The exponential law gives \(\Phi_{t+s}=\Phi_t\Phi_s\) and \(\Phi_0=\id\). This establishes the quantum Markov semigroup (QMS) \(\Phi_t=e^{t\LL}\) used in the main text.

The construction of Kraus operators from the Choi matrix gives the dilation \eqref{eq:dilation} of Section~\ref{supp:weights}. Apply it to \(\Phi_t\),  which gives Kraus operators \(K_\mu\)  with \(\Phi_t(X)=\sum_\mu K_\mu^*XK_\mu\), and put \(V_ty=\sum_\mu K_\mu y\otimes\ket\mu\). The adjoint \(V_t^*\) sends \(z\otimes\ket\nu\) to \(K_\nu^*z\), since \(\langle z\otimes\ket\nu,V_ty\rangle=\langle z,K_\nu y\rangle =\langle K_\nu^*z,y\rangle\) for every \(y\). Consequently
\begin{align}
V_t^*V_ty&=\sum_\mu K_\mu^*K_\mu y=y, \label{eq:V_isometry}\\
V_t^*(X\otimes I)V_ty&=\sum_\mu K_\mu^*XK_\mu y ,
\label{eq:V_compression}
\end{align}
here equation \eqref{eq:V_isometry} makes \(V_t\) an isometry, whereas equation \eqref{eq:V_compression} realizes the channel by compression to its image, the Stinespring form
\cite{doi:10.1090/S0002-9939-1955-0069403-4}.

One can also construct an interaction  \eqref{eq:access_collision} whose limit has  any given matrix generator \eqref{eq:gksl}. Introduce environment vectors \(\ket0,\ket1,\ldots,\ket m\), prepare \(\ket0\), and let  the coupling, via one environment level \(\ket k\) as the channel of each jump operator \(L_k\), be
\begin{equation}
G=\sum_{k=1}^m\left(L_k\otimes\ket k\bra0+L_k^*\otimes\ket0\bra k\right).
\end{equation}
For the preparation \(\ket0\) the access map is \(P(Z)=V_0^*ZV_0\) with \(V_0y=y\otimes\ket0\) (Section~\ref{sec:observer_expectation}). For a system vector \(y\), \(G(y\otimes\ket0)=\sum_kL_ky\otimes\ket k\). Applying \(X\otimes I\), then \(G\),
and finally taking the \(\ket0\) environment component, which is what \(V_0^*\) does, gives \(\sum_k L_k^*XL_ky\). The \(\ket0\) component of \(G(y\otimes\ket0)\) itself is zero, since each of its terms carries some \(\ket k\) with \(k\geq1\). These two calculations give \(P(GXG)=\sum_kL_k^*XL_k\) and \(P(G)=0\), as the derivation of \eqref{eq:gksl} requires. This construction uses a pure environment, as the preparation \(\ket0\) is pure, so \(P\) need not be faithful. The identities   \eqref{eq:access_defect}--\eqref{eq:access_memory} of Theorem~\ref{thm:partial_access} hold for this \(P\) as well; their proof uses faithfulness only for the statements about when the defect vanishes.

The scaling in \eqref{eq:access_collision}, a coupling of strength \(1/\sqrt h\) acting for a time \(h\), changes the order at which dissipation enters. If a single fixed bounded Hamiltonian \(K\) instead generates the full group,  \(\beta_t(Z)=e^{itK}Ze^{-itK}\), then \(\Phi'_0(X)=iP([K,X])=i[P(K),X]\). Also 
\(R_t(X)=it(\id-P)[K,X]+O(t^2)\), so  by \eqref{eq:access_defect} \(D_t(X,X)=O(t^2)\): the defect is of second order in time, and a generator sees only the first. If this particular reduction were a norm-continuous semigroup, its generator would be its derivative at zero, \(i[P(K),\cdot\,]\), and its evolution would be the Hamiltonian group generated by \(P(K)\), without dissipation. Independently prepared environments allow the quadratic term in \(G\), \(hP(GXG)\), to survive at first order in physical time.

\subsection{A fixed edge and a broken path on two spins}\label{sec:square}
A fixed edge can still have products that are not fixed.  Lemma~\ref{lem:fixed_operator_products} gives a sufficient condition, not a necessary one,  for its products to be fixed, and  a model of two spins shows that they can change when the condition fails.

\begin{lemma}[Products of a fixed observable]\label{lem:fixed_operator_products}
Suppose \(\Phi_t(X)=X\). The defects \(\Ds_t(X)\) and \(\Dr_t(X)\)  vanish exactly when \(X^*X\) and \(XX^*\) are fixed. A faithful state preserved by \(\Phi_t\) suffices for the two equalities  \(\Phi_t(X^*X)=X^*X\) and \(\Phi_t(XX^*)=XX^*\).
\end{lemma}
\begin{proof}
Since the hypothesis gives \(\Phi_t(X)=X\), the  defects \(\Ds_t(X)=\Phi_t(X^*X)-\Phi_t(X)^*\Phi_t(X)\) and \(\Dr_t(X)\) are precisely
\(A=\Phi_t(X^*X)-X^*X\) and \(B=\Phi_t(XX^*)-XX^*\).
Thus \(A=B=0\) is equivalent to the two fixed-product assertions. Now let \(\omega\) be a faithful state with \(\omega\Phi_t=\omega\). The Schwarz inequality after \eqref{eq:defect_factorization} gives \(A\geq0\) and \(B\geq0\), while stationarity gives
\(\omega(A)=\omega(\Phi_t(X^*X))-\omega(X^*X)=0\), and likewise \(\omega(B)=0\). 
A faithful state is strictly positive on nonzero positive matrices. Hence \(A=0\) and \(B=0\).
\end{proof}
\begin{example}[Two spins: a fixed edge and a broken path] Consider two spins with \(P_u=\ket u\bra u\), \(u\in\{00,10,01,11\}\), \(H=0\), and \(L=\sqrt\gamma\,\ket{01}\bra{10}\), with \(\gamma>0\). To solve the evolution from the generator \(\LL(X)=L^*XL-\tfrac12(L^*LX+XL^*L)\) of \eqref{eq:gksl}, let \(e_{uv}=\ket u\bra v\) and \(n_u=\delta_{u,10}\), the indicator of the decaying state \(\ket{10}\).
The jump term  \(L^*XL=\gamma\langle01|X|01\rangle P_{10}\) uses the \((01,01)\) entry of its argument, while \(L^*L=\gamma P_{10}\) gives \(P_{10}e_{uv}=n_ue_{uv}\) and \(e_{uv}P_{10}=n_ve_{uv}\). Hence
\begin{equation}
\LL(e_{uv})=\gamma\delta_{u,01}\delta_{v,01}P_{10}
             -\tfrac{\gamma}{2}(n_u+n_v) e_{uv}.
\end{equation}
Each off-diagonal unit  \(e_{uv}\), \(u\ne v\), is an eigenvector of \(\LL\) and evolves by  the factor \(e^{-\gamma t(n_u+n_v)/2}\). 
The diagonal equations are coupled by one term, \(\LL(P_{01})=\gamma P_{10}\).  Since \(\LL(P_{10})=-\gamma P_{10}\),  \(\Phi_t(P_{10})=e^{-\gamma t} P_{10}\), and integration of \(\tfrac{d}{dt}\Phi_t(P_{01})=\Phi_t(\LL(P_{01}))\) gives
\begin{equation}
\Phi_t(P_{01})=P_{01}+\int_0^t\gamma\Phi_s(P_{10})\,ds
              =P_{01}+(1- e^{-\gamma t})P_{10}.
\end{equation}
The remaining two diagonal units, \(P_{00}\) and \(P_{11}\), are fixed, since \(\LL\) vanishes on them. These formulas give \(\Phi_t =e^{t\LL}\) on every operator, since the matrix units form a basis. They also give  a convenient pair of Kraus operators.  Multiplication on both sides by
\(K_0=  I-(1-e^{-\gamma t/2})P_{10}\) multiplies \(e_{uv}\) by \(e^{-\gamma t(n_u+n_v)/2}\).  The one term of \(\Phi_t(e_{uv})\) not obtained in this way is \((1-e^{-\gamma t})\delta_{u,01}\delta_{v,01}P_{10}\), exactly \(K_1^* e_{uv}K_1\) for 
\begin{equation}
K_0=I-(1-e^{-\gamma t/2})P_{10},\qquad K_1=\sqrt{1-e^{-\gamma t}}\ket{01}\bra{10}.
\end{equation}
Here \(K_0^*K_0=I-(1-e^{-\gamma t})P_{10}\) and \(K_1^*K_1=(1-e^{-\gamma t})P_{10}\).
 The sum in the other order, \(K_0K_0^*+K_1K_1^*\), is \(I+(1-e^{-\gamma t})(P_{01}-P_{10})\), recording the direction of population transfer.

Multiplying initial edges gives
\( e_{11,10} e_{10,00}= e_{11,01} e_{01,00}= e_{11,00}\). The Kraus calculation yields
\begin{align}
\Phi_t(e_{10,00})&=e^{-\gamma t/2}e_{10,00},&\Phi_t(e_{11,10})&=e^{-\gamma t/2}e_{11,10},\notag\\
\Phi_t(e_{01,00})&=e_{01,00},&\Phi_t(e_{11,01})&=e_{11,01},\notag\\
\Phi_t(P_{10})&=e^{-\gamma t}P_{10},&
\Phi_t(P_{01})&=P_{01}+(1-e^{-\gamma t})P_{10},\notag\\
\Phi_t(P_{00})&=P_{00},&\Phi_t(P_{11})&=P_{11}.
\label{eq:square_evolution}
\end{align}
 The product itself is fixed, \(\Phi_t(e_{11,00})= e_{11,00}\). Thus the path through \(\ket{10}\), with its factors evolved separately, is \(\Phi_t(e_{11,10})\Phi_t(e_{10,00})=e^{-\gamma t/2}e_{11,10}\,e^{-\gamma t/2}e_{10,00}=e^{-\gamma t}e_{11,00}\). The path through \(\ket{01}\) still has amplitude \(e_{11,00}\). For example \(e_{01,00}e_{01,00}^*=P_{01}\) but \(\Phi_t(P_{01})- P_{01}=(1- e^{-\gamma t})P_{10}\). The edge \(e_{01,00}\) is fixed, so it stays a partial isometry, yet it has a nonzero range defect  \(\Dr_t(e_{01,00})=(1-e^{-\gamma t})P_{10}\). A stationary state must assign zero weight to \(P_{10}\), since invariance requires \(\omega(P_{10})= \omega(\Phi_t(P_{10}))= e^{-\gamma t}\omega(P_{10})\) for \(t>0\). No stationary state is therefore faithful, so the sufficient condition of  Lemma~\ref{lem:fixed_operator_products} is not available, and indeed the fixed edge \(e_{01,00}\) has a product that is not fixed.

For completeness, the source defect \(\Ds_t(S)=\Phi_t(S^*S)-\Phi_t(S)^*\Phi_t(S)\) of \( e_{10,00}\) is
\(\Phi_t(P_{00})- e^{-\gamma t}P_{00}=(1- e^{-\gamma t})P_{00}\), and its range defect \(\Phi_t(SS^*)-\Phi_t(S)\Phi_t(S)^*\) is
\(\Phi_t(P_{10})-e^{-\gamma t}P_{10} =0\). Substitution for the remaining three edges gives
\begin{align}
\Ds_t(e_{10,00})&=(1- e^{-\gamma t})P_{00},&\Dr_t( e_{10,00})&=0,\notag\\
\Ds_t(e_{01,00})&=0,&\Dr_t(e_{01,00})&=(1- e^{-\gamma t})P_{10},\notag\\
\Ds_t(e_{11,10})&=0,&\Dr_t(e_{11,10})&=(1- e^{-\gamma t})P_{11},\notag\\
\Ds_t(e_{11,01})&=(1- e^{-\gamma t})P_{10},&\Dr_t(e_{11,01})&=0 .
\end{align}
 For the path through \(\ket{10}\) the defect of the product is \(D_t(e_{11,10}^*,e_{10,00})=(1-e^{-\gamma t})e_{11,00}\), while for the path through \(\ket{01}\) \(D_t( e_{11,01}^*,e_{01,00})=0\). The two paths, with their factors evolved separately, therefore differ by \((1-e^{-\gamma t})e_{11,00}\), of norm \(1-e^{-\gamma t}\). Section~\ref{sec:relations} bounds this norm by the sum over the two paths of \(\sqrt{\norm{\Dr_t}\norm{\Ds_t}}\), taken for the two edges of each path. By the list of defects this sum is \((1-e^{-\gamma t})+0\). The bound \eqref{eq:path_bound} is saturated in this example. \end{example}

\section{Further record calculations}\label{supp:further_records}
\subsection{A given edge and its visibility}
\label{supp:chosen_edge}
Equation \eqref{eq:pair_information} permits the operator \(X\) to depend on the preparations \(x\), \(y\). For a  given \(S\) and unit \(y\in\HH_w\), the source weight gives
\begin{equation}
\langle y,\Ws_S(t)y\rangle =\norm{\Phi_t(S)y}^2
=\max_{\substack{x\in\HH_v\\\norm x=1}}|\langle x,\Phi_t(S)y\rangle|^2.
\label{eq:edge_visibility}
\end{equation}
The matrix element in \eqref{eq:edge_visibility} is measured as follows: prepare \(\zeta=(x+y)/\sqrt2\) and read the self-adjoint observable
\(O_\theta=e^{i\theta}S+e^{-i\theta}S^*\). The fixed-sector relation \(\Phi_t(S)=p_v\Phi_t(S)p_w\), from \eqref{eq:moving_corners}, with \(p_wx=0\) and \(p_vy=0\), gives \(\Phi_t(S)x=0\) and \(\Phi_t(S)^*y=0\). Expanding the four terms in \(\langle\zeta,\Phi_t(O_\theta)\zeta\rangle\) leaves
\begin{equation}
\tfrac12 e^{i\theta}\langle x,\Phi_t(S)y\rangle+
\tfrac12 e^{-i\theta}\langle x,\Phi_t(S)y\rangle^{*}
=\Rea\bigl(e^{i\theta}\langle x,\Phi_t(S)y\rangle\bigr).
\end{equation}
Thus the squared amplitude of this oscillation in \(\theta\), optimized over \(x\), is the expectation of the fixed edge weight on \(y\). Theorem~\ref{thm:information_weights} compares this expectation with the corresponding optimized fidelity, it identifies the
optimum over edges, and it bounds every edge with those endpoints.

\subsection{Reservoir cross evolution}\label{supp:reservoir_fidelity_proof}
Here we present the proof of Proposition~\ref{prop:environment_fidelity}.
\begin{proof}
First form the joint outer product for the input \(e_{21}\otimes q_a\).
By \eqref{eq:conditional_dilation} it is
\(e_{21}\otimes\ket{\Psi_{\kappa,a}}\bra{\Psi_{1,a}}\). Tracing over the environment keeps \(e_{21}\) and gives a matrix \(C_a(t)\) on the internal system, the trace over the environment of \(\ket{\Psi_{\kappa,a}}\bra{\Psi_{1,a}}\). To evolve this cross matrix directly, apply the generator for densities, \(\eta\mapsto-i[H,\eta]+\sum_\nu\bigl(L_\nu\eta L_\nu^*-\tfrac12L_\nu^*L_\nu\eta-\tfrac12\eta L_\nu^*L_\nu\bigr)\), to \(\eta=e_{21}\otimes C_a\). In the term \(L_\nu\eta L_\nu^*\) the sector multiplication is \(Fe_{21}F=\kappa e_{21}\). In \(L_\nu^*L_\nu\eta\) it is \(F^2e_{21}=\kappa^2 e_{21}\), and in \(\eta L_\nu^*L_\nu\) it is \(e_{21}F^2=e_{21}\). Thus the matrix \(C_a(t)\) of \eqref{eq:environment_cross_operator} is the trace over the environment of \(\ket{\Psi_{\kappa,a}}\bra{\Psi_{1,a}}\). By the proof of Corollary~\ref{prop:shared_jump_weight}, \(C_a(t)\) obeys the   equation \eqref{eq:shared_cross_generator} derived there for the cross matrix,  here with sector couplings \(\kappa\) and \(1\), \(H_0=\Omega q_1\) and the jumps \(\ell_\pm\). The three sector factors \(\kappa\), \(\kappa^2\) and \(1\) just computed are its coefficients:
\begin{equation}
\dot C_a=-i[\Omega q_1,C_a]+\kappa\sum_{\nu=\pm}\ell_\nu C_a\ell_\nu^*
-\tfrac12(\kappa^2RC_a+C_aR),\qquad C_a(0)=q_a .
\label{eq:cross_generator}
\end{equation} 
The equation maps diagonal matrices to diagonal matrices, and \(C_a(0)=q_a\) is diagonal.  Take \(C_a=c_{a,0}q_0+c_{a,1}q_1\). The positive jump term sends population in level \(1\) into level \(0\)
with coefficient \(\kappa\gamma_-\) and population in \(0\) into \(1\) with coefficient \(\kappa\gamma_+\). 
 The two loss terms, from \(-\tfrac12(\kappa^2RC_a+C_aR)\),  remain in the original level. In components,
\begin{equation}
 \dot c_{a,0}=-m\gamma_+c_{a,0}+\kappa\gamma_-c_{a,1},\qquad
 \dot c_{a,1}=\kappa\gamma_+c_{a,0}-m\gamma_-c_{a,1}.
\end{equation} 
  The two off-diagonal coefficients \(\kappa\gamma_-\) and \(\kappa\gamma_+\)  exchange positions relative to the equation \eqref{eq:selective_weight_ode} for the amplitudes \(f_a\), whose matrix is \(M_\kappa\) (Example~\ref{ex:reservoir_amplitudes}); here the matrix is its transpose. With \(\mathbf e_a\in\mathbb R^2\) the  unit column with entry one in position \(a\) and zero in the other, which lists the initial populations of \(q_a\),  and with \(\boldsymbol1=(1,1)^{\mathsf T}\) the vector of ones, the solution is
\begin{equation}
(c_{a,0}(t),c_{a,1}(t))^{\mathsf T}
=e^{tM_\kappa^{\mathsf T}}\mathbf e_a . \label{eq:cross_diagonal_evolution}
\end{equation}
All entries of \(e^{tM_\kappa^{\mathsf T}}\mathbf e_a\) are nonnegative,  because the off-diagonal entries of \(M_\kappa^{\mathsf T}\) are. So \(C_a(t)\geq0\) and its trace norm is its trace. Substituting \eqref{eq:cross_diagonal_evolution} gives the second equality, and trivially a number equals its transpose, which gives the third equality in 
\begin{equation}
\norm{C_a(t)}_1=\Tr C_a(t)
=\boldsymbol1^{\mathsf T}e^{tM_\kappa^{\mathsf T}}\mathbf e_a
=\mathbf e_a^{\mathsf T}e^{tM_\kappa}\boldsymbol1=f_a(t).
\end{equation}
The last equality  holds because \(e^{tM_\kappa}\boldsymbol1\) is the solution of \eqref{eq:selective_weight_ode}, a column of the two amplitudes, and \(\mathbf e_a^{\mathsf T}\) selects \(f_a(t)\) from it. The identity \eqref{eq:cross_trace_norm}  in the proof of Theorem~\ref{thm:information_weights} equates  \(\norm{C_a(t)}_1\) with the root fidelity of the two  reservoir records. This proves \eqref{eq:environment_fidelity_weight}.
\end{proof}

\subsection{Spectral records with a reference}\label{sec:supp_spectral}
For a general fixed-sector channel, the following construction  optimizes spectral sums over edges by retaining a reference in the preparations,   which extends Theorem~\ref{thm:information_weights} from the largest eigenvalue of an edge weight to its \(k\) largest, through the sum of their square roots.

Optimizing the matrix-element identity \eqref{eq:pair_information} of Theorem~\ref{thm:information_weights} over preparations selects a largest singular value. We can probe a sum of singular values by preparing several orthogonal inputs coherently with a reference: the reference keeps the \(i\)th input of one sector paired with the \(i\)th input of the other, which gives a sum \(\sum_i\langle x_i,\Phi_t(X)y_i\rangle\) in place of a single matrix element. Fix \(1\leq k\leq\min(d_v,d_w)\), where \(d_v=\dim\HH_v\). For an orthonormal frame \(x=(x_1,\ldots,x_k)\) in \(\HH_v\) introduce a \(k\)-dimensional reference  space \(R\) with  orthonormal basis \(\ket j_R\) and put
\begin{equation}
\Xi_v(x,t)=\frac1{\sqrt k}\sum_{j=1}^k V_{v,t}x_j\otimes\ket j_R,\qquad
\sigma_v^{E,k}(x,t)=\Tr_{\HH_v}\ket{\Xi_v(x,t)}\bra{\Xi_v(x,t)} .
\end{equation}
The vector \(\Xi_v\) lies in \(\HH_v\otimes\HH_E\otimes R\): the dilation \(V_{v,t}\) is applied to each of the \(k\) inputs, and the reference labels them. Here \(\sigma_v^{E,k}\) is a density on \(\HH_E\otimes R\). Orthonormality of the reference basis gives \(\|\Xi_v\|^2=k^{-1}\sum_j\|V_{v,t}x_j\|^2=1\).
Define \(\Phi^c_{v,t}(\eta)=\Tr_{\HH_v}(V_{v,t}\eta V_{v,t}^*)\), the output on the environment, so that \(\Phi^c_{v,t}(\ket x\bra x)=\sigma_v^E(x,t)\) is the record of \(x\). 
Expanding the outer product before tracing gives
\begin{equation}
 \sigma_v^{E,k}(x,t)  =\frac1k\sum_{i,j}\Phi^c_{v,t}(\ket{x_i}\bra{x_j})\otimes\ket i\bra j_R . \label{eq:reference_record}
\end{equation}
The trace over the output system leaves the reference matrix units in place, including the off-diagonal ones. Tracing out \(R\) would instead retain only the \(k\) diagonal terms.
Let \(\ff_{vw}^{(k)}(t)\) be the maximum root fidelity of the two \(E\otimes R\) records over both frames. For \(Y:\HH_w\to\HH_v\), list its singular values \(s_1(Y)\geq s_2(Y)\geq\cdots\) in decreasing order and set \(\norm Y_{(k)}=\sum_{j=1}^k s_j(Y)\), the Ky Fan norm.

\begin{proposition}[Spectral record  fidelities]\label{thm:spectral_records}
With \(S\) varying over all partial isometries in the corner,
\begin{equation}
\max_S\sum_{j=1}^k\sqrt{\lambda_j(\Ws_S(t))}
=\max_S\norm{\Phi_t(S)}_{(k)}
=k\,\ff_{vw}^{(k)}(t). \label{eq:spectral_records}
\end{equation}
The maximizing edge can depend on \(k\).
\end{proposition}
\begin{proof}
Set \(Y=\Phi_t(S)\), diagonalizing \(|Y|=(Y^*Y)^{1/2}\) gives 
\(\lambda_j(\Ws_S(t))=s_j(Y)^2\). Ky Fan's variational principle \cite{Fan1951} writes
\begin{equation}
\sum_{j=1}^k s_j(Y)=\max_{x,y}\left|\sum_{i=1}^k\langle x_i,Y y_i\rangle\right|.
\end{equation}

For fixed frames and a corner contraction \(X\), expansion of the reference inner product gives
\begin{align}
k\langle\Xi_v,(X\otimes I_E\otimes I_R)\Xi_w\rangle
&=\sum_{i,j}\langle V_{v,t}x_i,(X\otimes I)V_{w,t}y_j\rangle\langle i,j\rangle_R\notag\\
&=\sum_i\langle x_i,\Phi_t(X)y_i\rangle .
\end{align} 
For these fixed frames the rectangular cross operator is
\begin{equation}
 C^{(k)}=\Tr_{E\otimes R}\ket{\Xi_w}\bra{\Xi_v}
 =\frac1k\sum_i\Tr_E\ket{V_{w,t}y_i}\bra{V_{v,t}x_i}.
\end{equation}
The Kronecker delta from the reference trace enforces the same index in each summand.
The frame pairing \(\sum_i\langle x_i,\Phi_t(X)y_i\rangle\) is therefore \(k\Tr(XC^{(k)})\), its largest modulus over contractions \(X\) is \(k\|C^{(k)}\|_1\), attained at the partial isometry \(X=U^*\) of the polar decomposition \(C^{(k)}=U|C^{(k)}|\). The cross-operator calculation in the proof of Theorem~\ref{thm:information_weights}, applied to the vectors \(\Xi_v\) and \(\Xi_w\) with \(E\otimes R\) in place of \(E\), identifies that norm with the fidelity \(\ff(\sigma_v^{E,k}(x,t),\sigma_w^{E,k}(y,t))\) on \(E\otimes R\). Taking the maximum over frames gives \(k\ff_{vw}^{(k)}(t)\), by the definition of \(\ff_{vw}^{(k)}\). Maximizing first over frames and then over \(X\), Ky Fan's principle gives \(\norm{\Phi_t(X)}_{(k)}\) as the maximum of the pairing over frames, and its maximum over  contractions \(X\), which form the unit ball of the corner, is attained at a partial isometry, because the norm is convex in \(X\), so its maximum over this compact convex set is attained at an extreme point, and the extreme points of the unit ball of the corner are partial isometries. This proves
\eqref{eq:spectral_records}.
\end{proof}

At \(k=d_w\leq d_v\) the identity \eqref{eq:spectral_records} can be written with Choi states. Fix a basis of \(\HH_w\) and the vector \(\Omega_w=d_w^{-1/2}\sum_j\ket j\otimes\ket j_R\). The normalized Choi state of a channel \(\Lambda\), a completely positive trace-preserving map on the density matrices of \(\HH_w\), is \(\mathsf J(\Lambda)=(\Lambda\otimes\id_R)(\ket{\Omega_w}\bra{\Omega_w})\). Expanding \(\ket{\Omega_w}\bra{\Omega_w}\) gives the formula
\begin{equation}
 \mathsf J(\Lambda)=\frac{1}{d_w}\sum_{i,j}\Lambda(\ket i\bra j)\otimes\ket i\bra j_R.
 \label{eq:J(Lambda)}
\end{equation}
Given a full source frame \(y_j\) and target frame \(x_j\), put \(T=\sum_j\ket{x_j}\bra{y_j}\). Orthonormality gives \(T^*T=\sum_j\ket{y_j}\bra{y_j}=I_{\HH_w}\),
and \(T\) is determined by \(Ty_j=x_j\). In the basis \(y_j\), substitute first \(\Lambda=\Phi^c_{w,t}\), then \(\Lambda=\Phi^c_{v,t}\circ\Ad_T\), with \(\Ad_T(\eta)=T\eta T^*\). Since \(T\ket{y_i}\bra{y_j}T^*=\ket{x_i}\bra{x_j}\), the two Choi states are the records \(\sigma_w^{E,k}(y,t)\) and \(\sigma_v^{E,k}(x,t)\) at \(k=d_w\).
Indeed, write \eqref{eq:J(Lambda)} in the basis \(\ket j=y_j\). The first substitution gives \(\Lambda(\ket{y_i}\bra{y_j})=\Phi^c_{w,t}(\ket{y_i}\bra{y_j})\), and the second gives \(\Lambda(\ket{y_i}\bra{y_j})=\Phi^c_{v,t}(T\ket{y_i}\bra{y_j}T^*)=\Phi^c_{v,t}(\ket{x_i}\bra{x_j})\). With these entries \eqref{eq:J(Lambda)} is the expansion \eqref{eq:reference_record} of the record, of \(y\) in sector \(w\) and of \(x\) in sector \(v\), at \(k=d_w\).  To return to a fixed basis, expand \((U\otimes I)\Omega_w=d_w^{-1/2}\sum_{i,j}U_{ij}\ket i\otimes\ket j =(I\otimes U^{\mathsf T})\Omega_w\). A change of the source frame by a unitary \(U\) thus rotates the reference of both records by the same \(U^{\mathsf T}\), and their fidelity is unchanged. Varying the frames thus amounts to varying the isometry \(T\)  alone, and \eqref{eq:spectral_records} at \(k=d_w\), where its left side is \( \max_S\sum_{j=1}^{d_w}\sqrt{\lambda_j(\Ws_S(t))}=\max_S\Tr\sqrt{\Ws_S(t)}\), gives
\begin{equation}
\max_S\Tr\sqrt{\Ws_S(t)} =d_w\max_{T^*T=I_{\HH_w}}
\ff\bigl(\mathsf J(\Phi_{v,t}^c\circ\Ad_T),\mathsf J(\Phi_{w,t}^c)\bigr).
\label{eq:complementary}
\end{equation}
Equation \eqref{eq:complementary} compares the two complementary channels  \(\Phi^c_{v,t}\) and \(\Phi^c_{w,t}\), which send a preparation to its record, after aligning the input spaces by \(T\) and retaining a reference. The reference is what reaches beyond the largest eigenvalue of the weight: \(k=1\) is Theorem~\ref{thm:information_weights}, and a \(k\)-dimensional reference bounds the \(k\) largest eigenvalues of \(\Ws_S(t)\) together. The role of a reference in information--disturbance relations is discussed in~\cite{KretschmannSchlingemannWerner,HaydenWinter2012}. Thus records control the spectrum of an edge weight and not only its norm: the sum of the \(k\) largest singular values of an evolved edge is at most \(k\) times the fidelity of two records that keep a \(k\)-dimensional reference, with equality for the best edge.

\subsection{The initial slope from a norm derivative}\label{supp:record_rate_proof}
Here we present the proof of Proposition~\ref{thm:record_rate}. It uses   the logarithmic norm of a generator \cite{LumerPhillips1961,Soderlind2006} and the derivative of a maximum, and then evaluates the logarithmic norm of the corner generator.
\begin{proof}
We first compute the derivative of a norm at \(t=0^+\). By
\eqref{eq:corner_information}, \(\ff_{vw}(t)=\norm{e^{t\LL_{vw}}}\). The logarithmic norm of \(\LL_{vw}\) is the number  \(\mu(\LL_{vw})=\lim_{h\downarrow0}(\norm{\id+h\LL_{vw}}-1)/h\) \cite{LumerPhillips1961,Soderlind2006}, it is the slope at \(h=0\) of the convex function \(h\mapsto\norm{\id+h\LL_{vw}}\), so the limit exists. We show that \(\mu(\LL_{vw})\) is the derivative of \(\norm{e^{t\LL_{vw}}}\) at \(t=0^+\), that \(\norm{e^{t\LL_{vw}}}\leq e^{t\mu(\LL_{vw})}\) for every \(t\geq0\), and that \(\mu(\LL_{vw})=-\mathfrak r_{vw}/2\); together these are the two statements of \eqref{eq:record_rate_bound}. The induced norm is the maximum
\begin{equation}
\norm{\id+h\LL_{vw}}
=\max_{\norm X\leq1,\ \norm x=\norm y=1} \Rea\langle x,(X+h\LL_{vw}(X))y\rangle ,
\end{equation}
as the norm  \(\norm{\id+h\LL_{vw}}\) is the largest modulus of  this pairing \(\langle x,(X+h\LL_{vw}(X))y\rangle\)  over contractions \(X\) and unit vectors \(x,y\); multiplying \(x\) by  the phase that cancels the phase of this number makes the pairing real and nonnegative, so the real part has the same maximum  as the modulus.  At \(h=0\) the maximal value is one, attained exactly when \(Xy=x\), since \(\Rea\langle x,Xy\rangle=1\) saturates both the norm bound and Cauchy--Schwarz. Therefore
\begin{equation}
\mu(\LL_{vw})=\max_{\substack{\norm X\leq1,\ Xy=x\\\norm x=\norm y=1}}
\Rea\langle x,\LL_{vw}(X)y\rangle .
\end{equation}
Here we justify differentiating a maximum whose optimizer can change. On the compact set of triples \(w=(X,x,y)\) write \(\varphi(w)=\Rea\langle x,Xy\rangle\) and \(\psi(w)=\Rea\langle x,\LL_{vw}(X)y\rangle\). Let \(\psi_*=\max_{\varphi(w)=1}\psi(w)\) and \(M(h)=\max_w(\varphi(w)+h\psi(w))\). If \(w_*\) attains \(\psi_*\), then \(\varphi(w_*)=1\) and \(M(h)\geq \varphi(w_*)+h\psi(w_*)=1+h\psi_*\). Conversely any maximizer \(w_h\) satisfies \(\varphi(w_h)\leq1\), so
\begin{equation}
 \psi_*\leq \frac{M(h)-1}{h}\leq \psi(w_h),\qquad
 1\geq \varphi(w_h)\geq 1+h\bigl(\psi_*-\psi(w_h)\bigr).
\end{equation}
The last bound and boundedness of \(\psi\) give \(\varphi(w_h)\to1\) as \(h\downarrow0\). Along a sequence \(h_n\downarrow0\) realizing the upper limit  of \(\psi(w_h)\), compactness gives a convergent subsequence with \(\varphi(w_0)=1\). Then \(\psi(w_0)\leq \psi_*\), which forces the
lower and upper limits of \((M(h)-1)/h\) as \(h\downarrow0\) to agree at \(\psi_*\), so \(\mu(\LL_{vw})=\psi_*\).

At a maximizing triple for \(\varphi\), \(Xy=x\) and \(\|(I-X^*X)^{1/2}y\|^2=1-\|Xy\|^2=0\). Thus \(X^*x=X^*Xy=y\). For \(z\perp y\), \(\langle x,Xz\rangle=\langle y,z\rangle=0\), so \(Xz\perp x\). It follows that \(X=\ket x\bra y+Y\) with \(Y:y^\perp\to x^\perp\) a contraction, extended by zero on \(y\). Conversely every such \(Y\) is allowed, since \(\|X(\lambda y+z)\|^2=|\lambda|^2+\|Yz\|^2\leq |\lambda|^2+\|z\|^2\). Decompose the jump vectors as
\begin{equation}
L_kx=m_k(x)x+a_k,\qquad L_ky=m_k(y)y+b_k ,
\end{equation}
with  vectors \(a_k\perp x\) and \(b_k\perp y\), one pair for each jump operator \(L_k\). We evaluate \(\langle x,\LL_{vw}(X)y\rangle\) term by term, using \(Xy=x\) and \(X^*x=y\). For the Hamiltonian term,
\begin{equation}
\langle x,i[H,X]y\rangle=i\bigl(\langle x,Hx\rangle-\langle y,Hy\rangle\bigr)
\end{equation}
is purely imaginary, so it does not enter the real part. For the term \(L_k^*XL_k\) of a jump, two of the four products survive, since \(a_k\perp x\) and \(Yb_k\perp x\):
\begin{equation}
\langle x,L_k^*XL_ky\rangle =\langle L_kx,XL_ky\rangle
=m_k(x)^{*}m_k(y)+\langle a_k,Yb_k\rangle .
\end{equation}
For the first term of the anticommutator, \(\langle x,L_k^*L_kXy\rangle=\norm{L_kx}^2\). For the second, \(\langle x,XL_k^*L_ky\rangle=\langle X^*x,L_k^*L_ky\rangle=\norm{L_ky}^2\). Expanding the squared norms and collecting the means now gives
\begin{align}
\Rea\langle x, \LL_{vw}(X)y\rangle
= &-\tfrac12\sum_k\left(|m_k(x)-m_k(y)|^2+\norm{a_k}^2+\norm{b_k}^2\right) +\Rea\sum_k\langle a_k,Yb_k\rangle .
\end{align}
We maximize \(\Rea\sum_k\langle a_k,Yb_k\rangle\) over \(Y\) with \(x,y\) fixed.  Let \(A\) be the map \((c_k)\mapsto\sum_kc_ka_k\), whose columns are the \(a_k\), and \(B\) the same map for the \(b_k\), then \(BA^*=\sum_k\ket{b_k}\bra{a_k}\) and \(\sum_k\langle a_k,Yb_k\rangle=\Tr(YBA^*)\). As in the proof of Theorem~\ref{thm:information_weights}, a polar partial isometry attains \(\norm{BA^*}_1\). Moreover \(A^*A=\cov_x\) and \(B^*B=\cov_y\), so
\begin{equation}
\max_{\norm Y\leq1}\Rea\sum_k\langle a_k,Yb_k\rangle
=\norm{\sqrt{\cov_y}\sqrt{\cov_x}}_1 .
\end{equation}
The two covariance traces give \(\sum_k\norm{a_k}^2\) and \(\sum_k\norm{b_k}^2\). The maximum over  \(Y\) is therefore \(-\tfrac12\bigl(\sum_k|m_k(x)-m_k(y)|^2+d_{\mathrm B}(\cov_x,\cov_y)^2\bigr)\), and by \eqref{eq:record_rate} the maximum over all triples is \(\mu(\LL_{vw})=-\mathfrak r_{vw}/2\).

Since \(e^{h\LL_{vw}}=\id+h\LL_{vw}+O(h^2)\), the  norms \(\norm{e^{h\LL_{vw}}}\) and \(\norm{\id+h\LL_{vw}}\) have the same one-sided derivative  at \(h=0\), which is \(\mu(\LL_{vw})\) by definition. Thus, since \(\ff_{vw}(t)=\norm{e^{t\LL_{vw}}}\), 
\(\ff_{vw}(t)=1+t\mu(\LL_{vw})+o(t)\). With \(\mu(\LL_{vw})=-\mathfrak r_{vw}/2\) the square of \(\ff_{vw}(t)\) is \(1-\mathfrak r_{vw}t+o(t)\), which proves the first identity of \eqref{eq:record_rate_bound}.
Since \(e^{t\LL_{vw}}=\lim_n(\id+(t/n)\LL_{vw})^n\), submultiplicativity of the norm gives the all-time bound
\begin{equation}
\norm{e^{t\LL_{vw}}}\leq\lim_{n\to\infty}\norm{\id+(t/n) \LL_{vw}}^{\,n}
=\lim_{n\to\infty}\left(1+\frac{t\mu(\LL_{vw})}n+o(n^{-1})\right)^n=e^{t\mu(\LL_{vw})} .
\end{equation}
Hence \(\ff_{vw}(t)^2=\norm{e^{t\LL_{vw}}}^2\leq e^{2t\mu(\LL_{vw})}=e^{-\mathfrak r_{vw}t}\), the inequality in \eqref{eq:record_rate_bound}.

For \eqref{eq:record_rate_dissipation}, take an isometry on the source, \(S^*S=p_w\), and set \(x=Sy\). Then
\begin{equation}
[L_k,S]y=(m_k(x)-m_k(y))x+(a_k-Sb_k).
\end{equation}
The two summands on the right, the multiple of \(x\) and \(a_k-Sb_k\), are orthogonal since \(a_k\perp x\) and \(\langle x,Sb_k\rangle=\langle y,b_k\rangle=0\). Hence, by \eqref{eq:gamma},
\begin{equation}
\langle y,\Gamma(S,S)y\rangle
=\sum_k|m_k(x)-m_k(y)|^2 + \sum_k\bigl(\norm{a_k}^2+\norm{b_k}^2-2\Rea\langle a_k,Sb_k\rangle\bigr).
\end{equation}
It remains to find an isometry \(S\) for which the last sum is as small as for the optimal contraction \(Y\). If the polar optimizer for \(Y\), the partial isometry \(Y_0\) with \(\Tr(Y_0BA^*)=\norm{BA^*}_1\), has rank \(r\), its initial space uses \(r\) dimensions of \(y^\perp\) and its final space uses \(r\) dimensions of \(x^\perp\). An isometry on the source must be defined on all of \(y^\perp\), so we extend \(Y_0\) from its initial space to all of \(y^\perp\):  map  an orthonormal basis of the \(d_w-1-r\) unused source directions to distinct  orthonormal vectors among the \(d_v-1-r\) unused target directions. This is possible exactly  when \(d_v\geq d_w\). The extension agrees with
the polar optimizer on the range of \(BA^*\), so the pairing \(\Tr(YBA^*)\) stays maximal. Let the operator  \(S\) be this extension on \(y^\perp\) together with \(Sy=x\), it maps an orthonormal basis of \(\HH_w\) to orthonormal vectors of \(\HH_v\), so \(S^*S=p_w\). For this \(S\), the last sum is \(\sum_k(\norm{a_k}^2+\norm{b_k}^2)-2\norm{BA^*}_1 =\Tr\cov_x+\Tr\cov_y-2\norm{\sqrt{\cov_y}\sqrt{\cov_x}}_1 =d_{\mathrm B}(\cov_x,\cov_y)^2\), by \(A^*A=\cov_x\), \(B^*B=\cov_y\) and the definition of the Bures distance in Section~\ref{sec:record_rates}, so \(\langle y,\Gamma(S,S)y\rangle\) equals the expression minimized in \eqref{eq:record_rate}, and no isometry with \(Sy=x\) gives less, since its restriction to \(y^\perp\) is one of the contractions \(Y\). Thus, for every pair \(x,y\), the least value of \(\langle y,\Gamma(S,S)y\rangle\) over isometries \(S\) with \(Sy=x\) is the expression minimized in \eqref{eq:record_rate}, and every isometry \(S\) and unit \(y\) occur in this way, with \(x=Sy\). Minimizing over \(x,y\) therefore proves \eqref{eq:record_rate_dissipation}. The Hamiltonian contributed only a purely imaginary number before this minimization, which accounts for its absence from the initial rate, the last assertion of the proposition.
\end{proof}

Two comparisons help interpret the rate.

First, let \(\gamma_{vw}=-\max\{\Rea z:z\in\spec(\LL_{vw})\}\). The spectral radius of \(e^{t\LL_{vw}}\) is \(e^{-\gamma_{vw}t}\), which is bounded by its norm. Equation \eqref{eq:record_rate_bound} then gives
\begin{equation}
e^{-2\gamma_{vw}t}\leq\ff_{vw}(t)^2\leq e^{-\mathfrak r_{vw}t},
\qquad \mathfrak r_{vw}\leq 2\gamma_{vw}. \label{eq:record_spectral_bounds}
\end{equation}
In finite dimensions the norm of \(e^{t\LL_{vw}}\) is bounded above by a polynomial in \(t\) times \(e^{-\gamma_{vw}t}\), as the exponential of  the Jordan block form of \(\LL_{vw}\) shows. The spectral radius gives the reverse 
exponential bound, so that \(e^{-\gamma_{vw}t}\leq\norm{e^{t\LL_{vw}}}\leq p(t)e^{-\gamma_{vw}t}\) with \(p\) that polynomial, taking logarithms and dividing by \(t\) gives \(-\gamma_{vw}\leq t^{-1}\log\norm{e^{t\LL_{vw}}}\leq-\gamma_{vw}+t^{-1}\log p(t)\), with limit \(-\gamma_{vw}\) as \(t\to\infty\). If the two rates in \eqref{eq:record_spectral_bounds} agree, the bounds coincide at every time. Conversely, if \(\ff_{vw}(t)^2=e^{-ct}\) for all \(t\), then \(c=\mathfrak r_{vw}\) by \eqref{eq:record_rate_bound} and \(c=2\gamma_{vw}\) by this limit. For rank-one sectors \(\HH_v=\CZ\ket v\) write \(L_k\ket v=\ell_{v,k}\ket v\), \(H\ket v=h_v\ket v\), and similarly for \(w\).
The  corner unit \(e_{vw}=\ket v\bra w\) satisfies \(\LL(e_{vw})=z e_{vw}\), where
\begin{equation}
 z=i(h_v-h_w)+\sum_k\left(\ell_{v,k}^{*}\ell_{w,k}
                   -\tfrac12|\ell_{v,k}|^2-\tfrac12|\ell_{w,k}|^2\right).
\end{equation}
Its squared amplitude is \(e^{2t\Rea z}=\exp(-t\sum_k|\ell_{v,k}-\ell_{w,k}|^2)\). 
Here the covariance matrices vanish because \(L_k\ket v\) is a multiple of \(\ket v\), so the vectors \(a_k=L_kx-m_k(x)x\) are zero, and thus the jumps have no component perpendicular to their inputs. In higher rank a vanishing \(\mathfrak r_{vw}\) specifies, by \eqref{eq:record_rate_bound}, only the initial slope of \(\ff_{vw}(t)^2\), which can still decrease later through terms of higher order in \(t\).

Second, fix the vector \(y\in\HH_w\) and let \(\ff_y(t)=\max_x\ff(\sigma_v^E(x,t),\sigma_w^E(y,t))\). If \(d_v\geq d_w\), the polar optimizer in \eqref{eq:pair_information} extends to an isometry on the full source by the same basis completion. Interchanging the maxima  over \(x\) and over \(X\) in \(\ff_y(t)=\max_x\max_X|\langle x,\Phi_t(X)y\rangle|\) gives
\(\ff_y(t)=\max_{S^*S=p_w}\norm{\Phi_t(S)y}\) for \(S\) in the corner. Since \(p_w\) is fixed, \(1-\norm{\Phi_t(S)y}^2=\langle y,D_t(S,S)y\rangle\). Therefore \eqref{eq:integrated_kernel} yields
\begin{equation}
1-\ff_y(t)^2 =\min_{\substack{S\in p_v\A p_w\\S^*S=p_w}}
\int_0^t\langle y,\Phi_{t-u}\bigl(\Gamma(\Phi_uS,\Phi_uS)\bigr)y\rangle\,du .
\label{eq:integral_records}
\end{equation}
A further minimum over \(y\) gives \(1-\ff_{vw}(t)^2\), because \(\ff_{vw}(t)=\max_y\ff_y(t)\) by \eqref{eq:sector_record_overlap}. Thus the initial derivative \eqref{eq:record_rate_bound}, the finite-time fidelity deficit \eqref{eq:integral_records}, and the spectral decay rate \eqref{eq:record_spectral_bounds} can all be computed for the same invariant corner, with the optimizations specified in \eqref{eq:record_rate_dissipation} and \eqref{eq:integral_records}.

The proof of Proposition~\ref{thm:record_rate} gives the rate of each pair of records, and for that no sector is needed. For any unit vectors \(x,y\in\HH\), the records of \eqref{eq:unrestricted_pair_information} satisfy
\begin{equation}
\ff(\sigma^E(x,t),\sigma^E(y,t))^2 =1-t\Bigl(\sum_k|m_k(x)-m_k(y)|^2+d_{\mathrm B}(\cov_x,\cov_y)^2\Bigr)+o(t).
\label{eq:pair_record_rate}
\end{equation}
The expression minimized in \eqref{eq:record_rate} is thus the initial separation rate of the pair \(x,y\) itself.

 To obtain \eqref{eq:pair_record_rate}, keep the unit vectors \(x,y\in\HH\) fixed and let \(X\) range over all contractions in \(B(\HH)\). By \eqref{eq:unrestricted_pair_information} the record fidelity is \(\max_X|\langle x,\Phi_t(X)y\rangle|\). A phase of \(X\) makes the pairing real, so the fidelity is also \(\max_X\Rea\langle x,\Phi_t(X)y\rangle\), and \(\Phi_t(X)=X+t\LL(X)+O(t^2)\) uniformly in \(X\). At \(t=0\) the two records coincide and the maximum is one, attained exactly when \(Xy=x\). The argument for differentiating a maximum, given above, applies with \(x,y\) fixed. The argument gives the derivative of the fidelity at \(t=0^+\) as the maximum of \(\Rea\langle x,\LL(X)y\rangle\) over the contractions with \(Xy=x\). The evaluation of that maximum in the proof used only \(Xy=x\) and \(X^*x=y\), and no sector. Its value is \(-\tfrac12\bigl(\sum_k|m_k(x)-m_k(y)|^2+d_{\mathrm B}(\cov_x,\cov_y)^2\bigr)\), and squaring the fidelity gives \eqref{eq:pair_record_rate},  
 \begin{align} 
 \ff(\sigma^E(x,t),\sigma^E(y,t))& =\max_{\norm X\leq1}\Rea\langle x,\Phi_t(X)y\rangle \notag\\
 & =1+t\max_{\norm X\leq1,\ Xy=x}\Rea\langle x,\LL(X)y\rangle+o(t)\notag\\ 
 & =1-\tfrac t2\Bigl(\sum_k|m_k(x)-m_k(y)|^2+d_{\mathrm B}(\cov_x,\cov_y)^2\Bigr)+o(t).
 \end{align} 
 This subsection has thus proved Proposition~\ref{thm:record_rate} and shown that the minimum of the distance between the jump means and fluctuations of two preparations \eqref{eq:record_rate} is the initial separation rate of the records of the two sectors.

\subsection{The trace norm of the cross operator}\label{supp:cross_trace_norm}
Here we prove the identity \eqref{eq:cross_trace_norm} used in the proof of Theorem~\ref{thm:information_weights}. Recall the setting of that proof, the unit vectors \(\psi=V_{v,t}x\) and \(\chi=V_{w,t}y\) are expanded in a common environment basis, \(\psi=\sum_\mu a_\mu\otimes\ket\mu\) and \(\chi=\sum_\mu b_\mu\otimes\ket\mu\), with \(a_\mu\in\HH_v\) and \(b_\mu\in\HH_w\). Their environment marginals are the records \(\sigma_v^E(x,t)\) and \(\sigma_w^E(y,t)\) of \eqref{eq:conditional_environment}, and the cross operator is \(C=\Tr_E\ket\chi\bra\psi=\sum_\mu\ket{b_\mu}\bra{a_\mu}\). The identity \(\norm C_1=\ff(\sigma_v^E,\sigma_w^E)\) is the cross-operator form of Uhlmann's purification formula \cite{doi:10.1016/0034-4877(76)90060-4}~\cite[Lemma~20]{HaydenWinter2012}, obtained from the duality between the trace norm and the operator norm together with Uhlmann's theorem. The calculation here is direct: it uses polar decompositions  and does not use Uhlmann's theorem.

Define  two maps from \(\HH_E\) to the system space by \(A\ket\mu=a_\mu\) and \(B\ket\mu=b_\mu\), with the vectors \(a_\mu\), \(b_\mu\) of the two expansions. Then the cross operator is \(C=BA^*\), and the \((\mu,\nu)\) entry of the environment marginal of \(\psi\) is \(\Tr(\ket{a_\mu}\bra{a_\nu})=\langle a_\nu,a_\mu\rangle\).
Hence, with transpose in the chosen  environment basis,
\begin{equation}
\sigma_v^E(x,t)=(A^*A)^{\mathsf T},\qquad
\sigma_w^E(y,t)=(B^*B)^{\mathsf T}.
\end{equation}
Write \(A=U_A\sqrt{A^*A}\), \(B=U_B\sqrt{B^*B}\). The potentially different initial projections are
\(p_A=U_A^*U_A=\supp(A^*A)\) and \(p_B=U_B^*U_B=\supp(B^*B)\).
Put \(K=\sqrt{B^*B}\sqrt{A^*A}\). Multiplying by these projections gives \(p_BKp_A=K\). Consequently
\begin{equation}
C=U_BKU_A^*,\qquad
 C^*C=U_AK^*p_BKU_A^*=U_AK^*KU_A^* .
\end{equation}
Now \(U_A|K|U_A^*\) is positive and its square is \(C^*C\), since \(p_A|K|=|K|\). The uniqueness of the positive square root and cyclicity of trace give \(\Tr|C|=\Tr(|K|p_A)=\Tr|K|\). This also covers a singular \(A\) or \(B\). If \(K=U\Sigma W^*\) is a singular value decomposition, transposing gives \(K^{\mathsf T}=\overline W\Sigma^{\mathsf T}U^{\mathsf T}\), with the same singular values. The transpose of the positive square root is the positive square root of the transpose. Thus
\begin{align}
\norm C_1
&=\norm{\sqrt{B^*B}\sqrt{A^*A}}_1
 =\norm{(\sqrt{B^*B}\sqrt{A^*A})^{\mathsf T}}_1\notag\\
&=\norm{\sqrt{(A^*A)^{\mathsf T}}\sqrt{(B^*B)^{\mathsf T}}}_1
 =\ff(\sigma_v^E(x,t),\sigma_w^E(y,t)). \label{eq:cross_trace_norm_proof}
\end{align}
The equality of the two ends of \eqref{eq:cross_trace_norm_proof} is \eqref{eq:cross_trace_norm}.

 The calculation  of \eqref{eq:cross_trace_norm_proof} used the sectors only to name the spaces that contain \(a_\mu\) and \(b_\mu\). For arbitrary unit vectors \(x,y\in\HH\) put \(\psi=V_tx\) and \(\chi=V_ty\). As in the proof of Theorem~\ref{thm:information_weights}, \(\langle x,\Phi_t(X)y\rangle=\Tr(XC)\) with \(C=\Tr_E\ket\chi\bra\psi=\Phi_{t*}(\ket y\bra x)\). The maximum of \(|\Tr(XC)|\) over all contractions \(X\in B(\HH)\) is \(\norm C_1\), and \eqref{eq:cross_trace_norm_proof} identifies \(\norm C_1\) with the fidelity of the two records. Hence \eqref{eq:unrestricted_pair_information} holds.

The identity extends to sectors that move, which proves \eqref{eq:transported_corner_information}. Let \(q_v=\Phi_t(p_v)\) be a projection. Then \(D_t(p_v,p_v)=q_v-q_v^2=0\), so \(p_v\in\MD(\Phi_t)\) by \eqref{eq:md}. Since \(V_t^*(p_v\otimes I)V_t=q_v\) and \(V_t^*V_t=I\),
\begin{equation}
\bigl((p_v\otimes I)V_t-V_tq_v\bigr)^*\bigl((p_v\otimes I)V_t-V_tq_v\bigr)
=q_v-q_v^2-q_v^2+q_v=0 .
\label{eq:transported_intertwining}
\end{equation}
An operator \(Z\) with \(Z^*Z=0\) is zero, hence \((p_v\otimes I)V_t=V_tq_v\), and \(V_tx\in p_v\HH\otimes\HH_E\) for every \(x\in q_v\HH\). For \(x\in q_v\HH\) and \(y\in q_w\HH\) the cross operator therefore satisfies \(C=p_wCp_v\), and its polar optimizer \(U^*\) lies in the original corner \(p_v\A p_w\). So the maximum in \eqref{eq:unrestricted_pair_information} is already attained on that corner. Because \(\Phi_t(p_vXp_w)=q_v\Phi_t(X)q_w\), the norm of \(\Phi_t(X)\), for \(X\) in the corner, is attained on unit vectors \(x\in q_v\HH\) and \(y\in q_w\HH\). Maximizing over these vectors and over the corner in either order, as in the proof of Theorem~\ref{thm:information_weights}, gives \eqref{eq:transported_corner_information}. 

\subsection{Classical records for the shared jumps}\label{supp:shared_classical_records}
For the shared jumps \eqref{eq:shared_jumps} and the preparations of Corollary~\ref{prop:shared_jump_weight},  the record fidelity of the inputs \(\ket v\otimes\ket a\) and \(\ket w\otimes\ket a\) equals the classical fidelity of their jump records: the time-resolved record of the jumps is an optimal measurement of the environment in the sense of Section~\ref{sec:measured_records}. 
Let \(p_{v,a}\) and \(p_{w,a}\) be the   probability densities of the complete jump record  \(\xi\), i.e.,  the jump indices with their times up to time \(t\), for the inputs \(\ket v\otimes\ket a\) and \(\ket w\otimes\ket a\), with respect to a common reference measure \(\mu\) on records, the same for both inputs and written out in the proof. Then the record fidelity \(f_a(t)\) of the two inputs (Corollary~\ref{prop:shared_jump_weight}) is the classical root fidelity of the two densities,
\begin{equation}
f_a(t)=\int\sqrt{p_{v,a}(\xi)\,p_{w,a}(\xi)}\,d\mu(\xi),
\label{eq:shared_classical_fidelity}
\end{equation} 
so the bound \eqref{eq:classical_record_bound} holds with equality for the time-resolved jump measurement.

We now prove \eqref{eq:shared_classical_fidelity}. 
A record \(\xi=(k_1,s_1;\ldots;k_n,s_n)\) lists the jump indices and their times \(0<s_1<\cdots<s_n<t\); the empty record, \(n=0\), is implicitly included. An integral over records is a sum over the number of jumps and over their indices, and an integral over the ordered times, \(\int F\,d\mu=\sum_{n\geq0}\sum_{k_1,\ldots,k_n}\int_{0<s_1<\cdots<s_n<t}F(\xi)\,ds_1\cdots ds_n\),  where the term \(n=0\) is \(F\) at the empty record. Split the  generator on densities  as \(\LL_*(\eta)=K\eta+\eta K^*+\sum_kL_k\eta L_k^*\), with \(K=-iH-\tfrac12\sum_kL_k^*L_k\). Then \(e^{t\LL_*}(\eta)=e^{tK}\eta e^{tK^*}+\int_0^te^{(t-s)K}\bigl(\sum_kL_k\,e^{s\LL_*}(\eta)\,L_k^*\bigr)e^{(t-s)K^*}ds\), as differentiation in \(t\) shows. Substituting this identity into itself, once for each jump, gives   the expansion of \(e^{t\LL_*}\) in the number of jumps,  
\begin{equation}
\Phi_{t*}(\eta)=\int K(\xi)\,\eta\,K(\xi)^*\,d\mu(\xi),\qquad
K(\xi)=e^{(t-s_n)K}L_{k_n}e^{(s_n-s_{n-1})K}\cdots L_{k_1}e^{s_1K}.
\label{eq:jump_record_expansion}
\end{equation}
The operators \(K(\xi)\) are Kraus operators of \(\Phi_t\) labeled by the record, and \(\norm{K(\xi)\phi}^2\) is the probability density of the record \(\xi\) for the input \(\phi\). They give a dilation of the form \eqref{eq:dilation} whose environment basis is labelled by the records. Tracing over the environment sums over the records and returns \(\Phi_{t*}\), as in \eqref{eq:jump_record_expansion}. Reading the record keeps \(\xi\) and leaves the system in the unnormalized state \(K(\xi)\phi\); it is a measurement of the environment in this basis, in the sense of Section~\ref{sec:measured_records}.

The Hamiltonian and the jumps of \eqref{eq:shared_jumps} act on the internal levels separately in each sector: \(H=I_r\otimes H_0\) is the same internal Hamiltonian \(H_0\) in every sector, and \(L_k=\sum_vg_{v,k}e_{vv}\otimes\ell_k\) is the internal jump \(\ell_k\) with strength \(g_{v,k}\) in sector \(v\). Each factor of \(K(\xi)\) therefore preserves the sectors, and \(K(\xi)(\ket v\otimes\ket a)=\ket v\otimes K_v(\xi)\ket a\) with an internal operator \(K_v(\xi)\).

Between jumps the internal propagator in sector \(v\) is generated by \(-iH_0-\tfrac12\sum_kg_{v,k}^2\ell_k^*\ell_k\), which is diagonal in the basis \(\ket a\). We follow a basis vector \(\ket a\) through the factors of \(K_v(\xi)\). A propagator between jumps multiplies a basis vector by a positive damping factor, which depends on \(v\), and by a phase from \(H_0\), which does not. A jump \(g_{v,k}\ell_k\) sends the basis level \(a_k\) to \(b_k\) with the nonnegative amplitude \(g_{v,k}\sqrt{r_k}\) and annihilates the other levels. The vector thus stays a multiple of a basis vector, or becomes zero. With \(\ket{a(\xi)}\) the final basis vector and \(\phi_a(\xi)\) the accumulated phase,
\begin{equation}
K_v(\xi)\ket a=\sqrt{p_{v,a}(\xi)}\,e^{i\phi_a(\xi)}\ket{a(\xi)},\qquad
p_{v,a}(\xi)=\norm{K_v(\xi)\ket a}^2 .
\label{eq:record_conditioned_vector}
\end{equation} 
The final level \(a(\xi)\) and the phase \(\phi_a(\xi)\), which comes from \(H_0\) alone  because the damping factors and the jump amplitudes are nonnegative, are the same in every sector in which the vector is nonzero, because the transitions and \(H_0\) are common to the sectors. Only the modulus \(\sqrt{p_{v,a}(\xi)}\) depends on \(v\). Inserting \eqref{eq:record_conditioned_vector} for the sectors \(w\) and \(v\) into \eqref{eq:jump_record_expansion} for \(\eta=e_{wv}\otimes q_a\), with \(q_a=\ket a\bra a\) on the internal factor, the phases cancel:
\begin{equation}
\Phi_{t*}(e_{wv}\otimes q_a)=e_{wv}\otimes C_a(t),\qquad
C_a(t)=\int\sqrt{p_{v,a}(\xi)\,p_{w,a}(\xi)}\;q_{a(\xi)}\,d\mu(\xi).
\label{eq:record_cross_evolution}
\end{equation} 
The matrix \(C_a(t)\) is the internal matrix in the proof of Corollary~\ref{prop:shared_jump_weight}, where \(f_a(t)=\norm{C_a(t)}_1=\Tr C_a(t)\). Each \(q_{a(\xi)}\) has trace one, so the trace of \eqref{eq:record_cross_evolution} is \eqref{eq:shared_classical_fidelity}. The right side of \eqref{eq:shared_classical_fidelity} is the classical root fidelity of the two measured records, and by \eqref{eq:classical_record_bound} no measurement of the environment gives a smaller value.  Thus every eigenvalue \(f_a(t)^2\) of the weight of \(S_{vw}\) is the squared classical fidelity of two jump records, and the time-resolved record is an optimal measurement for every level, every pair of sectors and every time. 
With respect to fidelity, the environment holds nothing about the sector beyond this record: given the record, the internal state of the system is the same in both sectors, and the sectors differ only in how probable the record is. 
The equality is for preparations at the same level \(a\) in the two sectors; for two different levels the argument gives only the bound \eqref{eq:classical_record_bound}.

\section{Retained algebra: proofs and state limits}\label{supp:retained_proofs}
This section proves the statements in Section~\ref{sec:asymptotic_algebra} from the product calculus \eqref{eq:gamma}--\eqref{eq:integrated_kernel}. Here we give the commutator form of \(\Gamma\), the full survival and asymptotic proofs, followed by the state limit, the calculations for the common-reservoir model, the normalized trace of a matrix block, the retained dissipation integral, and the trajectory map.

\subsection{The commutator form of Gamma}\label{supp:gamma_expansion}
To obtain the second line in \eqref{eq:gamma}, first expand \([H,X^*Y]=[H,X^*]Y+X^*[H,Y]\). 
For one jump \(L\), write \(R=L^*L\). The remaining expansion is
\begin{align}
\Gamma_L(X,Y)
={}&L^*X^*YL-\tfrac12RX^*Y-\tfrac12X^*YR\notag\\
 &-L^*X^*LY+\tfrac12RX^*Y+\tfrac12 X^*RY\notag\\
 &-X^*L^*YL+\tfrac12X^*RY+\tfrac12X^*YR\notag\\
={}&L^*X^*YL-L^*X^*LY-X^*L^*YL+X^*L^*LY\notag\\
={}&(X^*L^*-L^*X^*)(LY-YL)=[L,X]^*[L,Y].
\end{align}

\subsection{Exact survival at positive times}\label{supp:exact_survival_proof}
The multiplicative domain \eqref{eq:md} is a unital \(C^*\)-algebra, as we now check. Write \(P=VV^*\) in a Stinespring representation \eqref{eq:dilation}, with \(\pi(X)=X\otimes I_E\). The two diagonal defects require
\((I-P)\pi(X)V=0\) and \((I-P)\pi(X^*)V=0\).
Multiplying each on the right by \(V^*\) and taking the adjoint of the second gives
\((I-P)\pi(X)P=0=P\pi(X)(I-P)\). 
These are just the off-diagonal blocks of \(\pi(X)\) on \(\ran P\oplus\ker P\). Conversely, if these blocks vanish then multiplication by \(V\), using \(PV=V\), gives both original conditions. Equivalently, \(P\pi(X)=P\pi(X)P=\pi(X)P\). Matrices commuting with \(P\) form an algebra closed under adjoints and norm limits. Their inverse image under the representation \(\pi\) has these same properties. This proves that \(\MD(\Phi)\) is a unital \(C^*\)-algebra.

Here we prove Proposition~\ref{prop:exact-survival}.
\begin{proof}
Proposition~\ref{prop:exact-survival} makes three claims. If \(X\) lies in the multiplicative domain of \(\Phi_t\) at one time \(t>0\), it lies in the multiplicative domain at every time. The algebra \(\N\) is mapped into itself. And the restricted maps are automorphisms that form a group. We prove them in four steps, starting from \eqref{eq:integrated_kernel}.

\emph{Step 1: from one time to the interval \([0,t]\).} Let \(X\in\MD(\Phi_t)\) for some \(t>0\), so that \(D_t(X,X)=0\). From \eqref{eq:integrated_kernel},
\begin{equation}
0=\int_0^t \Phi_{t-u}\bigl(\Gamma(\Phi_uX,\Phi_uX)\bigr)\,du.
\label{eq:survival_zero_integral}
\end{equation}
The integrand is a positive operator at every \(u\). For an abstract finite-dimensional algebra, positivity of \(\Gamma(A,A)\) follows directly by taking \(h\downarrow0\) in \(D_h(A,A)/h\). For matrices with \eqref{eq:gksl} it also follows term by term from
\eqref{eq:gamma}. At every intermediate time \(u\), the outer channel \(\Phi_{t-u}\) takes this positive matrix to a positive matrix. A continuous family of positive operators with zero integral vanishes identically. To see this, fix a vector \(y\) and write \(g_y(u) =\langle y,\Phi_{t-u}(\Gamma(\Phi_uX,\Phi_uX))y\rangle \) for the expectation of the integrand of \eqref{eq:survival_zero_integral} on \(y\).
It is continuous and nonnegative. If \(g_y(u_0)>0\), continuity gives a subinterval of positive length where \(g_y(u)>g_y(u_0)/2\), contradicting the zero integral. Thus \(g_y(u)=0\) for every \(y,u\).
A positive matrix with all these expectations zero vanishes, so \(\Phi_{t-u}\bigl(\Gamma(\Phi_uX,\Phi_uX)\bigr)=0\). Finally, \(e^{(t-u)\LL}\) has linear inverse \(e^{-(t-u)\LL}\), so  applying it to \(\Phi_{t-u}(\Gamma(\Phi_uX,\Phi_uX))=0\) shows that \(\Gamma(\Phi_uX,\Phi_uX)\) is zero too. We use the inverse only on an equality; it is not a positive map and would not preserve an inequality. The same argument for \(D_t(X^*,X^*)=0\) gives the conclusion for \(X^*\). The result of this step is
\begin{equation}
\Gamma(\Phi_uX,\Phi_uX)=\Gamma(\Phi_uX^*,\Phi_uX^*)=0\qquad(0\leq u\leq t).
\label{eq:survival_interval}
\end{equation}

\emph{Step 2: from the interval to all times.} The map \(u\mapsto\Phi_u=e^{u\LL}\) has real analytic matrix entries, because the exponential is a convergent power series. Products and adjoints give real analytic
entries for \(u\mapsto\Gamma(\Phi_uX,\Phi_uX)\) along the real axis. A real analytic function that vanishes on the interval \((0,t)\) vanishes throughout the connected time axis. Hence \eqref{eq:survival_interval} holds for every \(u\geq0\). Applying \eqref{eq:integrated_kernel} for any later time \(s\) now gives \(D_s(X,X)=D_s(X^*,X^*)=0\), so \(X\in\MD(\Phi_s)\) for every \(s\), that is, \(X\in\N\). The reverse inclusion \(\N\subseteq\MD(\Phi_t)\) in \eqref{eq:constant_domain} follows from the definition of \(\N\) as an intersection.

\emph{Step 3: \(\N\) is mapped into itself.} For \(X\in\N\) and any \(s,t\geq0\), the two-interval formula \eqref{eq:kernel_cocycle} gives \(D_t(\Phi_sX,\Phi_sX)=D_{s+t}(X,X)-\Phi_t(D_s(X,X))=0\). The same holds with \(X^*\) because \(X^*\in\N\). Keeping \(s\) fixed and allowing \emph{every} \(t\) shows that \(\Phi_sX\) belongs to the intersection defining \(\N\).

\emph{Step 4: the restrictions are automorphisms and form a group.} By Step 3 and the definition of the multiplicative domain, the restriction \(\alpha_s:\N\to\N\) is an endomorphism preserving products, the unit, and adjoints. Its kernel is zero by
injectivity of \(e^{s\LL}\). Its image consequently has the full finite dimension of \(\N\). It is bijective, its inverse preserves products and adjoints, and setting \(\alpha_{-s}=\alpha_s^{-1}\) gives the asserted group.

Steps 1 and 2 give \eqref{eq:constant_domain}, Step 3 the invariance of \(\N\) and Step 4 the group of automorphisms, thus proving Proposition~\ref{prop:exact-survival}.
\end{proof}

The commutant description of Section~\ref{supp:pointer_details} is derived as follows, with \(\RR\) the algebra \eqref{eq:pointer_noise_algebra}. We derive \(\N=\RR'\), which computes the retained algebra from the generator; this is the standard commutant description \cite{doi:10.1142/S0129055X16500033}. For \(X\in\N\),
invariance gives \(\Phi_tX\in\N\).   For \(Y=\Phi_tX\) the defects \(D_h(Y,Y)\) and \(D_h(Y^*,Y^*)\) vanish for all \(h\), hence so do \(\Gamma(Y,Y)\) and \(\Gamma(Y^*,Y^*)\). Equation \eqref{eq:gamma} yields \([L_k,\Phi_tX]=0\) and, by taking the adjoint of the calculation for \((\Phi_tX)^*\), \([L_k^*,\Phi_tX]=0\). Consequently \(L_k^*(\Phi_tX)L_k\) equals both \(L_k^*L_k\Phi_tX\) and \((\Phi_tX)L_k^*L_k\). As such, each dissipator vanishes, giving
\begin{align}
\tfrac d{dt}\Phi_tX=i[H,\Phi_tX],\qquad
\Phi_tX=e^{itH}Xe^{-itH},\qquad [e^{-itH}L_ke^{itH},X]=[e^{-itH}L_k^*e^{itH},X]=0 .
\label{eq:pointer_hamiltonian_orbit}
\end{align}
For example \(\frac{d}{dt}(e^{-itH}L_ke^{itH})=-i e^{-itH}[H,L_k]e^{itH}\). Iteration at \(t=0\) gives the derivatives \((-i)^j(\operatorname{ad}_H)^j(L_k)\).   These are the coefficients of the Taylor series in \(t\), which converges for every \(t\); with \(\operatorname{ad}_H(L)=[H,L]\),
\begin{equation}
e^{-itH}L_ke^{itH}=\sum_{j=0}^\infty\frac{(-it)^j}{j!}
(\operatorname{ad}_H)^j(L_k).
\end{equation}
Since the commutator with \(X\) vanishes at every \(t\), its derivatives at zero vanish as well, and the same argument applies to \(L_k^*\). This proves \(X\in\RR'\). Conversely let \(X\) commute with all those coefficients. The series for \(e^{-itH}L_ke^{itH}\) converges in norm, so \(X\) commutes with its sum, and likewise for \(L_k^*\); these are the last two commutators of \eqref{eq:pointer_hamiltonian_orbit}.  Conjugating \([e^{-itH}L_ke^{itH},X]=0\) by \(e^{itH}\) shows that 
\(Y_t=e^{itH}Xe^{-itH}\) commutes with each \(L_k,L_k^*\). Then \(L_k^*Y_t L_k=L_k^*L_kY_t=Y_tL_k^*L_k\), so  the dissipator in \eqref{eq:gksl} vanishes on \(Y_t\) and \(\LL(Y_t)=i[H,Y_t]=\dot Y_t\). Since \(Y_0=X\)  and \(\Phi_t(X)\) solves the same equation \(\dot Y=\LL(Y)\), uniqueness of a linear matrix ODE gives \(Y_t=\Phi_t(X)\). The commutant  \(\RR'\) is closed under products and adjoints, so  the same argument for \(X^*X\) and \(XX^*\) gives \(\Phi_t(X^*X)=Y_t^*Y_t\) and \(\Phi_t(XX^*)=Y_tY_t^*\). Thus  both defects of \(X\) vanish at every time, and \(X\in\N\).

\subsection{Faithful stationarity and the asymptotic expectation}\label{supp:asymptotic_proof}
Here we prove Proposition~\ref{prop:asymptotic_expectation}.
\begin{proof}
The proof has two tasks. We first identify the nondecaying part of the linear evolution, the term \(\alpha_t\EE(X)\) of \eqref{eq:asymptotic_decomposition}, and second show that the linear projection onto it has the positivity and multiplication properties required of a conditional
expectation.

\emph{Step 1: eigenoperators with imaginary eigenvalue lie in \(\N\).} We identify the eigenoperators of \(\LL\) on the imaginary axis, if \(\LL(X)=i\nu X\) for real \(\nu\), then
\(\Phi_tX=e^{i\nu t}X\).  The state \(\omega\) of \eqref{eq:faithful_stationary} is stationary; consequently
\begin{equation}
\omega(D_t(X,X)) 
=\omega(\Phi_t(X^*X))-\omega((e^{i\nu t}X)^*e^{i\nu t}X)
=\omega(X^*X)-\omega(X^*X)=0 .
\end{equation}
The operator \(D_t(X,X)\) is positive; by \eqref{eq:faithful_stationary} it is zero, and
applying the same calculation to \(X^*\) gives \(X\in\N\).

\emph{Step 2: such eigenoperators span \(\N\).} For the converse, a norm-continuous automorphism group on finite-dimensional \(\N\) is spanned by eigenoperators
with purely imaginary generator eigenvalues, that is, by \(X\) with \(\alpha_t(X)=e^{i\nu t}X\) and \(\nu\) real. This follows as continuity fixes each central summand:
an automorphism permutes the finite set of minimal central projections, and a continuous family of permutations starting at the identity is constant. Its generator \(\delta=\frac{d}{dt}\alpha_t\big|_{t=0}\) on a matrix block is a \(*\)-derivation, \(\delta(ab)=\delta(a)b+a\delta(b)\), as follows by differentiating \(\alpha_t(ab)=\alpha_t(a)\alpha_t(b)\). With the matrix units \(e_{ij}\) of one central block summand, put \(B=\sum_i\delta(e_{i1})e_{1i}\), the usual choice showing that a derivation of a matrix algebra is a commutator. We check \(\delta(e_{jk})=[B,e_{jk}]\) on each matrix unit, using the product rule for \(\delta\) and \(\sum_ie_{i1}e_{1i}=I\). Then
\begin{align}
B e_{jk}&=\delta(e_{j1})e_{1k},\notag\\
e_{jk}B&=\sum_i\bigl(\delta(e_{jk}e_{i1})-\delta(e_{jk})e_{i1}\bigr)e_{1i}
=\delta(e_{j1})e_{1k}-\delta(e_{jk}),
\end{align}
thus \(\delta(X)=[B,X]\).  Since \(\delta(X^*)=\delta(X)^*\), the adjoint of \(\delta(X^*)=[B,X^*]\) is \(\delta(X)=-[B^*,X]\), which yields \([B+B^*,X]=0\) for every \(X\) in the block, so \(B+B^*\) is a scalar on the block and replacing \(B\) by \((B-B^*)/2=iH_a\) gives the same derivation with \(H_a=H_a^*\). Let \(H_\N\in\N\) be the sum of these \(H_a\), then \(\alpha_t(n)=e^{itH_\N}n e^{-itH_\N}\). In an eigenbasis of \(H_a\), with eigenvalues \(h_i\), each matrix unit \(e_{ij}\) acquires the phase \(e^{it(h_i-h_j)}\). These matrix units span \(\N\), and each is an eigenoperator of \(\alpha_t\) with a phase.

\emph{Step 3: everything else decays.} The maps \(\Phi_t\) have norm one, therefore a generator eigenvalue \(\lambda\) of positive real part would contradict boundedness, since its eigenoperator has \(\norm{\Phi_tX}=e^{t\Rea\lambda}\norm X\). Nor can \(\LL\) have a Jordan block of size two or more at an imaginary eigenvalue: a Jordan chain of length two at \(i\nu\) would contain \(Y,X\) with \((\LL-i\nu)Y=X\ne0\) and
\((\LL-i\nu)X=0\), giving \(\Phi_tY=e^{it\nu}(Y+tX)\)  by the exponential series on the span of \(X\) and \(Y\). Its norm is at least \(t\|X\|-\|Y\|\), which is unbounded in \(t\) and again contradicts \(\norm{\Phi_t}=1\). On each remaining generalized eigenspace write \(\LL=\lambda I+N_\lambda\), where \(N_\lambda^k=0\) for some \(k\) and \(\Rea\lambda<0\).  The exponential  of \(t\LL\) on this eigenspace is
\begin{equation}
 e^{t\lambda}\sum_{j=0}^{k-1}\frac{t^j N_\lambda^j}{j!}.
\end{equation}
Every term tends to zero in norm, because \(t^je^{t\Rea\lambda}\to0\) when \(\Rea\lambda<0\); the imaginary eigenvalues were treated first and have no such blocks. The generator \(\LL\) is a linear map on the finite-dimensional space \(\A\), so by its Jordan decomposition \(\A\) is the direct sum of these generalized eigenspaces and the imaginary-axis eigenspaces, whose span is \(\N\) by Steps 1 and 2. Define the linear map \(\EE:\A\to\A\) to be zero on the former and the identity on the latter. This projection acts on observables. Each summand is invariant under \(\Phi_t\), so \(\EE\Phi_t=\Phi_t\EE=\alpha_t\EE\). The norm of the difference \(\Phi_t-\alpha_t\EE =\Phi_t(\id-\EE)\) tends to zero by summing the finitely many decaying blocks.

\emph{Step 4: \(\EE\) is a conditional expectation.} It remains to show that this spectral projection is a conditional expectation. We need times \(t_j\to\infty\) with \(\alpha_{t_j}\to\id_\N\). By Step 2, \(\alpha_t\) multiplies a basis of \(\N\) by finitely many phases \(e^{i\nu t}\). At integer times \(t=n\) their vector lies in a compact torus. Choose an increasing  sequence of integers \(n_1<n_2<\cdots\) for which those vectors converge. At  the positive differences \(t_j=n_{2j}-n_j\geq j\), the phase vectors are  \(e^{i\nu t_j}=e^{i\nu n_{2j}}/e^{i\nu n_j}\), quotients of two terms of this convergent sequence, so they approach the quotient of that limit by itself, the vector of ones. Therefore \(\alpha_{t_j}\to\id_\N\) and, since \(\Phi_{t_j}-\alpha_{t_j}\EE\to0\), \(\Phi_{t_j}\to \EE\). For any matrix size \(k\) and \(Z=[Z_{lm}]\geq0\) in \(M_k(\A)\), the matrices \([\Phi_{t_j}(Z_{lm})]_{lm}\), with \(\Phi_{t_j}\) applied to each entry, are positive, since \(\Phi_{t_j}\) is completely positive, and converge to \([\EE(Z_{lm})]_{lm}\).  A norm limit of positive matrices is positive, proving complete positivity of \(\EE\) at every \(k\). Evaluation at the identity and then in the state gives
\(\EE(I)=I\) and \(\omega \EE=\omega\). For \(a,b\in\N\), the multiplication identities at \(t_j\) give
\begin{equation}
\EE(aXb) =\lim_j\Phi_{t_j}(aXb)=\lim_j\alpha_{t_j}(a)\Phi_{t_j}(X)\alpha_{t_j}(b)=a\EE(X)b .
\end{equation}
Together with its range \(\N\) and \(\EE|_\N=\id\), these are the properties of a conditional expectation onto \(\N\) preserving \(\omega\), as Proposition~\ref{prop:asymptotic_expectation} asserts.
\end{proof}

\subsection{Conditional pointer densities and characters}\label{supp:pointer_states}
We derive the state description corresponding to Proposition~\ref{prop:commutative_pointer} and the observable expectation \eqref{eq:pointer_expectation}.

A pointer projection can have physical rank greater than one. Let \(\rho\)  be the density of the faithful stationary state \(\omega\) of \eqref{eq:faithful_stationary}, \(\omega(X)=\Tr(\rho X)\); its conditional
density in outcome \(a\) is
\begin{equation}
\rho_a=\frac{z_a\rho z_a}{\Tr(\rho z_a)} , \label{eq:pointer_sector_states}
\end{equation}
with \(z_a\) the minimal projections of \(\N\) in Proposition~\ref{prop:commutative_pointer}. For any initial density \(\eta\), evaluating the dual of \eqref{eq:pointer_expectation} on an arbitrary observable gives
\begin{align}
\Tr(\EE_*(\eta)X)
&=\sum_a\Tr(\eta z_a)\frac{\Tr(\rho z_aXz_a)}{\Tr(\rho z_a)}
=\sum_a\Tr(\eta z_a)\Tr(\rho_aX).
\end{align}
Since \(X\) is arbitrary, this identifies \(\EE_*(\eta) =\sum_a\Tr(\eta z_a)\rho_a\).
For any \(X\) with \(\|X\|\leq1\), \(\left|\Tr((\Phi_{t*}(\eta)-\EE_*(\eta))X)\right| =\left|\Tr(\eta(\Phi_t-\EE)(X))\right|\leq\|\eta\|_1\|\Phi_t-\EE\|\).
The maximum of the left side is the trace norm, by the polar trace formula. This gives
\begin{equation}
\norm{\Phi_{t*}(\eta)-\sum_a\Tr(\eta z_a)\rho_a}_1 \leq\norm{\Phi_t- \EE}\norm{\eta}_1\longrightarrow 0 .
\label{eq:pointer_state_limit}
\end{equation}
Since \(\EE\Phi_t=\EE\), the range of \(\EE_*\) consists of stationary states. Substituting \(\eta=\rho_a\) gives \(\EE_*(\rho_a)=\rho_a\). The probabilities \(\Tr(\eta z_a)\) are conserved because each \(z_a\) is fixed. For rank-one \(z_a\), \(\rho_a=z_a\), giving the usual mixture of orthogonal pointer states. Evaluation of a pointer observable \(X=\sum_a x_a z_a\) at one label, \(X\mapsto x_a\), is a multiplicative functional, or character, of the commutative algebra \(\mathcal P\); every state  on \(\mathcal P\) is a probability average of these characters, \(\omega(X)=\sum_a\omega(z_a)x_a\). 

\subsection{The common-reservoir model: stationary state, retained algebra and expectation}\label{supp:reservoir_retained_proofs}
Example~\ref{ex:reservoir_stationary} states the stationary density \(\rho=I_3/3\otimes\tau\) of \eqref{eq:selective_stationary} and verifies it. To obtain this density from the dynamics, let \(\eta_t\) be a density evolving under the common-reservoir  model and put \(u_{ia}(t):=\Tr(\eta_t(e_{ii}\otimes q_a))\), the population of level \(a\) in sector \(i\), with \(q_a=\ket a\bra a\) on the internal factor \(\CZ^2\) of Example~\ref{ex:common_reservoir}. 
The two level projections add to \(p_i\), so \(u_{i0}+u_{i1}=w_i\), the weight of sector \(i\), is constant. Multiplying
\eqref{eq:selective_populations} by \(p_i\) gives
\begin{equation}
 u'_{i1}=g_i^2(\gamma_+u_{i0}-\gamma_-u_{i1})
        =g_i^2(\gamma_+w_i-G_\gamma u_{i1}).
\end{equation}
For \(w_i>0\), solve this scalar equation by subtracting its equilibrium value \(\gamma_+w_i/G_\gamma\), at which the right side vanishes,
\begin{equation}
 \frac{u_{i1}(t)}{w_i}=\frac{\gamma_+}{G_\gamma}
  +e^{-g_i^2G_\gamma t}\left(\frac{u_{i1}(0)}{w_i}-\frac{\gamma_+}{G_\gamma}\right).
\end{equation}
Thus the stationary conditional populations are \(\tau_{11}=\gamma_+/G_\gamma\) and \(\tau_{00}=\gamma_-/G_\gamma\). Choose equal sector weights \(w_i=1/3\).

We turn to the family \eqref{eq:pointer_family_generator} of Example~\ref{ex:reservoir_retained} and derive its retained algebra. To identify the algebra \(\RR\) of Section~\ref{supp:pointer_details}, generated by the jump operators, their adjoints and their repeated commutators with \(H\), we use the two products  \(L_-^*L_-\) and \(L_+^*L_+\) in
\begin{equation}
\gamma_-^{-1}L_-^*L_-+\gamma_+^{-1}L_+^*L_+
=F^2\otimes(q_1+q_0)=F^2\otimes I_2.
\end{equation}
For each coupling value \(\lambda\in\spec F\), the polynomial
\begin{equation}
 P_\lambda:=\prod_{\mu\in\spec F,\ \mu\ne\lambda} \frac{F^2-\mu^2 I}{\lambda^2-\mu^2}
\end{equation}
acts as one on labels of coupling \(\lambda\) and zero on all others, because \(F^2\) is diagonal with entry \(g_i^2\) on label \(i\): each factor is one where \(g_i^2=\lambda^2\) and one of them is zero where \(g_i^2=\mu^2\). Thus \(P_\lambda\otimes I_2\in\RR\), and \(\sum_\lambda\lambda^{-1}P_\lambda\otimes I_2=F^{-1}\otimes I_2\in\RR\).
Multiplying  \(F^{-1}\otimes I_2\) by \(L_\pm\) produces \(I_r\otimes\ell_\pm\). Dividing by \(\sqrt{\gamma_\pm}\) supplies the two internal off-diagonal matrix units, and their products give \(I_r\otimes q_0,I_r\otimes q_1\), respectively. Multiplication by each central \(P_\lambda\otimes I_2\) now exhibits every \(P_\lambda\otimes e_{ab}\) in \(\RR\). In the other direction, \(L_\pm=\sum_\lambda\lambda P_\lambda\otimes\ell_\pm\) belongs to their span, and \([H,L_\pm]=\pm\Omega L_\pm\), with the same closure for adjoints. Thus every generator of \(\RR\) belongs to the direct sum \(\bigoplus_\lambda(\CZ P_\lambda\otimes M_2)\). Taking the commutant gives \eqref{eq:pointer_family_algebra}.

Next we compute the asymptotic expectation \(\EE\) on each \(e_{ij}\otimes B\), with \(B\) a matrix on the internal factor; these operators span \(\A\). For \(g_i\ne g_j\), the central projections \(P_\lambda\otimes I_2\) containing its source and its range are different, so  with \(z\), \(z'\) these two projections, bimodularity gives \(\EE(e_{ij}\otimes B)=z\,\EE(e_{ij}\otimes B)\,z'\), which is zero because \(\EE\) maps into \(\N\), where \(z\) is central, and \(zz'=0\). For \(i=j\), the image belongs to \(p_j\N p_j=\CZ p_j\),  so \(\EE(e_{jj}\otimes B)=c\,p_j\) for a number \(c\). Preservation of \(\omega(X)=\Tr((I_r/r\otimes\tau)X)\) gives \(c/r=\omega(cp_j)=\omega(e_{jj}\otimes B)=\Tr(\tau B)/r\), since \(\omega(p_j)=1/r\); hence \(c=\Tr(\tau B)\). For \(g_i=g_j\), the retained \(S_{ij}\) can be pulled outside the expectation by bimodularity,
\begin{equation}
 \EE(e_{ij}\otimes B)=\EE(S_{ij}(e_{jj}\otimes B))  =S_{ij}\EE(e_{jj}\otimes B)=\Tr(\tau B)S_{ij}.
\end{equation}
Writing \(X=\sum_{i,j}e_{ij}\otimes B_{ij}\) and adding the three cases by linearity gives \eqref{eq:pointer_family_expectation}.

\subsection{The normalized trace of a matrix block}\label{supp:block_trace}
Section~\ref{sec:sharp_retained_dimensions} defines the center-valued trace \(\vartheta_\N\) of \(\N=\bigoplus_a(M_{m_a}\otimes I_{n_a})\) through the normalized trace of each block. To derive the normalization in the trace formula, work within one such block. A tracial functional \(t\) has \(t(e_{ii})=t(e_{ij}e_{ji})=t(e_{ji}e_{ij})=t(e_{jj})\), normalizing \(t(I_{m_a})=1\) fixes each value at \(1/m_a\). For \(i\ne j\), \(t(e_{ij})=t(e_{ii}e_{ij})=t(e_{ij}e_{ii})=0\). Thus \(t(X_a)=\Tr_{m_a}(X_a)/m_a\). Multiplying each normalized trace \(t(X_a)\) by its central projection \(z_a\), the unit of the \(a\)th central summand, keeps the dimensions in distinct summands separate: \(\vartheta_\N(p)=\vartheta_\N(q)\) then says that the ranks of \(p\) and \(q\) agree in every summand, which is the condition for an edge in \(\N\) from \(p\) to \(q\) (Section~\ref{sec:sharp_retained_dimensions}).

\subsection{Accumulated retained dissipation}\label{supp:retained_dissipation_proof}
Here we prove Proposition~\ref{prop:retained_accumulation}.
\begin{proof}
For an endpoint \(p\in\N\), \(\Phi_t(p)=\alpha_t(p)\) and \(\EE(\alpha_t(p))=\alpha_t(p)\).  By \eqref{eq:edge_relations}, \(D_t(S,S)=\alpha_t(p)-\Phi_t(S)^*\Phi_t(S)\), so the definition \eqref{eq:retained_defects} gives \(H_t^{\mathrm s}=p-\alpha_{-t}\EE(\Phi_t(S)^*\Phi_t(S))\). Apply \(\alpha_{-t}\EE\) to the integral in \eqref{eq:integrated_kernel}. A bounded linear map  such as \(\alpha_{-t}\EE\) can be applied under the integral sign, because the integrand is norm continuous  and the map is linear and continuous. At a single integration time \(u\), \(\alpha_{-t}\EE\Phi_{t-u}=\alpha_{-t}\alpha_{t-u}\EE=\alpha_{-u}\EE\). The resulting integrand \(\alpha_{-u}\EE(\Gamma(\Phi_uS,\Phi_uS))\) depends on \(u\) alone. For \(0\leq s\leq t\),
\begin{equation}
 H_t^{\mathrm s}-H_s^{\mathrm s}  =\int_s^t\alpha_{-u}\EE\!\left(\Gamma(\Phi_u S,\Phi_u S)\right)\,du\geq0 .
\end{equation}
Positivity follows because the quadratic \(\Gamma\) is positive and the expectation and the automorphism preserve positivity. The zero-length integral gives \(H_0^{\mathrm s}=0\).

Write \(S(t)=\Phi_t(S)\), for the endpoint put \( \widehat S(t)=\alpha_t\EE(S)\). Since \(\widehat S(t)\in\N\), \(\alpha_{-t}\EE(\widehat S(t)^* \widehat S(t))=\EE(S)^*\EE(S)\). Both amplitudes are contractions, adding and subtracting \(S(t)^*\widehat S(t)\) in the difference of the two products gives
\begin{align}
\norm{H_t^{\mathrm s}-\Ds_{\EE}(S)}
&\leq\norm{S(t)^*S(t)-\widehat S(t)^*\widehat S(t)}\notag\\
&\leq(\norm{S(t)}+\norm{\widehat S(t)})\norm{S(t)-\widehat S(t)}
\leq2\norm{S(t)-\widehat S(t)}\longrightarrow 0 ,
\label{eq:retained_defect_convergence}
\end{align}
here the norm limit uses \eqref{eq:retained_dynamical_assumptions}. The same proof with \(S^*\) gives the range statement.
\end{proof}

\subsection{ The defects at all times as the defect of one map }\label{supp:initial_nonsplitting}
Here we construct the trajectory map of Section~\ref{sec:sharp_accumulated_dissipation}, read the defects and the asymptotic expectation from it, and show that, when \(\N\ne\A\),  no multiplicative map can take its place, i.e., no \(*\)-homomorphism sends each observable \(X\) to a function of time that starts at \(X\) and converges, as \(t\to\infty\), to an element of the retained algebra; the common-reservoir model shows this by a count of ranks, and a comparison of dimensions gives it in general.

Under the hypotheses of Proposition~\ref{prop:asymptotic_expectation}, a QMS on \(M_d\) with a faithful stationary state \eqref{eq:faithful_stationary}, send an observable \(X\) to its trajectory via the functional map \(t\mapsto\Phi_t(X)\),
\begin{equation}
j:\A\to C_b([0,\infty),\A),\qquad j(X)(t)=\Phi_t(X).
\label{eq:trajectory_map}
\end{equation} 
Here \(C_b([0,\infty),\A)\) is the algebra of bounded norm-continuous functions from \([0,\infty)\) to \(\A\), with products and adjoints taken at each time and the norm \(\sup_t\norm{F(t)}\), and \(C_0([0,\infty),\A)\) is the ideal of functions that vanish in norm at infinity. 
By \eqref{eq:asymptotic_decomposition}, \(j(X)\) is the sum of the rotating retained part \(t\mapsto\alpha_t(\EE(X))\) and a function vanishing at infinity.  Such functions form a unital \(C^*\)-algebra with a \(*\)-homomorphism onto \(\N\),
\begin{equation}
\Talg=\{\alpha_\cdot(n)+r:\ n\in\N,\ r\in C_0([0,\infty),\A)\},\qquad
q_\infty(\alpha_\cdot(n)+r)=n .
\label{eq:trajectory_algebra}
\end{equation}
Here \(\alpha_\cdot(n)\) is the function \(t\mapsto\alpha_t(n)\), and \(q_\infty\) returns the retained part \(n\). Indeed sums, adjoints and pointwise products are again of this form: the product of \(\alpha_\cdot(n)+r\) and \(\alpha_\cdot(m)+s\) is \(\alpha_\cdot(nm)\) plus three terms that vanish at infinity, and \(\norm n=\lim_t\norm{\alpha_t(n)+r(t)}\) gives closure and makes \(q_\infty\) well defined: applied to the difference of two such functions, it shows that \(n\) is determined by the function.

The map \(j\) is unital and completely positive, and
\begin{align}
\bigl(j(X^*Y)-j(X)^*j(Y)\bigr)(t)&{}=D_t(X,Y),&  q_\infty(j(X)) & =\EE(X) ,\notag\\
q_\infty\bigl(j(p)-j(S)^*j(S)\bigr)&{}=\Ds_{\EE}(S),&q_\infty\bigl(j(q)-j(S)j(S)^*\bigr)&{}=\Dr_{\EE}(S).
\label{eq:relative_defect_boundary}
\end{align}
In the second line \(S\) is an edge with \(S^*S=p\), \(SS^*=q\) and \(p,q\in\N\), and \(\Ds_{\EE}(S)\), \(\Dr_{\EE}(S)\) are its endpoint defects \eqref{eq:endpoint_defects}. The four identities read the defect at each time, the asymptotic expectation and the two endpoint defects from \(j\); the multiplicative domain of \(j\) is \(\N\) by Proposition~\ref{prop:exact-survival}.

For a fixed inclusion \(\N\subseteq\A\), the isomorphism class of \(\Talg\) does not depend on the rotation \(\alpha_t\) or on the decay rates: \(\alpha_t\) is implemented by a unitary group in \(\N\) (Section~\ref{supp:asymptotic_proof}), and removing this rotation and rescaling time to \([0,1]\) identifies \(\Talg\) with \(\{f\in C([0,1],\A):f(1)\in\N\}\), the mapping cylinder of the inclusion \(\N\subseteq\A\) \cite{AntoniniAzzaliSkandalis}. In detail, let \(F=\alpha_\cdot(n)+r\) be an element of \(\Talg\). Take \(H_\N\in\N\) from the proof of Proposition~\ref{prop:asymptotic_expectation}, so that \(\alpha_t(n)=e^{itH_\N}ne^{-itH_\N}\). For \(t_0>0\), set \(s=t/(t_0+t)\), so \(t=t_0s/(1-s)\) when
\(0\leq s<1\). Define
\begin{equation}
f(s)=e^{-itH_\N}F(t)e^{itH_\N}\quad(s<1),\qquad f(1)=n .
\end{equation}
Substitute \(F(t)=\alpha_t(n)+r(t)\). The first term becomes \(n\), and the norm of the second equals
\(\norm{r(t)}\to0\), so \(f\) is continuous at \(1\). Pointwise conjugation and change of variable preserve both products and the supremum norm. Conversely, for a continuous \(f\) with \(f(1)\in\N\) set \(F(t)=e^{itH_\N}f(s(t))e^{-itH_\N}\). Subtracting \(\alpha_t(f(1))\) leaves a function of norm \(\norm{f(s(t))-f(1)}\to0\). These operations are inverse to one another and give
\begin{equation}
\Talg\cong\mathcal C_i :=\{f\in C([0,1],\A):f(1)\in\N\},\qquad i:\N\hookrightarrow\A .
\label{eq:trajectory_mapping_cylinder}
\end{equation}

The functions  in \(\Talg\) vanishing at \(t=0\) form an ideal  \(J\), which contains the defect \(j(p)-j(S)^*j(S)\) of every edge.  It is the kernel of initial evaluation \(\ev_0(F)=F(0)\), a \(*\)-homomorphism of \(\Talg\) onto \(\A\) with \(\ev_0(j(X))=X\). The rank differences \(d_a=k_a(p)-k_a(q)\) of Theorem~\ref{thm:finite_record_obstruction} are a class in the group \(K_0(J)\), which we compute in three steps. 

First, the \(K\)-groups of matrix algebras count ranks \cite{doi:10.1017/CBO9780511623806}: \(K_0(\A)=\ZZ\) by the rank of a projection, and \(K_0(\N)=\bigoplus_a\ZZ\) by its retained ranks \((k_a)_a\). A projection of \(\N\) with retained ranks \((k_a)_a\) is \(\bigoplus_a(p_a\otimes I_{n_a})\), of rank \(\sum_an_ak_a\) in \(\A\), so the inclusion \(i:\N\hookrightarrow\A\) induces \(i_*((k_a)_a)=\sum_an_ak_a\), the ambient rank. Every unitary matrix is \(e^{iD}\) with \(D\) self-adjoint in the same matrix algebra, and \(s\mapsto e^{isD}\) joins it to the identity through unitaries. The same holds in every matrix algebra over \(\A\) or \(\N\), and \(K_1\) is the group of these unitaries modulo such paths, so \(K_1(\A)=K_1(\N)=0\). 

Second, \(\Talg\) has the \(K\)-groups of \(\N\). The algebra \eqref{eq:trajectory_mapping_cylinder} is the mapping cylinder of the inclusion \(i:\N\hookrightarrow\A\): the continuous paths in \(\A\), parametrized by \([0,1]\), that end in the image of \(i\).  In  \eqref{eq:trajectory_mapping_cylinder} the maps \((h_rf)(s)=f(r+(1-r)s)\), \(0\leq r\leq1\), are \(*\)-homomorphisms; \(h_0\) is the identity and \(h_1\) sends \(f\) to the constant function with value \(f(1)\). 
Write \(\ev_1(f)=f(1)\) and \(c(n)\) for the constant function with value \(n\in\N\). Then \(\ev_1\circ c=\id_\N\), and \(c\circ\ev_1=h_1\) is homotopic to \(h_0=\id\) through the family \(h_r\), \(0\leq r\leq1\); homotopic \(*\)-homomorphisms induce the same maps on \(K\)-groups. Evaluation at \(s=1\) therefore gives \(K_j(\Talg)\cong K_j(\N)\), with inverse induced by \(c\). Under this identification \(\ev_0\) induces \(i_*\), because \(\ev_0\circ c=i\): \(\ev_0\) of the constant function \(c(n)\) with value \(n\) is \(i(n)\).

Third, \(J\) is the kernel of \(\ev_0\) and \(\ev_0\) maps \(\Talg\) onto \(\A\), so the extension \(0\to J\to\Talg\to\A\to0\) has a six-term exact sequence in \(K\)-theory~\cite{doi:10.1017/CBO9780511623806}, which relates the \(K\)-groups of \(J\), \(\Talg\) and \(\A\).  
Two of its six groups, \(K_1(\A)\) and \(K_1(\Talg)\), are zero, which leaves
\begin{equation}
0\longrightarrow K_0(J)\xrightarrow{\ \eta\ }K_0(\N)
 \xrightarrow{i_*}K_0(\A)\xrightarrow{\ \partial_0\ }K_1(J)\longrightarrow0 .
\label{eq:relative_k_sequence}
\end{equation}
Here \(\eta\) is induced by the inclusion \(J\subset\Talg\cong\mathcal C_i\) followed by evaluation at \(s=1\). Exactness  at \(K_0(J)\) says that \(\eta\) is injective, and exactness at \(K_0(\N)\) says \(\ran\eta=\ker i_*\), so every retained difference of ambient rank zero has exactly one preimage in \(K_0(J)\).
Thus the \(K_0\) group  of \(J\) is the kernel of \(K_0(\N)\to K_0(\A)\), the integer vectors \((d_a)\) with \(\sum_an_ad_a=0\), so the rank differences of Theorem~\ref{thm:finite_record_obstruction} are a class in it: the two endpoints of an edge have the same ambient rank, so \(\sum_an_ad_a=0\). In the matrix setting this class contains no more than those integers. At the other end of the sequence, exactness gives \(K_1(J)\cong\ZZ/g\ZZ\) with \(g=\gcd_an_a\), the quotient of \(K_0(\A)=\ZZ\) by the image \(g\ZZ\) of \(i_*\).

The trajectory map \(j\) of \eqref{eq:trajectory_map} sends each observable to a trajectory that starts at the observable, \(j(X)(0)=X\), and \(j\) is in general not multiplicative. For the common-reservoir model with \(\kappa\ne1\) we show, by counting ranks, that no \(*\)-homomorphism \(\sigma:\A\to\Talg\) has this property. The count is the one made by \(K_0\).
\begin{example}[Common-reservoir model: no multiplicative trajectory map]
By \eqref{eq:selective_surviving_algebra} the retained algebra is \(\N=(M_2\otimes I_2)\oplus\CZ I_2\subset\A=M_6\), and \(\alpha_t=\id\). A projection \(p\in\N\) has retained ranks \(k_1(p)\in\{0,1,2\}\) and \(k_2(p)\in\{0,1\}\) in the two block central summands, and both  summands have multiplicity \(n_1=n_2=2\). Its rank in \(M_6\) is the ambient rank \(\sum_an_ak_a(p)\) of Section~\ref{sec:sharp_retained_dimensions},
\begin{equation}
\rank p=2k_1(p)+2k_2(p),
\label{eq:retained_even_rank}
\end{equation}
which is even. In the coordinates of \eqref{eq:trajectory_mapping_cylinder}, with time rescaled to \([0,1]\),
\begin{equation}
\Talg\cong\{f\in C([0,1],M_6):f(1)\in\N\},\qquad \ev_0(f)=f(0),
\label{eq:initial_evaluation}
\end{equation}
where \(\ev_0\) is evaluation at the initial time, and \(\ev_0\circ j=\id\). Suppose a \(*\)-homomorphism \(\sigma:M_6\to\Talg\) satisfied \(\ev_0\circ\sigma=\id\). Let \(e\in M_6\) be a projection of rank one and put \(f=\sigma(e)\). Since \(\sigma\) preserves products and adjoints, \(f^2=\sigma(e^2)=f\) and \(f^*=\sigma(e^*)=f\), that is, \(f\) is a projection of \(\Talg\), so every   \(f(t)\) is a projection in \(M_6\), \(f(0)=e\), and \(t\mapsto\Tr f(t)\) is continuous with integer values, because the trace of a projection is its rank, and hence constant. So \(f(1)\) would be a projection of rank one in \(\N\), which \eqref{eq:retained_even_rank} excludes. Initial evaluation therefore has no \(*\)-homomorphic right inverse, while \(j\) is a unital completely positive one.

In \(K\)-theory the same count is the map induced by the inclusion \(i:\N\hookrightarrow M_6\),
\begin{equation}
i_*:K_0(\N)=\ZZ^2\to K_0(M_6)=\ZZ,\qquad i_*(k_1,k_2)=2k_1+2k_2 .
\label{eq:inclusion_k0}
\end{equation}
Evaluation along \([0,1]\) is a homotopy from \(\ev_0\) to \(i\circ\ev_1\),  and homotopic maps induce the same map on \(K_0\), so a right inverse \(\sigma\) would give \(\id_{\ZZ}=i_*\circ(\ev_1\circ\sigma)_*\), whereas the image of \(i_*\) is \(2\ZZ\). The \(K_0\) obstruction and the rank count \eqref{eq:retained_even_rank} are the same statement.
\end{example}
For any proper inclusion \(\N\subsetneq M_d\) a comparison of dimensions gives the same conclusion.  Suppose \(\sigma:M_d\to\Talg\) is a \(*\)-homomorphism with \(\ev_0\circ\sigma=\id\). The projections \(\sigma(I)(t)\) form a continuous path that starts at \(I\), so their rank is always \(d\) and \(\sigma(I)(t)=I\). Thus \(\ev_1\circ\sigma\) would be a unital, hence injective, \(*\)-homomorphism of \(M_d\) into the smaller algebra \(\N\).

\section{Preservation of  required operations}\label{supp:do-proof}
\subsection{Preserving a given family without dimension counting}\label{sec:do-family}
The endpoint bounds of Section~\ref{sec:sharp_loss} (Theorems~\ref{thm:sharp_endpoint_defects} and~\ref{thm:finite_record_obstruction}) compare dimensions in the retained algebra. There is also a way to test preservation of a  given family of operations that does not count dimensions. It keeps track of the same positive multiplication defects, but asks whether they vanish on a common reducing part of the family, a subspace that every operation of the family leaves invariant together with its orthogonal complement. Related representation-theoretic correction and complementarity results also hold for von Neumann algebras~\cite{doCKLT}; here we examine the preservation of the required operations under the  given partial access.

Let \(\Phi:\A\to\mathcal C\subseteq B(\HH)\) be a faithful normal unital completely
positive map between von Neumann algebras; in the application it is \(\Phi_t\) at one fixed time. Fix the required operations \(S_j\in\A\), \(j\) in an index set \(J\); together they are the family. The family is given once as part of the data; the operations  need not be invariant under \(\Phi\), that is, \(\Phi(S_j)=S_j\) is not assumed. Write \(\mathcal G=W^*(S_j:j\in J)\) for the unital von Neumann algebra they generate, the smallest von Neumann algebra containing them. We denote  the commutant algebra \(\mathcal G'=\{X:XY=YX\text{ for all }Y\in\mathcal G\}\). An accessible reducing restriction has the form \(\mathcal Gf\), where \(f\in\A\cap\mathcal G'\) is a projection: since \(f\) commutes with every \(S_j\) and \(S_j^*\), every required operation leaves the subspace selected by \(f\) and its orthogonal complement invariant. Preserving this restriction means that \(\Phi|_{\mathcal Gf}\) is a homomorphism,
with unit \(\Phi(f)\); such a homomorphism is injective because \(\Phi\) is faithful.

Write \(D_\Phi(X,Y)=\Phi(X^*Y)-\Phi(X)^*\Phi(Y)\) for the multiplication defect of \(\Phi\) and \(\MD(\Phi)\) for its multiplicative domain \eqref{eq:md}. Normality of \(\Phi\), its continuity in the ultraweak topology, makes \(\MD(\Phi)\) ultraweakly closed, hence a von Neumann algebra: for each fixed \(Y\in\A\), the identities \(\Phi(XY)=\Phi(X)\Phi(Y)\) and \(\Phi(YX)=\Phi(Y)\Phi(X)\) are ultraweakly closed conditions on \(X\), because both sides are ultraweakly continuous in \(X\).

\begin{proposition}[Preservation of a  given operation family]\label{prop:do-family}
Put
\begin{equation}
\mathcal B:=\mathcal G'\cap\MD(\Phi),
\label{eq:do-direct-B}
\end{equation}
the operators that commute with every required operation and are themselves preserved by \(\Phi\). Among the projections \(f\in\A\cap\mathcal G'\) for which \(\Phi|_{\mathcal Gf}\) is a homomorphism there is a largest one, \(f_*\), and it lies in the center of \(\mathcal B\), the elements of \(\mathcal B\) that commute with all of \(\mathcal B\). For every projection \(f\in\A\cap\mathcal G'\),
\begin{equation}
\Phi|_{\mathcal Gf}\text{ is a homomorphism}
\quad\Longleftrightarrow\quad
f\in\mathcal B\ \text{ and }\ f\leq f_* .
\label{eq:do-family-criterion}
\end{equation}
If \(f\notin\MD(\Phi)\), the relation \(f^2=f\) of the unit of \(\mathcal Gf\) has the nonzero positive defect \(\Phi(f)-\Phi(f)^2\). If \(f\in\mathcal B\) but \(f\not\leq f_*\), some \(X\in\{S_j,S_j^*:j\in J\}\) satisfies
\begin{equation}
D_\Phi(Xf,Xf)=\Phi(f)D_\Phi(X,X)\Phi(f)\ne0.
\label{eq:do-restricted-witness}
\end{equation}
This positive defect is detected by a normal state supported by \(\Phi(f)\).
In particular, if \(\mathcal B\) is a factor, i.e., an algebra whose center is \(\CZ I\),  and \(\mathcal G\not\subseteq\MD(\Phi)\), then \(f_*=0\): no nonzero accessible reducing restriction of the whole family is preserved.
\end{proposition}

\begin{proof}
The proof works inside \(\A\), the algebra on which \(\Phi\) is defined, and has three steps.

\emph{Step 1: what preservation means.} The set \(\mathcal Gf\) is closed under products and adjoints and has unit \(f\). A unital completely positive map is a homomorphism on such an algebra exactly when the algebra lies in its multiplicative domain \eqref{eq:md}. Hence
\begin{equation}
\Phi|_{\mathcal Gf}\text{ is a homomorphism}\quad\Longleftrightarrow\quad\mathcal Gf\subseteq\MD(\Phi).
\label{eq:do-direct-criterion}
\end{equation}
In particular the unit \(f\) must itself lie in \(\MD(\Phi)\). Since \(f\) also commutes with the family  (\(f\in\mathcal G'\)), it then belongs to the algebra \(\mathcal B\) of \eqref{eq:do-direct-B}. For a projection, \(f\in\MD(\Phi)\) means \(D_\Phi(f,f)=\Phi(f)-\Phi(f)^2=0\). So if \(f\notin\MD(\Phi)\), the unit relation \(f^2=f\) has the defect \(\Phi(f)-\Phi(f)^2\), which is nonzero, and positive by the Schwarz inequality.

\emph{Step 2: the preserved projections lie under one central projection.} Collect the elements of \(\mathcal B\) that keep the whole family inside the multiplicative domain,
\begin{equation}
\mathcal J:=\{b\in\mathcal B:\ Xb\in\MD(\Phi)\text{ for every }X\in\mathcal G\}.
\label{eq:do-direct-ideal}
\end{equation}
By \eqref{eq:do-direct-criterion}, a projection \(f\in\mathcal B\) gives a preserved restriction exactly when \(f\in\mathcal J\). The set \(\mathcal J\) is a two-sided ideal of \(\mathcal B\). To check this, let \(b\in\mathcal J\), \(c\in\mathcal B\) and \(X\in\mathcal G\). Then \(X(bc)=(Xb)c\) is a product of two elements of \(\MD(\Phi)\), and \(X(cb)=c(Xb)\) because \(c\) commutes with \(X\); both lie in \(\MD(\Phi)\). Also \(Xb^*=(X^*b)^*\) lies in \(\MD(\Phi)\), so \(\mathcal J\) is closed under adjoints. The ideal \(\mathcal J\) is ultraweakly closed, because \(\MD(\Phi)\) and \(\mathcal G'\) are and \(b\mapsto Xb\) is ultraweakly continuous. An ultraweakly closed two-sided ideal of a von Neumann algebra consists of all elements of the algebra multiplied by one central projection~\cite{Takesaki2003}: there is a projection \(f_*\) in the center of \(\mathcal B\) with \(\mathcal J=\mathcal Bf_*\). A projection \(f\in\mathcal B\) lies in \(\mathcal Bf_*\) exactly when \(f\leq f_*\). Hence \eqref{eq:do-family-criterion} holds, and \(f_*\), which itself lies in \(\mathcal J\), is the largest preserved reducing projection. If \(\mathcal B\) is a factor, its center is \(\CZ I\), so the only projections in its center are \(0\) and \(I\), and \(f_*\) is one of the two. The case \(f_*=I\) says \(I\in\mathcal J\), that is, \(\mathcal G\subseteq\MD(\Phi)\). So if \(\mathcal G\not\subseteq\MD(\Phi)\), the case \(f_*=I\) is excluded; then \(f_*=0\) and by \eqref{eq:do-family-criterion} no nonzero restriction is preserved. 

\emph{Step 3: a failure shows on a generator.} Let \(f\) be a projection in \(\mathcal B\) with \(\mathcal Gf\not\subseteq\MD(\Phi)\). Consider the operators \(X\in\mathcal G\) with \(Xf\in\MD(\Phi)\), they include \(I\). These operators are closed under products because \((XY)f=(Xf)(Yf)\), under adjoints because \(X^*f=(Xf)^*\), and under ultraweak limits, because \(X\mapsto Xf\) is ultraweakly continuous and \(\MD(\Phi)\) is ultraweakly closed. If they included every \(S_j\), they would be all of \(\mathcal G\). Hence, since \(\mathcal Gf\not\subseteq\MD(\Phi)\), \(S_jf\notin\MD(\Phi)\) for some \(j\), which means \(D_\Phi(Xf,Xf)\ne0\) for \(X=S_j\) or \(X=S_j^*\). Since \(f\) lies in \(\MD(\Phi)\) and commutes with \(X\), it can be taken out of \(\Phi\): \(\Phi(fX^*Xf)=\Phi(f)\Phi(X^*X)\Phi(f)\) and \(\Phi(Xf)=\Phi(X)\Phi(f)\). This gives \eqref{eq:do-restricted-witness}. Because \(f\) is a projection in \(\MD(\Phi)\), \(\Phi(f)\) is a projection, and the defect \eqref{eq:do-restricted-witness} is a nonzero positive element of the corner \(\Phi(f)\mathcal C\Phi(f)\). Some unit vector \(\xi=\Phi(f)\xi\) gives the defect a positive expectation, as for every nonzero positive operator in the corner \(\Phi(f)\mathcal C\Phi(f)\). The vector state \(c\mapsto\langle\xi,c\xi\rangle\) of \(\xi\) on \(\mathcal C\) is normal and supported by \(\Phi(f)\), it has a positive value on the defect \eqref{eq:do-restricted-witness}, which is the sense in which the proposition says that the defect is detected.

Step 1 gives the defect of the unit when \(f\notin\MD(\Phi)\), Step 2 the criterion \eqref{eq:do-family-criterion} and the statement for a factor, and Step 3 the witness \eqref{eq:do-restricted-witness}; this proves the proposition.
\end{proof}

The common-reservoir model shows why requiring a family matters. Take \(g_0=g_1=1\), \(g_2=\kappa\ne1\), and \(t>0\). The map \(\Phi_t=e^{t\LL}\) is injective as a linear map, hence faithful, and \eqref{eq:constant_domain} gives \(\MD(\Phi_t)=\N\). For \(\mathcal G_0=W^*(S_{21})\), the projection \(p_0\) is reducing and \(\mathcal G_0p_0=\CZ p_0\) is fixed. Now enlarge the family by \(S_{10}\),  so that the required operations are \(S_{21}\) and \(S_{10}\);  the added edge \(S_{10}\) is itself fixed, \(\Phi_t(S_{10})=S_{10}\) by \eqref{eq:selective_surviving_edge}. Their products and adjoints generate all label matrix units, so
\begin{align}
\mathcal G&=W^*(S_{21},S_{10})=M_3(\CZ)\otimes I_2,\notag\\
\mathcal G'\cap\MD(\Phi_t)
&=(I_3\otimes M_2(\CZ))\cap
  \bigl((M_2(\CZ)\otimes I_2)\oplus\CZ I_2\bigr)
 =\CZ I_6.
\label{eq:do-reservoir-family}
\end{align}
The retained algebra is the one computed in Section~\ref{sec:distinct_pointer_couplings}. Since  \(\mathcal B=\CZ I_6\) is a factor and \(S_{21}\notin\N\)\({}=\MD(\Phi_t)\),  Proposition~\ref{prop:do-family} excludes every nonzero preserved reducing restriction of \(\mathcal G\). The edge \(S_{10}\) remains fixed; adding it to the family removes \(p_0\) as a reducing projection because it connects sectors \(0\) and \(1\). Thus an individually preserved operation can make preservation of a larger required family impossible.

\subsection{Faithful partial access and type III algebras}\label{supp:faithful_partial_access}
Return to \(\Phi=P\beta|_\A\), where \(\A\subseteq\M\), \(\beta\) is a normal automorphism, and \(P:\M\to\A\) is a faithful normal conditional expectation. The remainder calculation in Theorem~\ref{thm:partial_access} and Section~\ref{supp:von_neumann_extensions} gives
\begin{equation}
D_\Phi(X,X)=P\bigl((\beta(X)-\Phi(X))^*(\beta(X)-\Phi(X))\bigr)
\label{eq:do-partial-access-defect}
\end{equation}
and 
\begin{equation}
\MD(\Phi)=\A\cap\beta^{-1}(\A).
\label{eq:do-partial-access}
\end{equation}
To obtain \eqref{eq:do-partial-access}, faithfulness of \(P\) makes \eqref{eq:do-partial-access-defect} zero exactly when \(\beta(X)=\Phi(X)\in\A\). Applying the same argument to \(X^*\) gives the second of the two multiplicative-domain conditions  in \eqref{eq:md}, \(D_\Phi(X,X)=0\) and \(D_\Phi(X^*,X^*)=0\). Thus a restriction that is not preserved has \(\mathcal Gf\not\subseteq\MD(\Phi)=\A\cap\beta^{-1}(\A)\), and since \(\mathcal Gf\subseteq\A\) this means \(\beta(\mathcal Gf)\not\subseteq\A\): the  required restricted operations cannot all remain accessible. Equation~\eqref{eq:do-family-criterion} shows that  the largest preserved reducing projection \(f_*\)  is determined by \(\mathcal G\) and \(\MD(\Phi)\), and by \eqref{eq:do-partial-access} \(\MD(\Phi)=\A\cap\beta^{-1}(\A)\), in which \(P\) does not appear. Hence \(f_*\) is the same for every faithful normal expectation onto \(\A\).

Proposition~\ref{prop:do-family} and its proof use no trace or projection-dimension count, so they apply also to type III algebras whenever the stated faithful normal expectation \(P\) exists; no resolution of the identity into sectors is needed.  For example, if \(\MD(\Phi)\) is a type III factor and the required family contains \(\MD(\Phi)\) and one operation that is not in \(\MD(\Phi)\), then, with \(Z(\cdot)\) the center, an element of \(\mathcal G'\cap\MD(\Phi)\) lies in \(\MD(\Phi)\) and commutes with it, because \(\MD(\Phi)\subseteq\mathcal G\); so
\begin{equation}
\mathcal G'\cap\MD(\Phi)\subseteq Z(\MD(\Phi))=\CZ I.
\end{equation} 
Thus \(\mathcal B\) is a factor and \(\mathcal G\not\subseteq\MD(\Phi)\), so Proposition~\ref{prop:do-family} excludes every nonzero preserved accessible reducing restriction of the  family. The products inside \(\MD(\Phi)\) are still preserved, the obstruction involves no dimension, and the state of \eqref{eq:do-restricted-witness}, the normal state of Proposition~\ref{prop:do-family} supported by \(\Phi(f)\), has a positive value on that defect: the obstruction shows in an expectation value.

\section{Extensions to von Neumann algebras}\label{supp:von_neumann_extensions}
The main text works in a matrix algebra. This section demonstrates which of its statements hold for an inclusion of von Neumann algebras, where there may be no trace and no density matrix, including the partial-access identities and the weights (Section~\ref{supp:vn_access}) and the record fidelity (Section~\ref{supp:vn_record}). The dimension comparison of Theorem~\ref{thm:sharp_endpoint_defects} is already stated for finite von Neumann algebras, in \eqref{eq:sharp_central_defects} with the center-valued trace \(\vartheta_\N\).

\subsection{Partial access and weights}\label{supp:vn_access}
A von Neumann algebra is a unital \(*\)-algebra of bounded operators closed in the weak operator topology, in which \(X_i\to X\) means \(\langle u,X_i v\rangle\to\langle u,Xv\rangle\) for every \(u,v\). Inclusions and multiplication still make sense without choosing a matrix size or a density matrix. Let \(\A\subseteq\M\) be such a unital inclusion and \(P:\M\to\A\) a normal conditional expectation.
Normality means that for a bounded increasing family of positive operators directed by an index,
\(P(\sup_i X_i)=\sup_i P(X_i)\).  The directed index set, in which any two indices have a common later one, permits two approximants to have a common later approximant, even when there is no countable enumeration.

For a group of automorphisms \(\beta_t\) on \(\M\), put \(A_X=P\beta_t(X)\) and 
\(R_X=\beta_t(X)-A_X\). Idempotence gives \(P(R_X)=0\). To see which steps of the proof of Theorem~\ref{thm:partial_access} carry over, apply \(P\) to a single product in \(\M\),
\begin{equation}
 P\bigl((A_X+R_X)^*(A_Y+R_Y)\bigr)
  =A_X^*A_Y+A_X^*P(R_Y)+P(R_X)^*A_Y+P(R_X^*R_Y).
\end{equation}
The two middle terms on the right vanish by bimodularity and \(P(R_X)=P(R_Y)=0\), leaving \(P(R_X^*R_Y)\) as the multiplication defect.  Replacing \(A_X\) by \(A_X-B\) in the corresponding square gives \eqref{eq:access_prediction}. Inserting \(\id=P+(\id-P)\) between two consecutive automorphisms gives \eqref{eq:access_memory}. If \(P\) is faithful, \(P(R_X^*R_X)=0\) forces \(R_X=0\). These identities apply to type II and type III algebras. When the expectation must preserve a given faithful normal state \(\Omega\), Takesaki's theorem gives existence exactly when
\(\sigma_s^\Omega(\A)=\A\) for all real \(s\) \cite{Takesaki1972}.
The computation in \eqref{eq:modular_group_intro} was the matrix realization of the modular group \(\sigma_s^\Omega\)  that appears in Takesaki's theorem.

For a normal UCP map the source and range weights are still \(W^{\mathrm s}_S=\Phi(S)^*\Phi(S)\) and 
\(W^{\mathrm r}_S=\Phi(S)\Phi(S)^*\), respectively. In a Stinespring representation the insertion of
\(I=VV^*+(I-VV^*)\) between \(\pi(X)^*\) and \(\pi(Y)\) gives \eqref{eq:defect_factorization}. For a uniformly continuous semigroup, \(e^{t\LL}\) has a bounded generator on the whole algebra. In norm the first-order expansion gives
\begin{equation}
 \Gamma(X,X)=\lim_{h\downarrow0}\frac{\Phi_h(X^*X)-\Phi_h(X)^*\Phi_h(X)}{h}\geq 0 ,
\end{equation}
as in the matrix case of Section~\ref{sec:dissipation}. The quotient on the right is positive by the Schwarz inequality, and  the positive operators form a norm-closed cone, so a norm limit of positive operators is positive. The product rule therefore gives \eqref{eq:weight_evolution}, with \(\Gamma\) defined by the difference in \eqref{eq:gamma}. For fixed \(t\), the map \(u\mapsto\Phi_{t-u}((\Phi_uX)^*\Phi_uY)\) is differentiable  for the operator norm, because \(\Phi_u=e^{u\LL}\) is differentiable. Its derivative is \(-\Phi_{t-u}(\Gamma(\Phi_uX,\Phi_uY))\), as calculated in Section~\ref{sec:accumulation}.  The derivative is norm continuous, hence integrable as a continuous function with values in a Banach space. Integrating  it from \(0\) to \(t\) gives \(F(t)-F(0)\) for \(F(u)=\Phi_{t-u}((\Phi_uX)^*\Phi_uY)\), which is \(\Phi_t(X)^*\Phi_t(Y)-\Phi_t(X^*Y)=-D_t(X,Y)\), and proves \eqref{eq:integrated_kernel} for a uniformly continuous semigroup of normal UCP maps on a von Neumann algebra.

\subsection{How to define the complementary record without a partial trace}\label{supp:vn_record}
Represent \(\A\) faithfully and normally, by an injective normal \(*\)-homomorphism into the bounded operators, on a Hilbert space \(\HH\), here \(\A\) is now any von Neumann algebra,   its image need not be all of \(B(\HH)\). Let \(\Phi:\A\to\A\) be normal UCP, represented as \(\Phi(X)=V^*\pi(X)V\) with \(\pi\) normal on a dilation space \(\mathcal K\). The complementary algebra is defined as the commutant \(\mathcal B_E=\pi(\A)'\), so that every \(B\in\mathcal B_E\) commutes with every operator \(\pi(X)\), i.e., with every represented output observable, as \(I\otimes B\) commutes with \(X\otimes I\) in the matrix case. For fixed sectors \(p_v,p_w\), which means \(\Phi(p_v)=p_v\), expanding the product gives \((\pi(p_v)V-Vp_v)^*(\pi(p_v)V-Vp_v)=\Phi(p_v)-\Phi(p_v)p_v-p_v\Phi(p_v)+p_v=0\), hence \(\pi(p_v)V=Vp_v\). For a unit input \(x\in p_v\HH\), define its environment record as a functional on \(\mathcal B_E\),
\begin{equation}
\sigma_v^E(x)(B)=\langle Vx,BVx\rangle,\qquad B\in\mathcal B_E .
\end{equation}
At \(B=I\) this functional has value \(\norm{Vx}^2=1\), and on \(B^*B\) its value is \(\norm{BVx}^2\geq0\).  Vector functionals \(B\mapsto\langle\zeta,B\zeta\rangle\)  are normal, being continuous in the weak operator topology, so it defines a normal state.

Now we give the commutant argument that replaces the finite-dimensional matrix trace calculation, in the proof of Theorem~\ref{thm:information_weights}, for the case without a trace. For vector states  \(B\mapsto\langle\zeta,B\zeta\rangle\), \(\zeta\in\mathcal K\), of an algebra \(\mathcal B_E\) represented on \(\mathcal K\), the root fidelity is the supremum of absolute overlaps in common representations implementing the two states \cite{doi:10.1016/0034-4877(76)90060-4,Alberti1983}. Alberti~\cite{Alberti1983} showed that, when  two states of an algebra are given by vectors \(\xi,\eta\) in one representation, this supremum equals the supremum of \(|\langle\xi,U\eta\rangle|\) over the unitaries \(U\) in the commutant of the represented algebra. We use the same characterization with contractions in place of unitaries: for vectors \(\xi,\eta\in\mathcal K\) the root fidelity equals
\begin{equation}
\sup_{\substack{T\in\mathcal B_E'\\\norm T\leq1}}|\langle \xi,T\eta\rangle| ,
\label{eq:commutant_fidelity}
\end{equation}
where \(T\) runs over the contractions that commute with \(\mathcal B_E\), i.e.,  operators \(T\) on \(\mathcal K\) with \(\norm T\leq1\) and \(TB=BT\) for every \(B\in\mathcal B_E\). We prove \eqref{eq:commutant_fidelity} directly, in the form needed here, by two inequalities. Two features of \eqref{eq:commutant_fidelity} differ from Alberti's statement. First, the algebra is the commutant \(\mathcal B_E=\pi(\A)'\) of a Stinespring representation, so the contractions \(T\) in \eqref{eq:commutant_fidelity}, which lie in \(\mathcal B_E'=\pi(\A)''=\pi(\A)\), come from observables of \(\A\); this is what turns a fidelity of environment records into a supremum over system operators. Second, the supremum is over contractions, because the compression of an observable to a corner \(p_v\A p_w\) is a contraction and need not be unitary. To verify one inequality, take any contraction \(T\in\mathcal B_E'\). In the doubled representation 
\(B\mapsto B\oplus B\), the vectors \(\xi\oplus0\) and \(T\eta\oplus(I-T^*T)^{1/2}\eta\) implement the same two states as
\(\xi,\eta\). Indeed the expectation on the second vector is
\begin{equation}
\langle\eta,(T^*BT+(I-T^*T)^{1/2}B(I-T^*T)^{1/2})\eta\rangle
=\langle\eta,B\eta\rangle ,
\end{equation}
using commutation of \(B\) with \(T\) and with \((I-T^*T)^{1/2}\) in both summands, which turns the bracket into \((T^*T+I-T^*T)B=B\). The overlap of these two vectors is \(\langle\xi,T\eta\rangle\), and the fidelity is by definition \cite{doi:10.1016/0034-4877(76)90060-4} the supremum of the absolute values of such overlaps over all common representations; so \(|\langle\xi,T\eta\rangle|\) is at most the fidelity.

For the opposite inequality, that every overlap in a common representation is at most the supremum \eqref{eq:commutant_fidelity}, let \(\pi'\) be a representation of \(\mathcal B_E\) on a Hilbert space \(\mathcal K'\) and let \(\xi',\eta' \in\mathcal K'\) implement the same states as \(\xi\) and \(\eta\): \(\langle\xi',\pi'(B)\xi'\rangle=\langle\xi,B\xi\rangle\) for all \(B\in\mathcal B_E\), and likewise for \(\eta\). 
On vectors \(B\xi\), define \(U_\xi(B\xi)=\pi'(B)\xi'\), a map from the subspace \(\mathcal B_E\xi\) of \(\mathcal K\) into \(\mathcal K'\). This prescription is independent of the chosen
\(B\). Indeed, if \((B-C)\xi=0\), equality of the two states on \((B-C)^*(B-C)\) implies \(\|\pi'(B-C)\xi'\|=0\). More generally
\begin{equation}
 \langle B\xi,C\xi\rangle=\sigma_\xi(B^*C)  =\langle\pi'(B)\xi',\pi'(C)\xi'\rangle,\qquad
 \sigma_\xi(D)=\langle\xi,D\xi\rangle ,
\end{equation}
so \(U_\xi\) preserves inner products. It therefore extends by continuity isometrically to the closure \(\overline{\mathcal B_E\xi}\) of this subspace (here the bar is the norm closure of a subspace of \(\mathcal K\)). Extend by zero on its orthogonal complement, this complement is invariant under every \(B\in\mathcal B_E\), because for \(z\) in the complement, \(\langle C\xi,Bz\rangle=\langle B^*C\xi,z\rangle=0\). It follows on both subspaces that \(U_\xi B=\pi'(B)U_\xi\).
Construct \(U_\eta\) in the same way. Taking the adjoint of the identity for \(B^*\) gives \(B U_\xi^*=U_\xi^*\pi'(B)\), and hence
\begin{equation}
 B U_\xi^*U_\eta=U_\xi^*\pi'(B)U_\eta=U_\xi^*U_\eta B .
\end{equation}
Thus \(U_\xi^*U_\eta\) is a contraction in \(\mathcal B_E'\). Because \(U_\xi\xi=\xi'\) and \(U_\eta\eta=\eta'\), its pairing equals \(\langle\xi',\eta'\rangle\). Taking suprema over the common representations shows that the fidelity is at most the supremum in \eqref{eq:commutant_fidelity}. With the first inequality this proves \eqref{eq:commutant_fidelity}.

Set \(\xi=Vx,\eta=Vy\). By von Neumann's bicommutant theorem a von Neumann algebra is the commutant of its commutant, so \(\mathcal B_E'=\pi(\A)''=\pi(\A)\), since the image \(\pi(\A)\) of a normal representation of a von Neumann algebra is weakly closed and is therefore a von Neumann algebra. To pass from a contraction \(T\) in this image to one in \(\A\), use the fact that \(\ker\pi=z\A\) for a central projection \(z\) of \(\A\), one that commutes with all of \(\A\); the kernel of a normal representation always has this form. Restriction to \((1-z)\A\) is an injective \(*\)-homomorphism onto \(\pi(\A)\) and is therefore isometric.
The unique preimage there has norm \(\|T\|\), so it supplies a contraction \(X\in\A\).
The dilation vectors have sector supports \(\pi(p_v)Vx=Vx,\ \pi(p_w)Vy=Vy\). Hence
\begin{equation}
\langle Vx,\pi(X)Vy\rangle =\langle Vx,\pi(p_vXp_w)Vy\rangle =\langle x,\Phi(p_vXp_w)y\rangle .
\end{equation}
Compressing \(X\) this way preserves the pairing and cannot increase its norm.  By \eqref{eq:commutant_fidelity} with \(\xi=Vx\) and \(\eta=Vy\), the fidelity of the two records is the supremum of \(|\langle Vx,\pi(X)Vy\rangle|\) over contractions \(X\in\A\), and each of these equals \(|\langle x,\Phi(p_vXp_w)y\rangle|\). We have thus proved the normal-state version of \eqref{eq:pair_information},
\begin{equation}
\sup_{\substack{X\in p_v\A p_w\\\norm X\leq1}}|\langle x,\Phi(X)y\rangle|
=\ff(\sigma_v^E(x),\sigma_w^E(y)).
\end{equation}
When \(\pi(X)=X\otimes I_E\) on \(\HH\otimes\HH_E\), the complementary algebra is the commutant of the operators \(X\otimes I_E\), \(X\in\A\). By the commutation theorem for tensor products of von Neumann algebras \cite{RieffelVanDaele1975} it is \(\A'\,\bar\otimes B(\HH_E)\): the von Neumann algebra generated by the operators \(a\otimes b\), with \(a\) in the commutant \(\A'\) of \(\A\) in \(B(\HH)\) and \(b\in B(\HH_E)\). 

For \(\A=B(\HH)\),  where the commutant is \(\A'=\CZ I\), the complementary algebra is \(I\otimes B(\HH_E)\), because \(\CZ I\,\bar\otimes B(\HH_E)\) consists of the operators \(I\otimes b\), and its vector states are exactly the environment marginals used in 
\eqref{eq:conditional_environment}. For a proper observable subalgebra, the complementary algebra contains \(\A'\otimes I\) as well, since \(a\otimes I\) commutes with every \(X\otimes I_E\) when \(a\in\A'\): the records then include the operators on the system itself that commute with every accessible observable.

The optimized corner-weight identity \eqref{eq:corner_information}  follows with suprema. To check that the supremum over corner contractions equals the supremum over corner partial isometries, let \(X=U|X|\) be a contraction in \(p_v\A p_w\), in its polar decomposition. We approximate \(|X|\) by an average of its spectral projections. The spectral projections \(e_{j,n}=\one_{[j/n,1]}(|X|)\), \(j=1,\ldots,n\) (\(n\) fixes the step \(1/n\) and \(j\) labels the threshold \(j/n\)), satisfy \(e_{j,n}\leq\supp |X|\leq p_w\), because \(|X|\) vanishes on \(\ker X\), which contains \((I-p_w)\HH\). Each \(Ue_{j,n}\) is a corner partial isometry because \((Ue_{j,n})^*(Ue_{j,n})=e_{j,n}\).   For a scalar \(\lambda\in[0,1]\), there are exactly \(\lfloor n\lambda\rfloor\) integers \(1\leq j\leq n\) with \(j/n\leq\lambda\). The average \(\tfrac1n\sum_j\one_{[j/n,1]}(\lambda)\) of the indicator functions \(\one_{[j/n,1]}\), i.e., one on the interval  \([j/n,1]\) and zero off it, is \(\lfloor n\lambda\rfloor/n\), within \(1/n\) of \(\lambda\).  The functional calculus of the positive contraction \(|X|\) satisfies \(\norm{g(|X|)}\leq\sup_{\lambda\in[0,1]}|g(\lambda)|\) for a bounded Borel function \(g\).  Take \(g\) to be the difference between \(\lambda\) and the average just computed, \(g(\lambda)=\lambda-\tfrac1n\sum_j\one_{[j/n,1]}(\lambda)\), its supremum is at most \(1/n\), and \(g(|X|)=|X|-\tfrac1n\sum_je_{j,n}\), since \(\one_{[j/n,1]}(|X|)=e_{j,n}\). Hence \(\norm{|X|-\tfrac1n\sum_je_{j,n}}\leq1/n\), and multiplication by \(U\), which has norm at most one and satisfies \(X=U|X|\),  gives
\begin{equation}
\norm{X-\frac1n\sum_{j=1}^n Ue_{j,n}}\leq\frac1n .
\end{equation}
 Hence \(\norm{\Phi(X)}\leq\frac1n\sum_j\norm{\Phi(Ue_{j,n})}+\frac1n\) for every \(n\):  linearity and contraction of \(\Phi\) bound \(\norm{\Phi(X)}\) by the supremum on corner partial isometries. Taking the supremum of \(|\langle x,\Phi(X)y\rangle|\) over unit preparations, which is \(\norm{\Phi(X)}\), then yields \eqref{eq:corner_information} in this representation.
 Thus we have shown that the two identities of Theorem~\ref{thm:information_weights}, \eqref{eq:pair_information} and \eqref{eq:corner_information}, hold for a normal unital completely positive map on a von Neumann algebra with fixed sectors, with the record defined as a state on the commutant \(\pi(\A)'\) in place of a partial trace.

\bibliographystyle{unsrt}
\bibliography{references}

\end{document}